\documentclass[sigconf, 9pt]{acmart}
\usepackage[utf8]{inputenc}
\usepackage{amsmath}
\usepackage{amsfonts}
\usepackage{algorithmic}
\usepackage{graphicx}
\usepackage{textcomp}
\usepackage{xcolor}
\usepackage{epsfig}
\usepackage{fixmath}
\usepackage{multirow}
\usepackage{siunitx}
\usepackage{algorithm}
\usepackage[algo2e]{algorithm2e}
\usepackage{setspace}
\usepackage[final, commandnameprefix=always]{changes}
\usepackage{soul}
\usepackage{newunicodechar}
\newunicodechar{、}{,}
\usepackage{booktabs}
\usepackage{url}
\usepackage{cancel}
\usepackage{tabularx}
\usepackage{environ} 
\usepackage{placeins} 
\usepackage{caption}
\usepackage{subcaption}
\usepackage{etoolbox}
\usepackage{stfloats}
\usepackage{pifont} 
\newcommand{\cmark}{\ding{51}}%
\newlength{\figw}
\newlength{\figsep}
\AtBeginDocument{%
  \setlength{\abovedisplayskip}{3pt plus 1pt minus 1pt}%
  \setlength{\belowdisplayskip}{3pt plus 1pt minus 1pt}%
  \setlength{\abovedisplayshortskip}{1pt plus 1pt}%
  \setlength{\belowdisplayshortskip}{1pt plus 1pt}%
}

\makeatletter
\patchcmd{\section}{-3.5ex \@plus -1ex \@minus -.2ex}{-1.8ex \@plus -0.5ex \@minus -.1ex}{}{}
\patchcmd{\section}{2.3ex \@plus.2ex}{0.8ex \@plus .1ex}{}{}
\patchcmd{\subsection}{-3.25ex\@plus -1ex \@minus -.2ex}{-1.2ex \@plus -0.5ex \@minus -.1ex}{}{}
\patchcmd{\subsection}{1.5ex \@plus .2ex}{0.6ex \@plus .1ex}{}{}
\patchcmd{\subsubsection}{-3.25ex\@plus -1ex \@minus -.2ex}{-1.0ex \@plus -0.3ex \@minus -.1ex}{}{}
\patchcmd{\subsubsection}{1.5ex \@plus .2ex}{0.4ex \@plus .1ex}{}{}
\makeatother

\usepackage{array}
\usepackage{enumitem} 
\setlist[itemize]{labelsep=0.6em} 
\usepackage{fancyhdr}
\usepackage{indentfirst}
\usepackage{amsmath}
\usepackage{comment}

\setcopyright{none}
\renewcommand\footnotetextcopyrightpermission[1]{}

\newcommand{\red}[1] {\textcolor[rgb]{0.00,0.00,0.00}{#1}}
\newcommand{\blue}[1] {\textcolor[rgb]{0.00,0.00, 0.00}{#1}}

\newcommand{\green}[1]{\textcolor[rgb]{0, 0, 0}{#1}}
\newcommand{\greenn}[1]{\textcolor[rgb]{0, 0, 0}{#1}}

\definecolor{red}{rgb}{0,0,0}
\newcommand{\bluee}[1]{\textcolor[rgb]{0.00,0.00,0.00}{#1}}
\newlength{\layoutw}
\newlength{\mapw}
\newlength{\driftw}
\newlength{\setupw}
\newlength{\stftw}

\begin{document}
\sloppy
\raggedbottom

\title{RadarCom: Unlocking Simultaneous Communication and Sensing on Commodity Automotive mmWave Radars}

\author{Yi Wei}
\affiliation{%
  \institution{Saint Louis University}
  \city{St. Louis}
  \state{MO}
  \country{USA}}
\email{yi.wei.1@slu.edu}

\author{Yue Zhang}
\affiliation{%
  \institution{Saint Louis University}
  \city{St. Louis}
  \state{MO}
  \country{USA}}
\email{yue.zhang.5@slu.edu}

\author{Peijun Hou}
\affiliation{%
  \institution{Saint Louis University}
  \city{St. Louis}
  \state{MO}
  \country{USA}}
\email{peijun.hou@slu.edu}

\author{Qiuyu Liao}
\affiliation{%
  \institution{Saint Louis University}
  \city{St. Louis}
  \state{MO}
  \country{USA}}
\email{qiuyu.liao@slu.edu}

\author{Nan Cen}
\affiliation{%
  \institution{Saint Louis University}
  \city{St. Louis}
  \state{MO}
  \country{USA}}
\email{nan.cen@slu.edu}

\noindent \begin{abstract}
\noindent 
\red{Vehicle-to-vehicle (V2V) communication can improve road safety and traffic efficiency, yet its deployment remains limited by the need for dedicated communication hardware and supporting infrastructure. Meanwhile, production vehicles are already widely equipped with millimeter-wave radars for sensing. 
To unlock their communication potential, this paper presents RadarCom, a software-defined system that enables direct radar-to-radar communication on unmodified commercial off-the-shelf (COTS) automotive mmWave radars while preserving their native sensing functionality.
The key insight is that mutual inter-radar interference, traditionally viewed as harmful, can be repurposed as a communication carrier. Based on this insight, RadarCom introduces slope-mismatch-based discovery, adaptive chirp synchronization, differential coding on range-FFT, and sensing-aware interference suppression to enable simultaneous communication and sensing without hardware modification, waveform redesign, or infrastructure support. We implement RadarCom on TI AWR2243 automotive radars and evaluate it under static, dynamic, and challenging real-world conditions. Results show an average effective data rate of $5.55$~kbps with symbol error rates below $1.5$\%, while maintaining sensing F1 scores of up to $80.5$\% during concurrent communication. These results demonstrate the feasibility of transforming commodity automotive radars into dual-purpose communication and sensing devices.}

\end{abstract}


\maketitle
\pagestyle{plain}
\pagestyle{empty}


\section{Introduction}\label{sec: intro}

\begin{figure}[t]
    \centering
    \includegraphics[width=0.70\columnwidth]{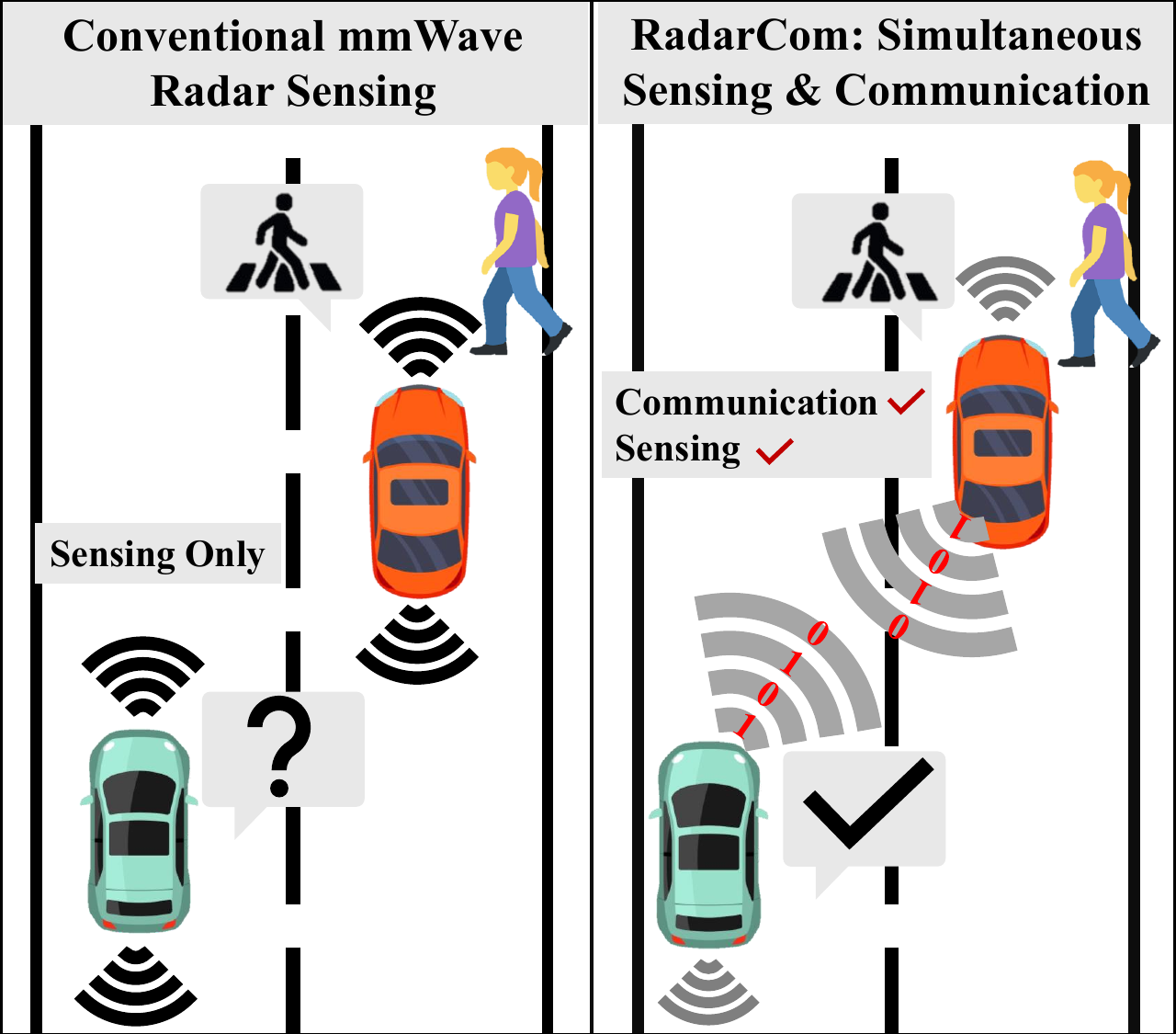}
    \caption{RadarCom enables simultaneous sensing and communication on unmodified COTS mmWave radars without additional hardware or infrastructure support.}\vspace{-2mm}
    \label{fig:overview}
\end{figure}

\noindent Vehicle-to-vehicle (V2V) communication has long been recognized as a key technology for improving road safety and traffic efficiency. By allowing nearby vehicles to exchange safety-critical information such as position, velocity, and hazard alerts in real time, V2V communication can help prevent accidents and support cooperative driving \cite{wang2025cooperative,zhang2025vehicle,wang2020demystifying}. However, despite decades of research and development, V2V communication has not been widely deployed in commercial vehicles. Its adoption has been slowed by the long-standing competition between different communication standards, such as Dedicated Short-Range Communications (DSRC) and Cellular V2X (C-V2X), as well as the economic cost of deploying and maintaining supporting communication infrastructure \cite{wang2020demystifying,usman2020business,cucor2022physical,cigno2022cooperative}. 

Nowadays, most vehicles are already widely equipped with millimeter-wave (mmWave) radars for perception~\cite{han20234d, peng20254d}, given their accurate ranging and velocity estimation, robustness to adverse weather and lighting, and low cost \cite{lu2020see,zhang2023push,xue2021mmmesh}. \blue{As a result, multiple mmWave radars are now commonly integrated into 
production vehicles to support advanced driver assistance systems (ADAS), such as adaptive cruise control, blind-spot detection, and collision avoidance \cite{soltanaghaei2021millimetro,kong2024survey,zhao2020m}. This widespread deployment has motivated researchers to extend mmWave radar beyond perception and to explore the possibility of radar-based inter-vehicle communication. Several recent works have investigated joint radar communication or integrated sensing and communication (ISAC) on mmWave platforms \cite{gaofeng2024co,tao2025mmwave,8828030,bae2022omniscatter,okubo2024integrated,aydogdu2019radchat,liang2023ieee}. However, existing approaches generally suffer from one or more of the following limitations: \textbf{(i)} \red{they require customized waveform design that is difficult to realize on COTS automotive radars} \cite{gaofeng2024co,tao2025mmwave,8828030},
\textbf{(ii)} 
\red{they rely on dedicated additional hardware to enable communication, increasing both deployment cost and integration overhead}
\cite{bae2022omniscatter,okubo2024integrated}, \textbf{(iii)} 
\red{they fail to preserve the radar's native sensing performance during communication, leading to degraded perception accuracy}
\cite{sanchez2025radar}, or \textbf{(iv)} 
\red{they require centralized synchronization or infrastructure support, limiting practical deployment in decentralized V2V environments}
\cite{aydogdu2019radchat,liang2023ieee}. }

As summarized in Table~\ref{tab:comparison}, existing systems cannot simultaneously satisfy all of these requirements,  
limiting their practicality in decentralized vehicular deployments. 
This naturally raises the following question: 
\textit{\textbf{Can we enable direct radar-to-radar communication on unmodified COTS automotive radars, without additional hardware,  customized waveform design, or centralized coordination, while preserving sensing performance?}} To answer this question, we propose \textbf{RadarCom}, a software-defined system that unlocks simultaneous sensing and direct radar-to-radar communication \red{using a unified waveform} on unmodified COTS automotive mmWave radars. 
RadarCom is motivated by a simple but overlooked observation. 
As automotive radars become increasingly ubiquitous, 
mutual interference among nearby radars is inevitable: signals transmitted by one radar are naturally received by neighboring radars and appear as interference in their sensing output. 
While this effect is traditionally regarded as harmful and has motivated extensive work on interference mitigation \cite{alland2019interference, 9337437}, RadarCom takes a fundamentally different perspective.
\red{Our {key insight} is that these naturally received radar signals can serve as a communication carrier. Built on this insight,the {core idea} of RadarCom is to encode information through software-level reconfiguration of standard FMCW chirp parameters, enabling one radar directly convey information to another using the same waveform employed for sensing, without compromising the radar's native sensing functionality.}

\newlength{\panelw}
\newlength{\pairw}

However, realizing such direct radar-to-radar communication on COTS mmWave radars remains highly non-trivial. Automotive radars are designed 
for monostatic sensing: they transmit chirps and receive and process their own reflected echoes, 
rather than coordinating with neighboring radars~\cite{moussa2021enhanced,9337437}. Without centralized synchronization, a radar's transmitted chirps may not fall within another radar's narrow Intermediate Frequency (IF) observation window \green{(e.g., 5 to 25 MHz)} \cite{yang2022interference}, making radar discovery and synchronization unreliable and challenging. Moreover, once a signal is captured, communication and sensing must share the same hardware, spectrum, and raw measurements. As a result, communication signals can introduce unintended interference into the sensing pipeline, potentially degrading sensing performance. To enable direct radar-to-radar communication on unmodified COTS radars while preserving native sensing functionality and avoiding additional hardware or infrastructure support, RadarCom must overcome the following three key challenges:

\begin{table}[t]
\vspace{+2mm}
\centering
\resizebox{\columnwidth}{!}{%
\begin{tabular}{l@{\hspace{8pt}}ccccccc}
\toprule
\multirow{2}{*}{System} 
& Radar-to & COTS & Sensing & No extra & No central & No extra & Real-world \\
& -radar & Radar & preserv. & HW & sync & infra. & deploy. \\
\midrule
OmniScatter~\cite{bae2022omniscatter}       &        & \cmark & \cmark &        & \cmark & \cmark & \cmark \\
BiScatter~\cite{okubo2024integrated}        &        & \cmark & \cmark &        & \cmark & \cmark & \cmark \\
Cui et al.~\cite{gaofeng2024co}             &        &        & \cmark & \cmark & \cmark &        &        \\
Liang et al.~\cite{liang2023ieee}           &        &        & \cmark & \cmark &        &        &        \\
S{\'a}nchez-A.~\cite{sanchez2025radar}      & \cmark & \cmark &        & \cmark & \cmark & \cmark & \cmark \\
RadChat~\cite{aydogdu2019radchat}           &        &        & \cmark &        &        & \cmark &        \\
Tao et al.~\cite{tao2025mmwave}             &        &        & \cmark &        & \cmark & \cmark &        \\
Mishra et al.~\cite{8828030}                &        &        & \cmark &        & \cmark & \cmark &        \\
\midrule
\textbf{Ours}                               & \cmark & \cmark & \cmark & \cmark & \cmark & \cmark & \cmark \\
\bottomrule
\end{tabular}%
}
\caption{Comparison with prior radar communication and ISAC systems.}
\vspace{-2mm}
\label{tab:comparison}
\end{table}

\textbf{I. How can independent mmWave radars discover one another and establish an initial communication channel while sensing?}
As discussed above, our key insight is to repurpose mutual interference among independent automotive radars as a communication carrier, rather than regarding it solely as interference. However, before communication can take place, independent radars must first discover one another and establish an initial communication channel while continuing their normal sensing operation. This is challenging because automotive radars are designed for monostatic sensing and lack mechanisms for inter-radar discovery or coordination. Without prior synchronization, signals transmitted by one radar may not consistently appear within another radar's IF observation window, making discovery unreliable. Moreover, the receiving radar must distinguish the intended radar from environmental reflections and interference generated by other nearby radars.
To address this challenge, RadarCom introduces a slope-mismatch-based discovery mechanism. A transmitting radar periodically emits beacon chirps with predefined parameters, while a receiving radar searches for their distinctive frequency-sweeping signatures. \green{This design} enables reliable inter-radar discovery and initial communication channel establishment on unmodified COTS platforms, laying the foundation for subsequent synchronization and data communication.

\textbf{II. How to establish and maintain a reliable communication link between independent radars in dynamic environments?}
After discovery and establishing an initial communication channel, independent radars must maintain synchronization and reliably exchange data without centralized coordination. This is challenging because even small timing misalignments can cause received chirps to fall outside the observable IF band. Furthermore, due to inevitable hardware imperfections introduced during manufacturing, practical radar systems are affected by sampling-frequency offset (SFO) and carrier-frequency offset (CFO) \cite{ti_awr2243_datasheet, zhang2022integration, alemdar2021rfclock}, while vehicle mobility continuously changes the propagation delay between communicating radars.
In FMCW radars, these effects manifest as spectral peak shifts across chirps, disrupting synchronization and distorting symbol decoding. To address this challenge, we design an adaptive synchronization mechanism and a differential coding scheme operating on range-FFT outputs. The synchronization mechanism continuously estimates and compensates for timing and frequency offsets to maintain chirp alignment, while the differential coding scheme encodes and decodes information using the relative spectral shifts between adjacent chirps, thus enabling reliable communication despite hardware imperfections, synchronization uncertainty, and dynamic mobility.

\textbf{III. How to preserve sensing performance while enabling simultaneous communication?} 
\red{Once communication is enabled, the communication-bearing radar signals are directly captured by the receiving radar and can dominate the sensing measurements, introducing substantial interference into the sensing pipeline. These signals are often significantly stronger than environmental reflections and can therefore elevate the noise floor, generate false detections, and mask weak targets in the sensing output.}

To address this challenge, we exploit a key observation: communication-induced interference vary sharply across chirps, while real target reflections remain relatively stable within a frame. Building on this observation, we design a novel sensing-aware interference suppression mechanism that reconstructs corrupted range-FFT samples through cross-chirp \green{reconstruction} and removes residual false detections through sub-frame consistency verification, thereby preserving native sensing performance during simultaneous communication.

\begin{figure*}[t]
    \centering

    \setlength{\figsep}{0.016\linewidth}
    \setlength{\layoutw}{0.181\linewidth} 
    \setlength{\mapw}{0.190\linewidth}     
    \setlength{\driftw}{0.190\linewidth} 

    \setlength{\setupw}{\dimexpr 2\layoutw+\figsep\relax}
    \setlength{\stftw}{\dimexpr 2\mapw+\figsep\relax}

    \begin{minipage}[t]{\setupw}
        \centering
        \vspace{0pt}

        \begin{minipage}[t]{\layoutw}
            \centering
            \includegraphics[width=\linewidth]{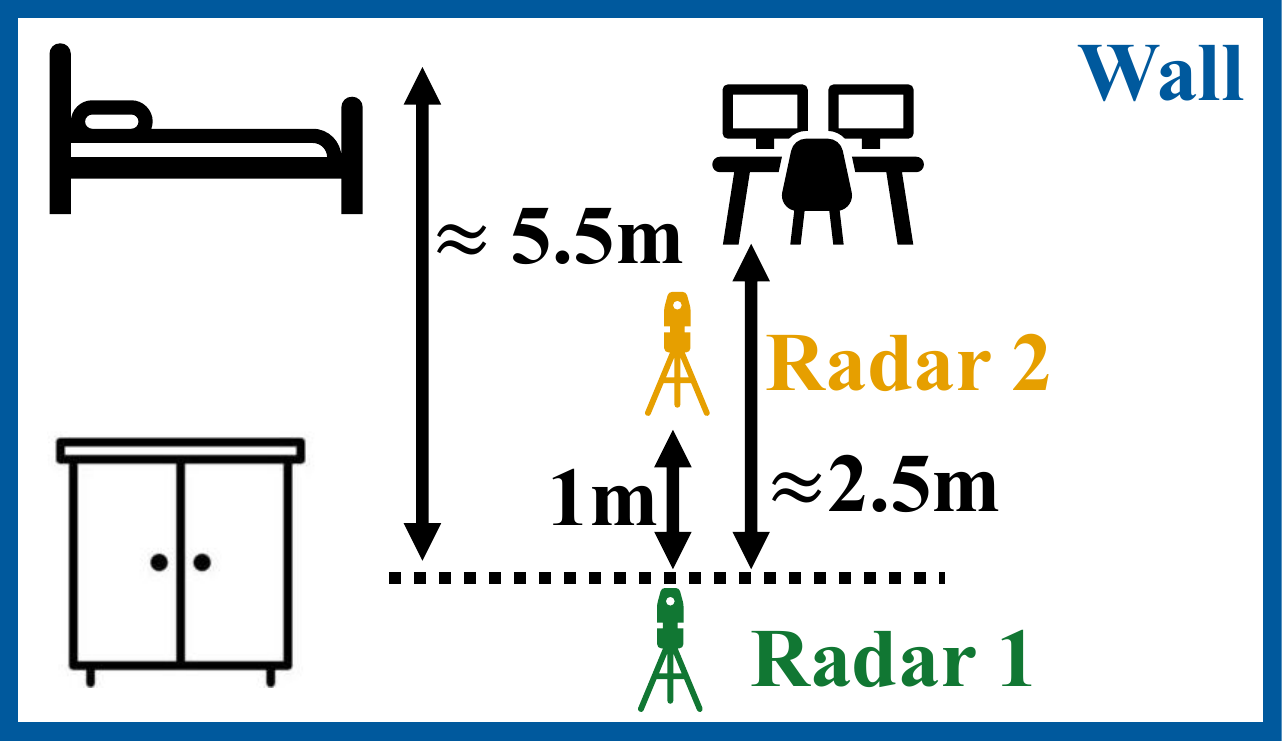}
            \par\vspace{-0.5mm}
            {\small\textbf{(a)} \textbf{Layout Diagram}\par}
        \end{minipage}%
        \hspace{\figsep}%
        \begin{minipage}[t]{\layoutw}
            \centering
            \includegraphics[width=\linewidth]{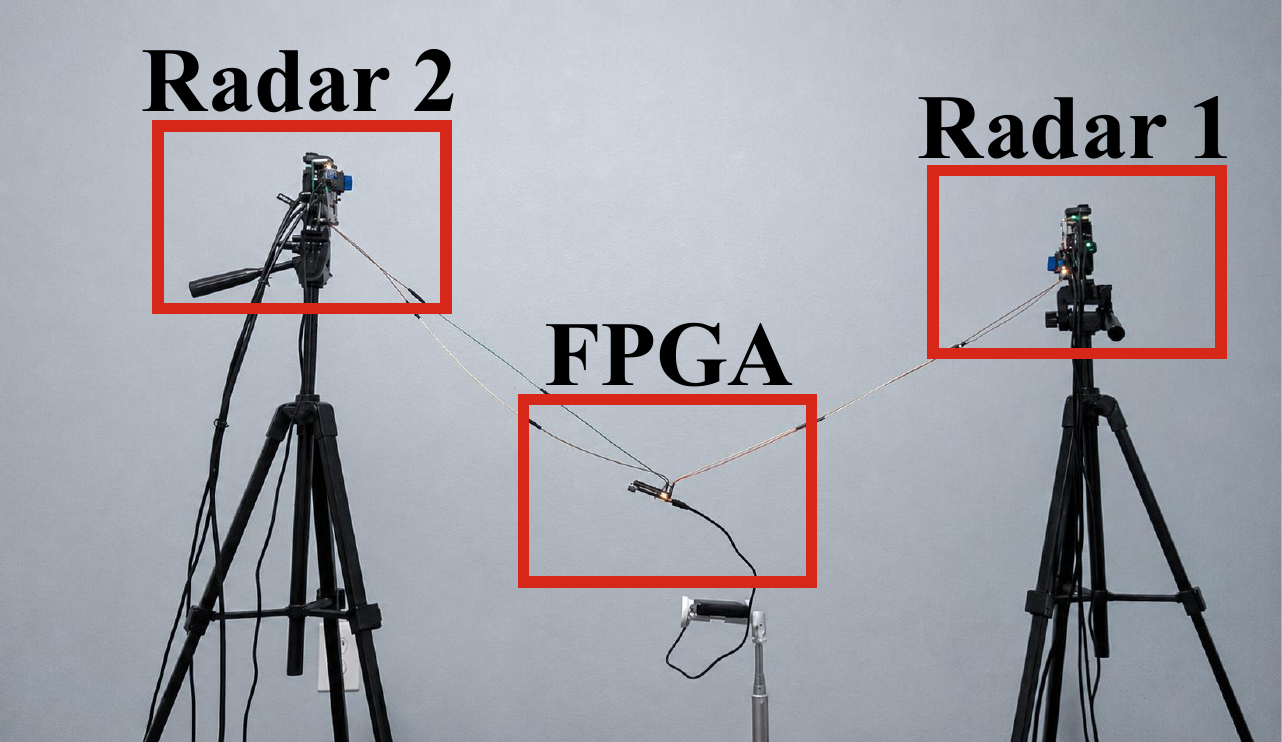}
            \par\vspace{-0.5mm}
            {\small\textbf{(b)} \textbf{Device Photo}\par}
        \end{minipage}

        \vspace{-0.5mm}
        \caption{\small\textbf{Preliminary Experimental Setup.}}
        \label{fig:experiment_setup}
    \end{minipage}%
    \hspace{\figsep}%
    \begin{minipage}[t]{\stftw}
        \centering
        \vspace{0pt}

        \begin{minipage}[t]{\mapw}
            \centering
            \includegraphics[width=\linewidth]{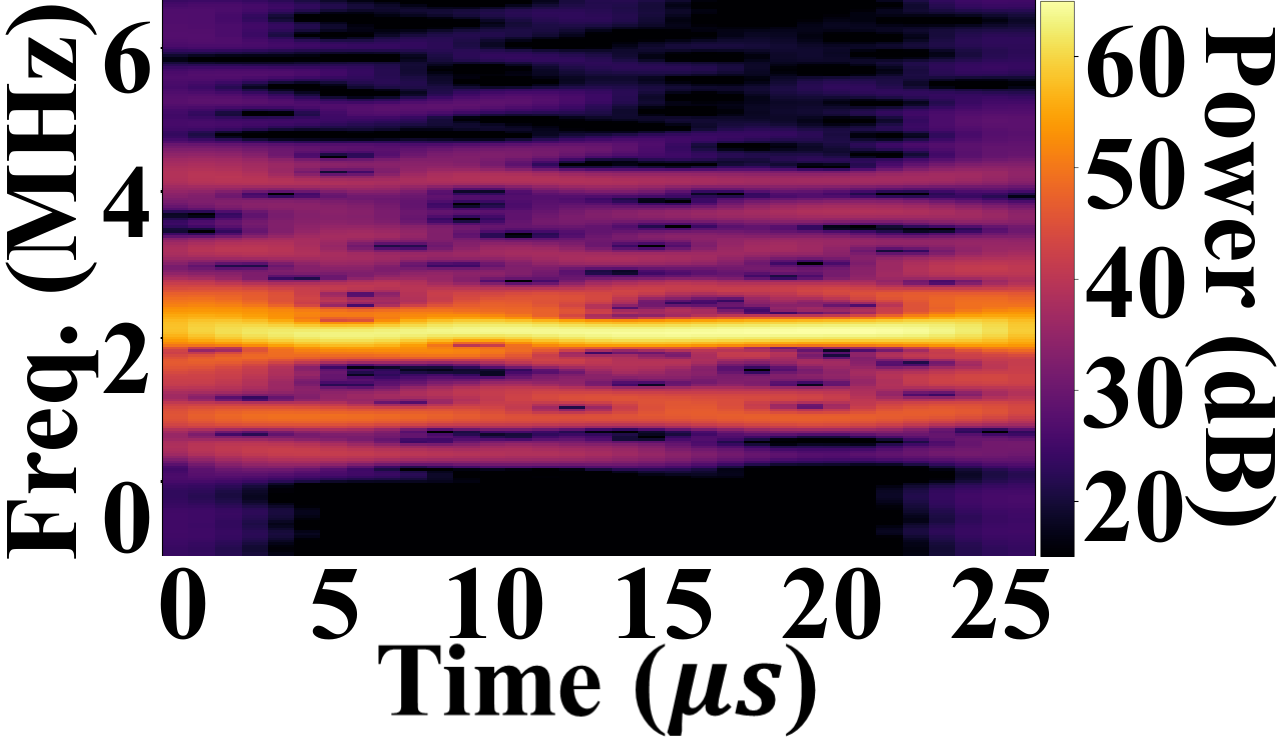}
            \par\vspace{-1.5mm}
            {\small\textbf{(a)} \textbf{Radar 1 active}\par}
        \end{minipage}%
        \hspace{\figsep}%
        \begin{minipage}[t]{\mapw}
            \centering
            \includegraphics[width=\linewidth]{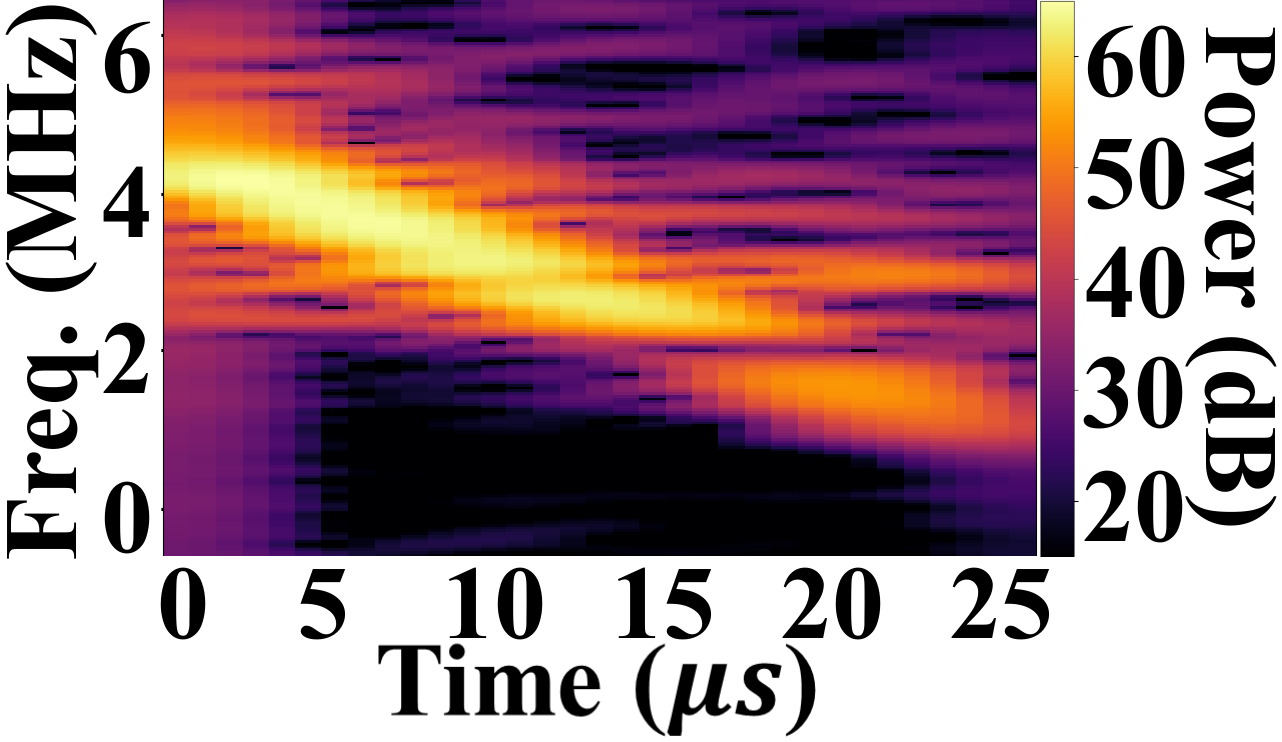}
            \par\vspace{-1.5mm}
            {\small\textbf{(b)} \textbf{Both radars active}\par}
        \end{minipage}

        \vspace{-0.5mm}
        \caption{\small One-chirp STFT maps.}
        \label{fig:stft_analysis}
    \end{minipage}%
    \hspace{\figsep}%
    \begin{minipage}[t]{\driftw}
        \centering
        \vspace{0pt}

        \includegraphics[width=\linewidth]{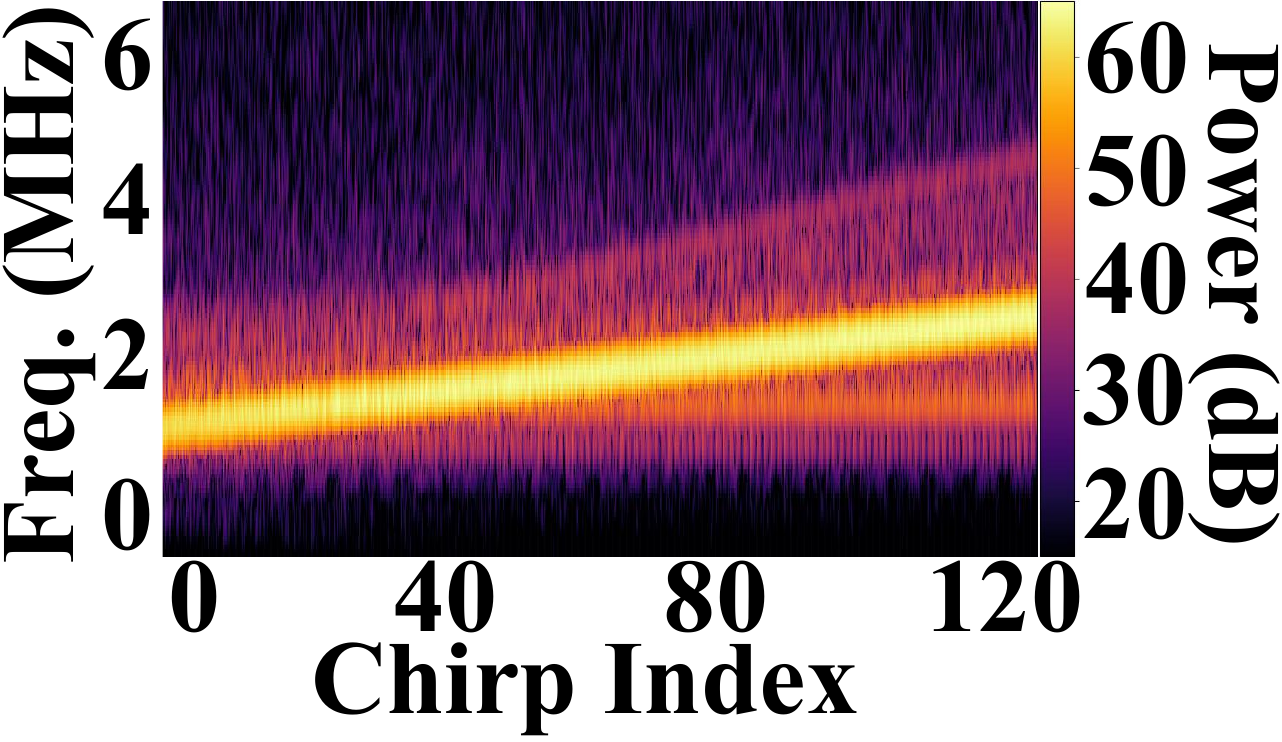}

        \vspace{1pt}
        {\small\phantom{\textbf{(a)} Drift}\par}

        \par\vspace{-6mm}
        \caption{\small Frequency drift caused by SFO.}
        \label{fig:chirp_drift}
    \end{minipage}

    \vspace{-3.5mm}
\end{figure*}

\begin{figure*}[t]
\setlength{\figsep}{0.008\linewidth}
\setlength{\figw}{0.244\linewidth}
    \centering

    \begin{minipage}[t]{\dimexpr3\figw+2\figsep\relax}
        \centering
        \vspace{0pt}

        \begin{subfigure}[t]{\figw}
            \centering
            \vspace{0pt}
            \includegraphics[width=0.9\linewidth]{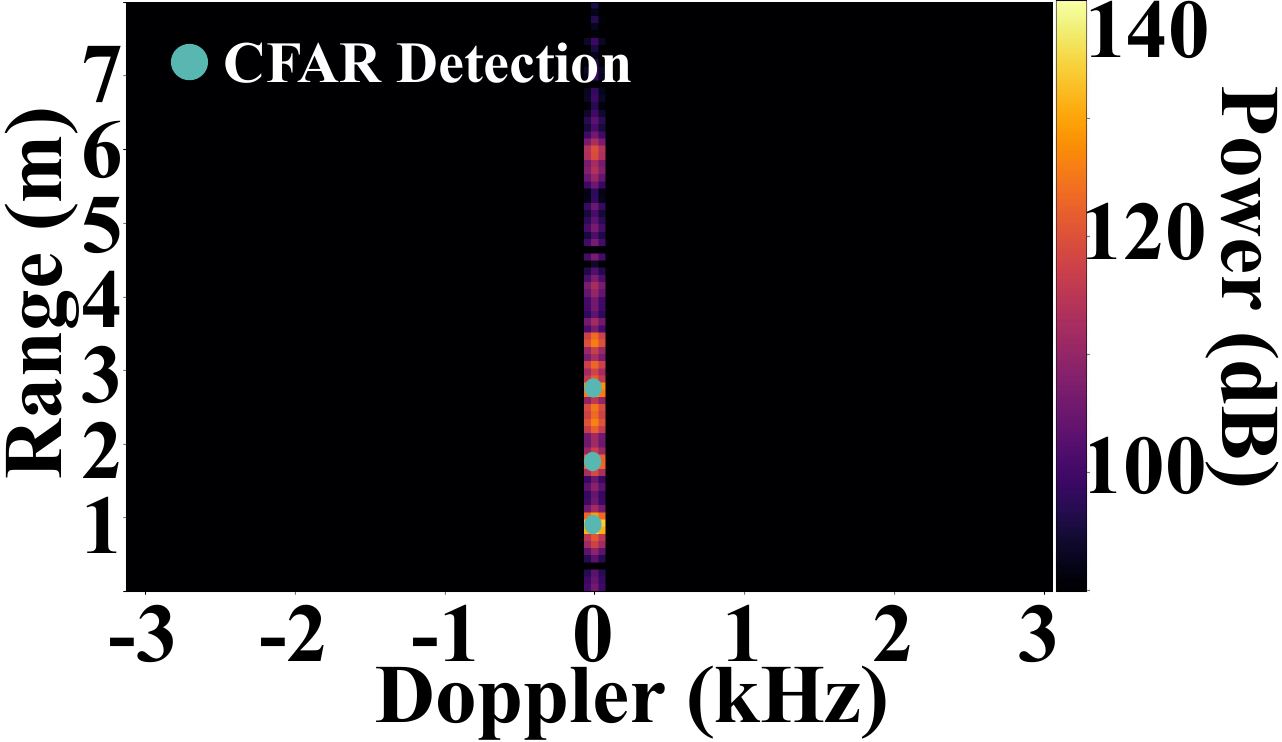}
            \caption{RD (Radar~2 inactive)}
            \label{fig:rd_clean}
        \end{subfigure}%
        \hspace{\figsep}%
        \begin{subfigure}[t]{\figw}
            \centering
            \vspace{0pt}
            \includegraphics[width=0.9\linewidth]{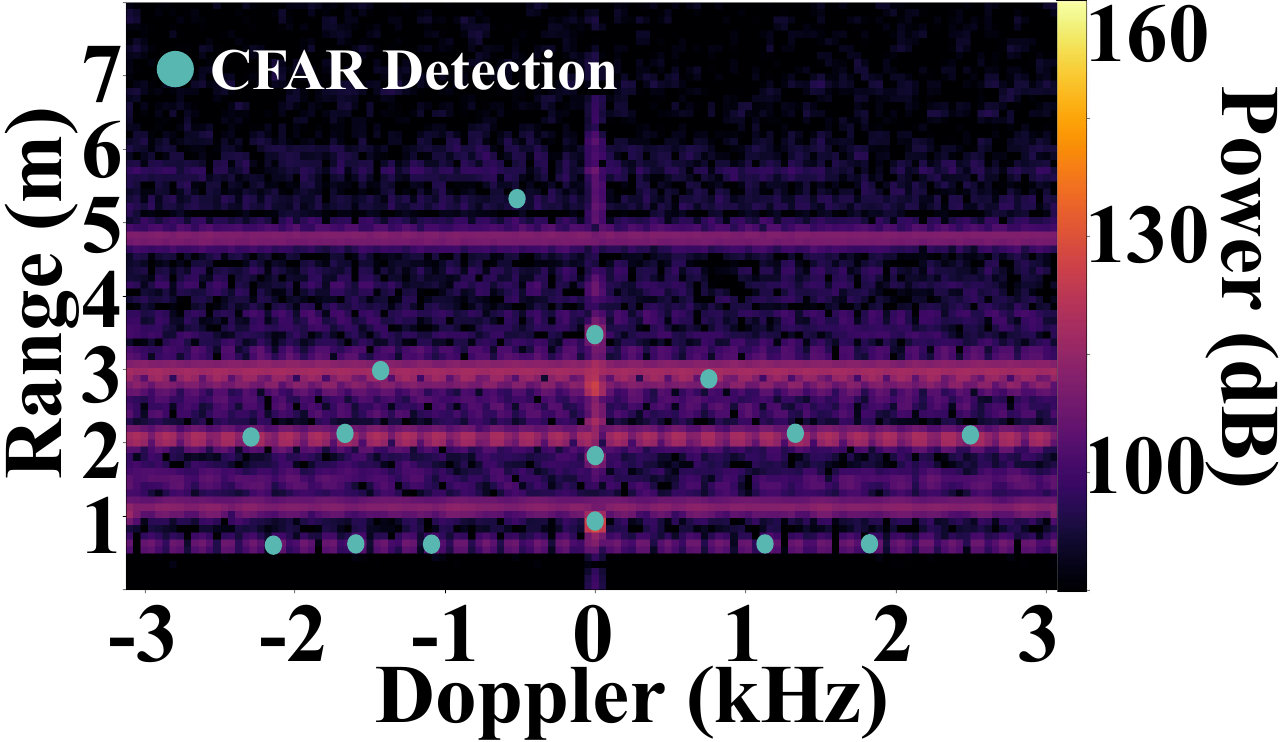}
            \caption{RD (Radar~2 active)}
            \label{fig:rd_interference}
        \end{subfigure}%
        \hspace{\figsep}%
        \begin{subfigure}[t]{\figw}
            \centering
            \vspace{0pt}
            \includegraphics[width=0.9\linewidth]{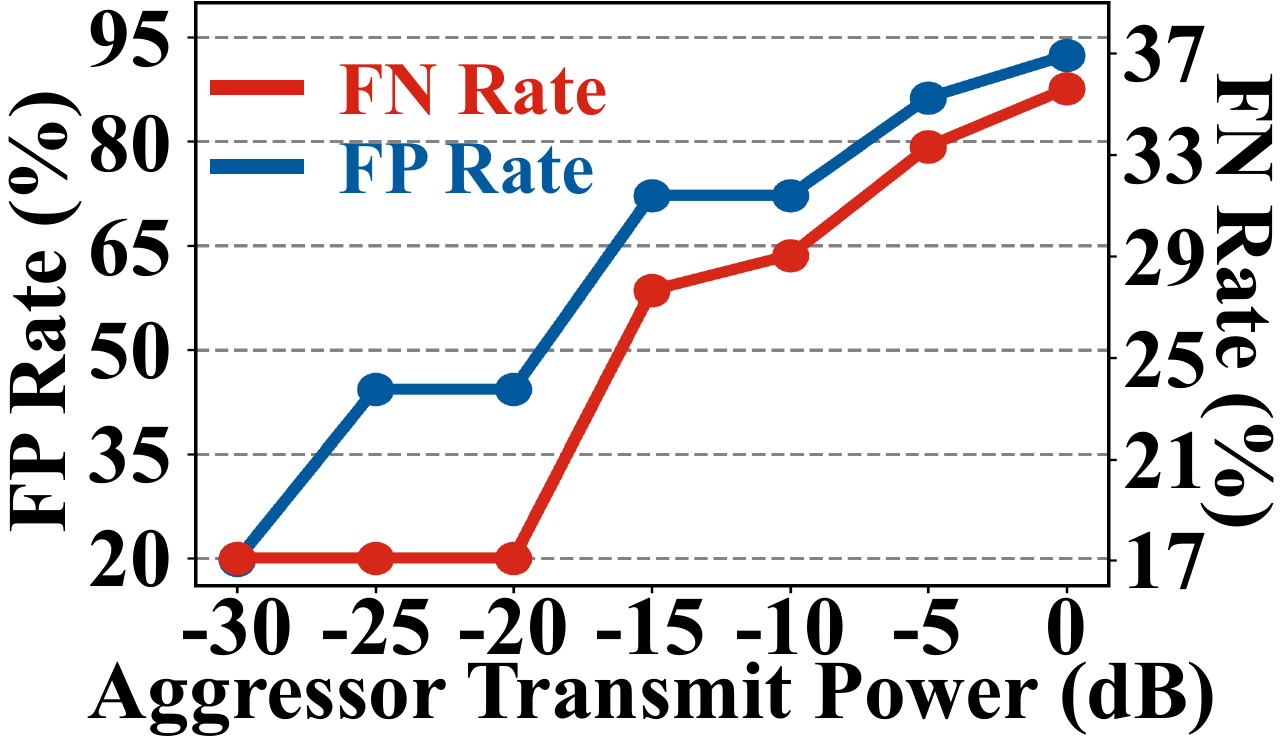}
            \caption{Error Rates vs. Power}
            \label{fig:error_rates}
        \end{subfigure}
        \vspace{-1mm}
        \caption{Impact of inter-radar interference on sensing.}
        \vspace{-1mm}
        \label{fig:combined_analysis}
    \end{minipage}%
    \hspace{\figsep}%
    \begin{minipage}[t]{\figw}
        \centering
        \vspace{0pt}
        \includegraphics[width=0.9\linewidth]{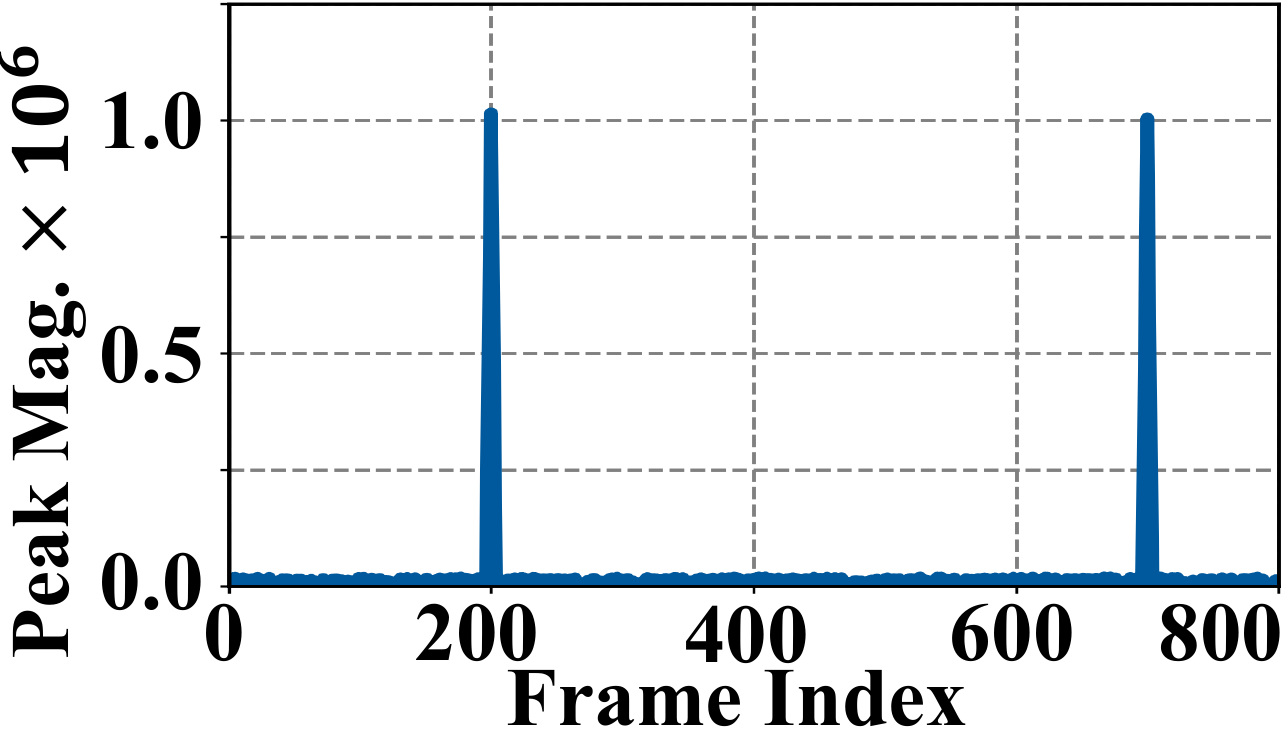}
        \vspace{+3mm}
        \caption{Discovery Rate.}
        \label{fig:discovery_rate}
    \end{minipage}%
    \vspace{-1.5mm}
\end{figure*}

We summarize our key contributions as follows:

$\bullet$ To the best of our knowledge, RadarCom is the \textbf{first} system to realize simultaneous direct radar-to-radar communication and sensing on unmodified COTS $77$~GHz automotive radars. By operating entirely within the native FMCW sensing framework, RadarCom achieves communication through software-level reconfiguration of standard chirp parameters while preserving normal sensing operation, without requiring dedicated hardware, customized waveform design, or centralized coordination and infrastructure.

$\bullet$ \red{We propose a new communication paradigm that repurposes naturally occurring inter-radar interference as an information carrier for direct radar-to-radar communication. To make this feasible on unmodified COTS radars, we design novel radar-aware mechanisms, including slope-mismatch-based discovery, adaptive chirp synchronization, differential coding on range-FFT outputs, and sensing-aware interference suppression, thus enabling reliable inter-radar discovery, synchronization, data communication, and sensing preservation.}

$\bullet$ \red{We fully implement RadarCom on COTS TI AWR2243 automotive mmWave radars and experimentally evaluate its effectiveness in  static, dynamic, and robustness scenarios. Comprehensive results show that RadarCom achieves reliable direct radar-to-radar communication at an effective data rate of up to 5.55~kbps under the current API-limited prototype configuration, while preserving sensing performance with an F1 score of up to 80.5\% during simultaneous communication and sensing. }

\vspace{-2mm}
\section{Preliminary Analysis and Observations}
\label{sec:preliminary}
\noindent To understand whether direct inter-radar communication is physically feasible on COTS mmWave radars and to identify the key practical obstacles, we conduct a set of controlled experiments using two TI AWR2243BOOST radars in an indoor setup. As shown in Fig.~\ref{fig:experiment_setup}, the platform consists of two radars and a Tang Nano FPGA, where the FPGA is used only to provide hardware-level synchronization for controlled signal capture and analysis. 
\red{Through these preliminary experiments, we seek to answer three questions: (\textbf{i}) Can one radar capture another radar's signals? (\textbf{ii}) How reliably can such signals be captured without synchronization? (\textbf{iii}) What challenges remain after successful signal capture?}

$\bullet$ \textbf{Inter-radar Communication Feasibility and Synchronization Challenge.}
We first examine whether one radar can capture another radar's signal under controlled timing alignment. In FMCW mmWave radars, the received signal is mixed with the radar's own transmitted chirp, producing an IF signal whose frequency content is used for sensing. We collect data from Radar~1 under two scenarios: (i) only Radar~1 is active, and (ii) \red{both radars are active and synchronized through a common trigger}. We configure Radar~2 with a slight chirp-slope mismatch relative to Radar~1. 
\red{As a result, Radar~2's transmitted chirps produce a time-varying beat frequency after dechirping at Radar~1, making them distinguishable from environmental reflections. Specifically, environmental echoes appear as relatively stable IF components, whereas Radar~2's signal manifests as a distinctive frequency-sweeping pattern within each chirp.}
As shown in Fig.~\ref{fig:stft_analysis}(a) and Fig.~\ref{fig:stft_analysis}(b), results from these controlled experiments confirm that, with proper timing alignment, one COTS automotive radar can capture another radar's transmission, \red{demonstrating the physical feasibility of direct inter-radar communication.}

\red{Next, we investigate whether inter-radar signals can still be captured when the two radars operate without hardware synchronization by independently starting the radars while keeping their chirp parameters fixed.}
We apply dechirp processing to detect whether Radar~2's signal falls into Radar~1's IF observation window. \green{A frame is considered successfully captured when a prominent peak corresponding to the received chirp is observed in the dechirped FFT output.} 
As shown in Fig.~\ref{fig:discovery_rate}, only $0.75$\% of \green{$800$} frames satisfy this condition. 
This result indicates that inter-radar signal capture is physically feasible but highly unreliable without synchronization, even when the radar parameters are known in advance. 
\red{These findings reveal a fundamental gap between communication feasibility and practical operation, motivating the need for dedicated mechanisms for inter-radar discovery and synchronization.}

\vspace{0mm}
$\bullet$ \textbf{From Signal Capture to Practical Communication.} Our experiments reveal that successfully capturing another radar's signal does not immediately enable reliable communication. Through experiments, we observe two key challenges that arise after signal capture: hardware-induced timing drift and communication-induced sensing interference.

\emph{\textbf{Hardware-Induced Timing Drift:}} \red{As mentioned previously, due to unavoidable manufacturing imperfections, even for two COTS radars using the same hardware model, their internal oscillators are not perfectly identical~\cite{ti_awr2243_datasheet, ti_awr2243_cascade}. We next investigate whether the relative timing between two communicating radars remains stable after signal capture. To this end, we run two synchronized radars with identical chirp parameters, including matched chirp slopes and timing settings. Under this matched-slope configuration, Radar~2's signal should appear as a stable single-tone peak in Radar~1's FFT spectrum if the relative timing remains fixed. In practice, as shown in Fig.~\ref{fig:chirp_drift}, we observe an approximately linear drift of the spectral peak across chirps, indicating that the relative timing between the two radars continuously changes over time. These results indicate that practical radar-to-radar communication cannot rely on one-time synchronization alone and instead requires continuous tracking and compensation of timing drift.}

\emph{\textbf{Impact on Native Sensing:}} \red{Beyond timing drift, our experiments reveal another important observation. Once a neighboring radar's signal is successfully captured, the signal itself can significantly interfere with the sensing process. To investigate this effect, we use the same experimental setup as above and keep the two radars synchronized via an external trigger, ensuring that Radar~2's signal is consistently observed by Radar~1.} The two radars are configured with matched chirp slopes, creating a strong interference case for Radar~1's sensing output. As shown in Fig.~\ref{fig:combined_analysis}(a) and Fig.~\ref{fig:combined_analysis}(b), Radar~2's signal significantly raises the noise floor in Radar~1's Range-Doppler (RD) map. We further quantify this impact in Fig.~\ref{fig:combined_analysis}(c) by measuring OS-CFAR~\cite{18645} detection errors while increasing Radar~2's transmit gain from $-30$~dB to $0$~dB in $5$~dB steps. Both false positives (FPs) and false negatives (FNs) increase with increasing interference strength. Specifically, FPs increase from 19.7\% to 92.4\%, while FNs increase from 17.1\% to 35.6\%. These results show that inter-radar signals can both introduce false detections and mask real targets, motivating a sensing-aware mechanism to suppress communication-induced interference.

\begin{figure}[t] %
    \centering
    \includegraphics[width=0.75\linewidth]{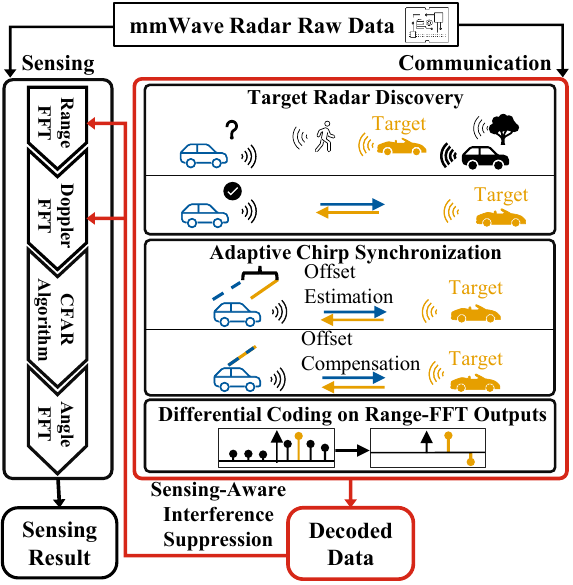}
    \vspace{-1mm}
    \caption{RadarCom design overview.}
    \vspace{-3mm}
    \label{fig:system_overview}
\end{figure}

\vspace{-4mm}
\section{System Overview}
\noindent RadarCom aims to realize simultaneous direct radar-to-radar communication and sensing on unmodified COTS automotive mmWave radars. It operates within the native FMCW radar workflow and requires only software-level reconfiguration of standard chirp parameters, without hardware modification, customized waveform design, or centralized coordination/infrastructure support. Guided by the observations in Sec.~\ref{sec:preliminary}, \red{we design RadarCom with four novel and tightly integrated modules, }
as shown in Fig.~\ref{fig:system_overview}: \textbf{(i) Target Radar Discovery}, \textbf{(ii) Adaptive Chirp Synchronization}, \textbf{(iii) Differential Coding on Range-FFT Outputs}, and \textbf{(iv) Sensing-Aware Interference Suppression}.

$\bullet$ \textbf{Target Radar Discovery}: 
\red{We refer to the radar initiating communication as the \textbf{local radar} and its intended communication peer as the \textbf{target radar}. This module enables the local radar to discover and identify the target radar in the presence of environmental reflections and signals from nearby radars.}
It consists of two stages: \textbf{(i)} \textit{Interference-Resilient Beacon Detection}, the target radar broadcasts beacon chirps with predefined parameters, allowing the local radar to identify their distinctive frequency-sweeping signature after dechirping; and \textbf{(ii)} \textit{SFO-Based Target Identification}, the local radar distinguishes radars using the device-specific spectral drift induced by  SFO, even when multiple radars use the same beacon parameters.

$\bullet$ \textbf{Adaptive Chirp Synchronization}: After discovery, the two radars must maintain chirp-level alignment without centralized synchronization support \red{despite hardware-induced drift and relative motion}.  \green{This module addresses this challenge through a dual-slope synchronization beacon design.} 
\red{By alternately transmitting beacons with two different slopes, the target radar provides sufficient information for the local radar to jointly estimate the initial timing offset, per-chirp timing drift, and carrier-frequency offset. These estimates enable accurate chirp-level synchronization without requiring external timing references.}

$\bullet$ \textbf{Differential Coding on Range-FFT Outputs}: \red{This module is designed to enable data transmission using the native radar sensing pipeline. It employs differential coding on the range-FFT outputs already generated during FMCW sensing and encodes information by varying chirp start frequencies, resulting in controlled shifts of the inter-radar peak observed in the receiving radar's range-FFT output. By decoding information from the relative peak shifts between adjacent chirps, this design suppresses common timing and distance-dependent effects, enabling lightweight and robust data transmission while reusing the existing sensing pipeline.}

$\bullet$ \textbf{Sensing-Aware Interference Suppression}: 
\red{As observed in Fig.~\ref{fig:combined_analysis}, the inter-radar communication signal introduces strong interference into the sensing pipeline, which elevates the sensing noise floor and increases both false positives and false negatives. This module is designed to mitigate such communication-induced interference and preserve native sensing performance, where a new mechanism is proposed to suppress the interference by exploiting the} \green{temporal inconsistency with real target reflections. Specifically, RadarCom detects abrupt power jumps across chirps to remove dominant interference in the range-FFT domain, and further checks detection consistency within each frame to remove residual false detections.}

\vspace{-2mm}
\section{Detailed Design}
\label{Sec:Detailed Design}
\subsection{Target Radar Discovery}
\label{discovery}
\noindent
Before data transmission, the local radar must first identify its intended communication peer, referred to as the target radar, in a complex mmWave environment. In vehicular networking deployments, the local/target radar roles are naturally determined by driving geometry: the rear-facing radar of the front vehicle acts as the target radar, while the front-facing radar of the following vehicle acts as the local radar. A vehicle can therefore serve as the local radar for the vehicle ahead and as the target radar for the vehicle behind. However, discovery is challenging because the local radar's IF signals may contain environmental reflections and transmissions from other nearby radars operating in the same $77$--$81$ GHz band. Moreover, without prior synchronization, the target radar's chirps may only intermittently fall within the local radar's IF observation window\footnote{During discovery, the local radar adjusts the idle times of the chirps across frames to shift its timing and scan for beacons more efficiently} as shown in Fig.~\ref{fig:discovery_rate}.

To address these challenges, RadarCom introduces a beacon-based discovery mechanism inspired by WiFi~\cite{nxp_wifi_connection} and Bluetooth~\cite{SilabsBLEBeacon}. The target radar periodically transmits standard FMCW beacon chirps with a predefined chirp-parameter pattern, while the local radar monitors its IF samples and applies dechirp-FFT processing to detect the beacon's distinctive frequency-sweeping signature. 
\red{To enable reliable target radar discovery, we design two new modules:  interference-resilient beacon detection and SFO-based target identification to achieve \textbf{(i)} \red{reliable beacon detection in the presence of } 
environment reflections and non-target radar transmissions, and \textbf{(ii)} \red{target radar identification when multiple nearby radars use the same beacon parameters.}}

\begin{figure}[t]
    \centering

    \setlength{\figsep}{0.006\linewidth}

    \begin{subfigure}[t]{0.32\linewidth}
        \centering
        \includegraphics[width=\linewidth]{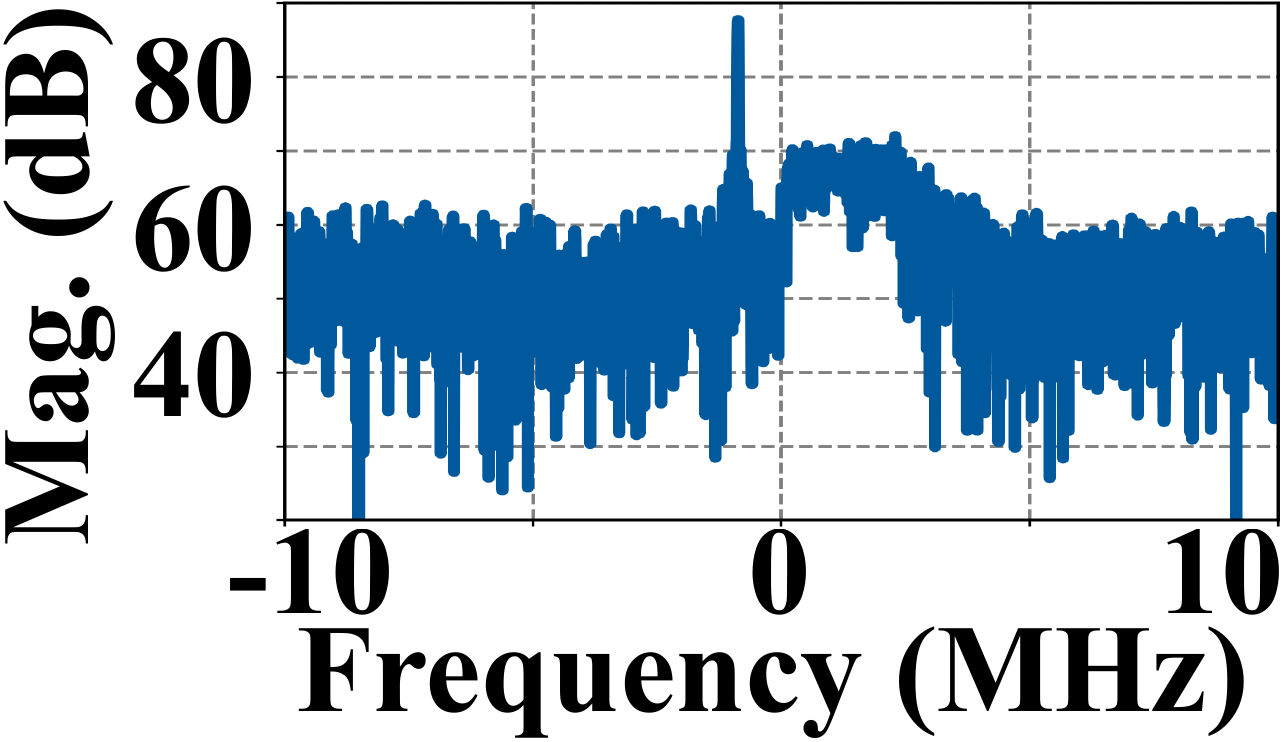}
        \caption{Matched Beacon}
        \label{fig:slope_match}
    \end{subfigure}%
    \hspace{\figsep}%
    \begin{subfigure}[t]{0.32\linewidth}
        \centering
        \includegraphics[width=\linewidth]{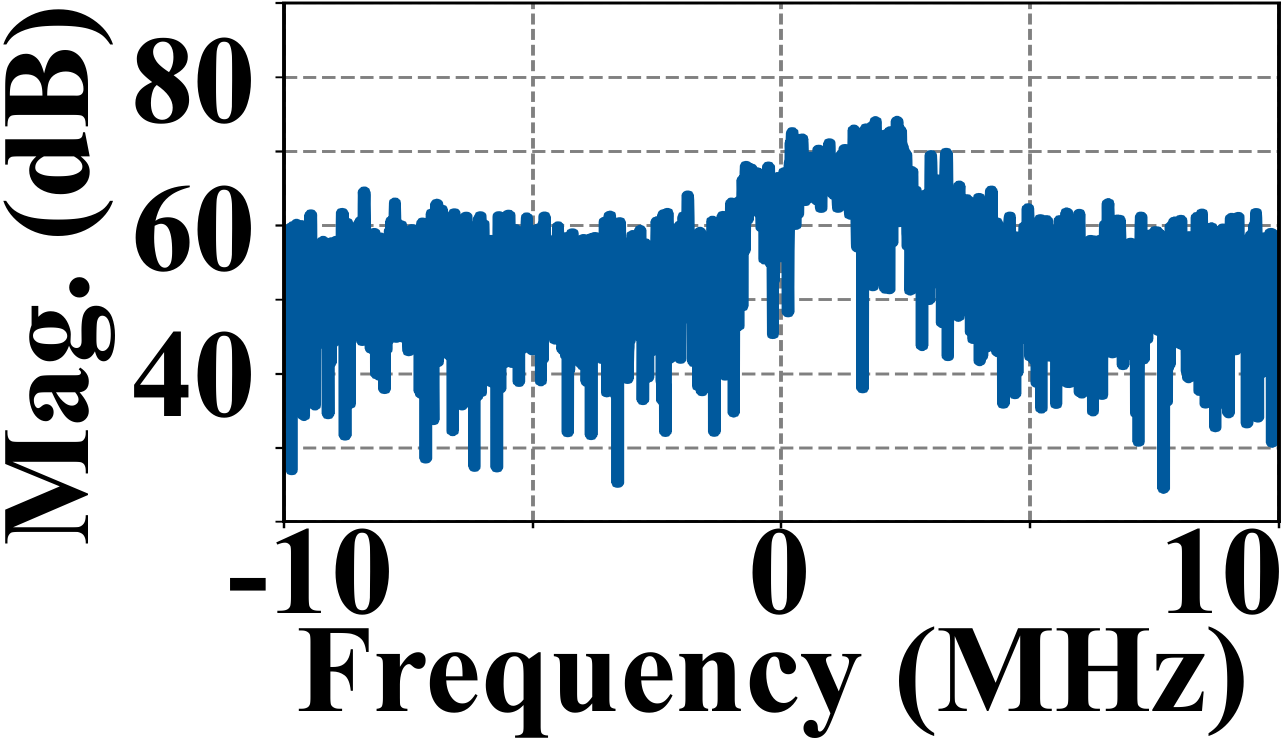}
        \caption{Mismatched Slope}
        \label{fig:slope_mismatch}
    \end{subfigure}%
    \hspace{\figsep}%
    \begin{subfigure}[t]{0.32\linewidth}
        \centering
        \includegraphics[width=\linewidth]{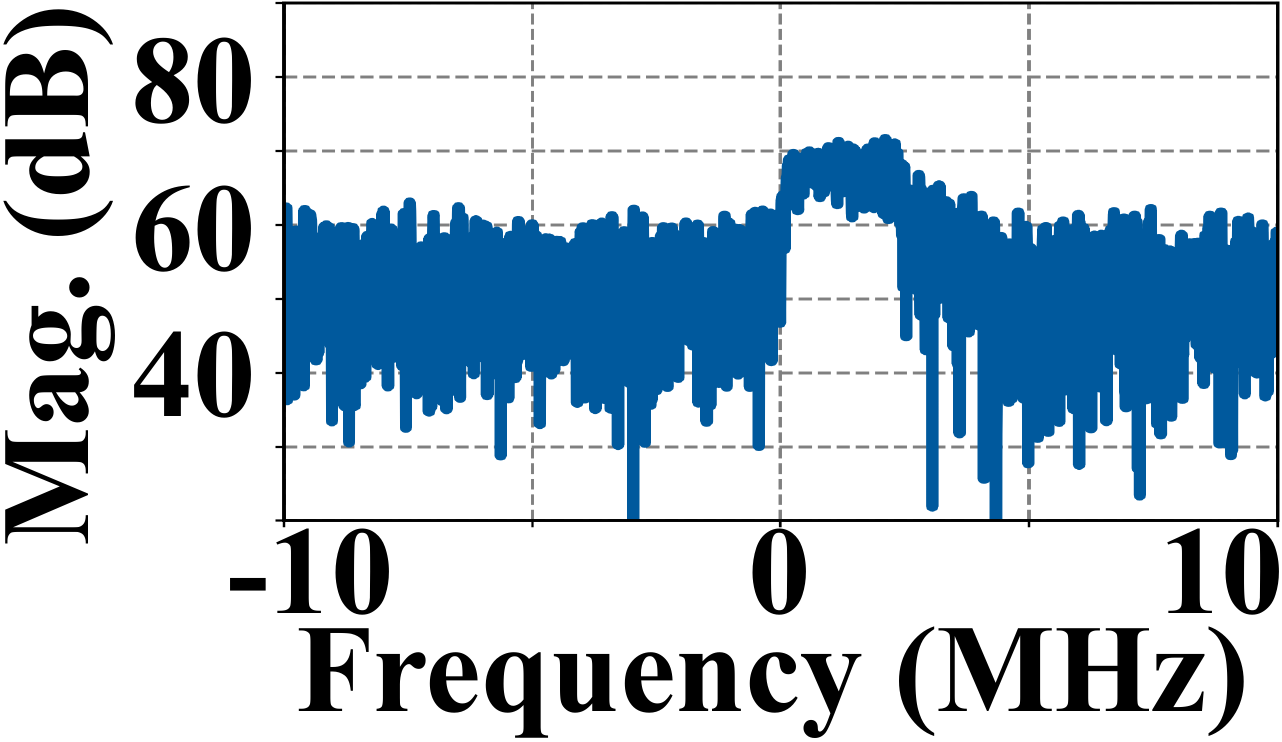}
        \caption{Env. Echoes}
        \label{fig:static_reflections}
    \end{subfigure}
    \vspace{-1mm}
    \caption{Dechirping Spectral Analysis.}
    \vspace{-1mm}
    \label{fig:spectral_analysis}
\end{figure}

\begin{figure}[t]
  \centering
  \begin{subfigure}{0.48\columnwidth}
    \centering
    \includegraphics[width=\linewidth]{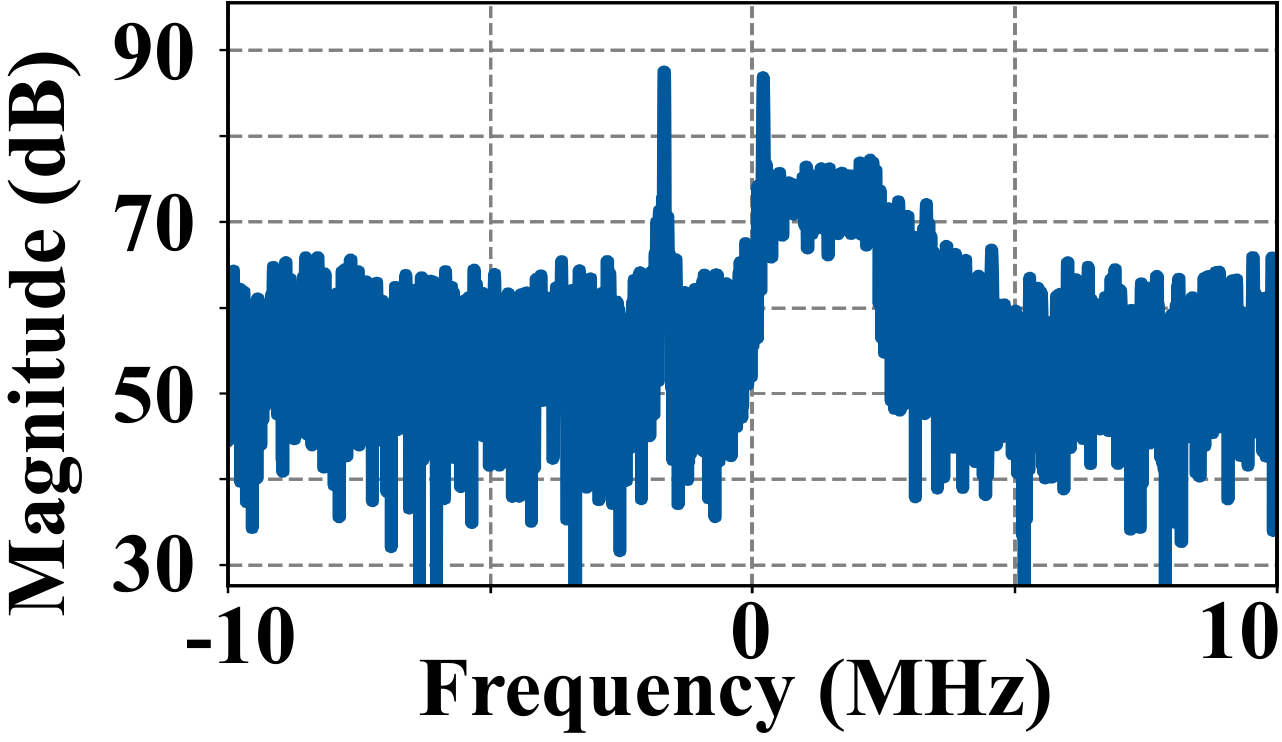}
    \caption{Multiple peaks from co-existing target radars.}
    \label{fig:two_peaks}
  \end{subfigure}
  \hfill
  \begin{subfigure}{0.48\columnwidth}
    \centering
    \includegraphics[width=\linewidth]{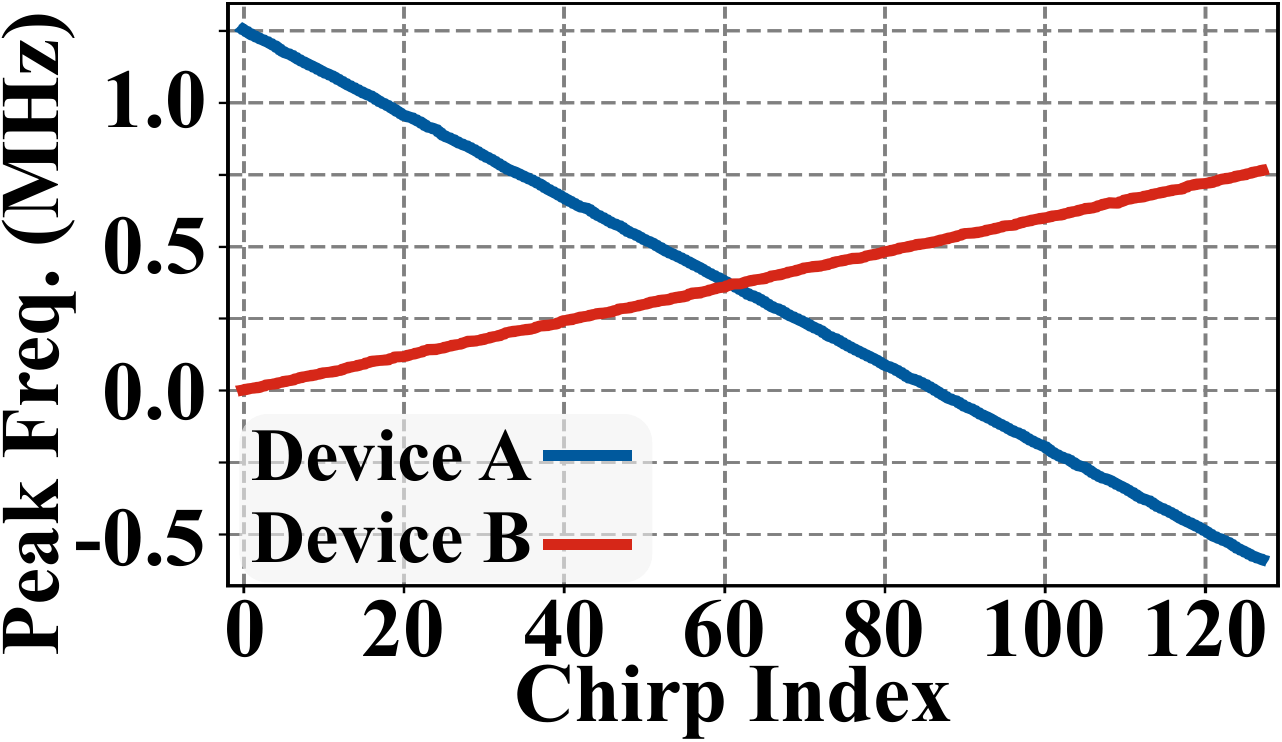}
    \caption{Distinct drift rates of different devices.}
    \label{fig:different_drift}
  \end{subfigure}
  \vspace{-1mm}
  \caption{SFO-based target identification.}
  \vspace{-2mm}
  \label{fig:sfo_identification}
\end{figure}
$\bullet$ \textbf{Interference-Resilient Beacon Detection:}
We design a slope-mismatch-based beacon detection mechanism. The key idea is to intentionally introduce a slight chirp-slope mismatch between the local radar \green{slope ($S_l$)} and the target radar's \green{beacon chirp slope ($S_t$)}.  The local radar configures its own chirps to match the beacon's timing parameters (i.e., ramp time and idle time), but sets its chirp slope to $S_l \neq S_t$.

\green{When the local radar receives a target-radar beacon chirp and mixes it with its own chirp, the intentional slope mismatch ($S_l-S_t$) produces a residual IF chirp}\footnote{In Sec.~\ref{sec:synchronization}, RadarCom uses two predefined beacon slopes, $S_{t1}$ and $S_{t2}$, transmitted alternately for synchronization; the \red{following} derivation applies to either slope.}:
\vspace{-0.5mm}
\begin{equation}
\label{if_chirp}
s_{IF}(t)= \exp[j(\pi(S_l-S_t) t^2-2\pi(S_t\Delta t+\Delta f_c)t-2\pi (f_c + \Delta f_c)\Delta t)],
\end{equation}
where $\Delta t$ denotes the time offset, including the chirp timing offset and propagation delay, and $\Delta f_c$ is the carrier-frequency offset between the local and target radars. The local radar then generates a dechirping reference signal $s_d(t)$ to cancel the quadratic phase term in Eq.~\eqref{if_chirp}, given as:
\vspace{-0.5mm}
\begin{equation}
\label{dechirp}
s_d(t) = \exp[-j\pi (S_l-S_t)t^2].
\end{equation}
Multiplying the received IF signal $s_{IF}(t)$ by $s_d(t)$ converts the target radar's beacon signal into a stable single-tone:
\vspace{-0.5mm}
\begin{equation}
\label{single_tone}
s'_{IF}(t)=\exp[j(-2\pi(S_t\Delta t+\Delta f_c)t-2\pi (f_c + \Delta f_c)\Delta t)].
\end{equation}
\red{After FFT, the beacon forms a concentrated spectral peak, as shown in Fig.~\ref{fig:spectral_analysis}(a). In contrast, environmental reflections and non-target radar signals do not match the dechirping reference in Eq.~\eqref{dechirp}. As a result, their energy is dispersed across multiple frequency bins rather than concentrated into a single dominant peak, as illustrated in Fig.~\ref{fig:spectral_analysis}(b) and Fig.~\ref{fig:spectral_analysis}(c), respectively.}

Therefore, the proposed slope-mismatch-based dechirping process can effectively distinguish beacon signals from 
environmental reflections and non-target radar transmissions, enabling robust beacon detection.

$\bullet$ \textbf{SFO-Based Target Identification:}
\red{While slope-mismatch-based dechirping enables beacon detection, it cannot distinguish multiple nearby radars using the same beacon pattern. To overcome this limitation, we design a new SFO-based target identification mechanism that transforms hardware-induced timing drift from a synchronization impairment into a device-specific signature for radar identification. This mechanism is inspired by the observation that hardware-induced SFO creates a unique device-specific peak-drift 
pattern across chirps within a frame, showing a linear movement of the dechirped beacon peak over chirp index. Therefore, even when multiple radars transmit the same beacon pattern, their beacon peaks follow different trajectories and can be distinguished accordingly. }

For a given beacon slope $S_t$, based on Eq.~\eqref{single_tone}, 
the dechirped IF signal becomes a single-tone whose peak frequency is
\vspace{0mm}
\begin{equation}
\label{eq}
f_{peak} = -(S_t\Delta t + \Delta f_c),
\end{equation}
where $\Delta f_c$ is the carrier-frequency offset. When multiple nearby radars transmit the same beacon pattern, the local radar may observe multiple spectral peaks, as illustrated in Fig.~\ref{fig:sfo_identification}(a). Within one frame, the propagation-delay variation caused by relative motion is negligible, following the standard FMCW assumption that targets remain in the same range bin over a measurement interval~\cite{9380893}, so the timing offset evolves mainly due to SFO:
\vspace{-0.5mm}
\begin{equation}
\label{eq:drift}
f_{peak}(n)=-S_t(t_o+n\cdot t_d)-\Delta f_c,
\end{equation}
where $t_o$ is the initial effective timing offset, including the initial propagation delay, $n\in\{0,1,\ldots,N-1\}$ is the chirp index within the frame, N denotes the total number of chirps in one frame and $t_d$ denotes the per-chirp timing drift induced by SFO. Eq.~\eqref{eq:drift} shows that the beacon peak frequency changes linearly with the chirp index. The intercept is determined by the initial timing offset and CFO, while the slope is $-S_t t_d$. Since $S_t$ is predefined, \red{different relative SFOs yield different values of $t_d$  	
and thus distinct peak-drift slopes. }
As illustrated in Fig.~\ref{fig:sfo_identification}(b), \red{RadarCom exploits these trajectories as device-specific signatures to identify the intended target radar.}

\vspace{-3mm}
\subsection{Adaptive Chirp Synchronization}
\label{sec:synchronization}

\noindent Temporal synchronization between independent COTS mmWave radars is challenging because these devices are designed for sensing rather than inter-device coordination. RadarCom operates in a decentralized vehicular environment without centralized synchronization support such as GPS or base stations. Therefore, synchronization must be achieved directly between independent radars with minimal disruption to normal sensing.

This is particularly stringent in FMCW radar systems because the inter-radar signal is observable only when its beat frequency falls within the receiver's IF band which is bounded by the ADC sampling rate. For a sampling rate of $F_s$, the  observable IF band is bounded by $F_s/2$~\cite{TI_SWRA553}. Thus, for the local radar to capture the target radar's signal, the resulting beat frequency must satisfy:
\begin{equation}
|f_{beat}(t)| = |(S_l - S_t)\cdot t - S_t \cdot \Delta t - \Delta f_c| \leq \frac{F_s}{2},
\label{eq:beat_freq_constraint}
\end{equation}
where $t \in [0, T_{ADC}]$ is the time within the local IF sampling window. This constraint results in only a narrow timing-visibility window: even a microsecond-level time offset can move the target signal outside the local radar's IF band~\cite{yang2022interference}. In our configuration, $F_s = 20$ MHz and the local chirp slope is $S_l = 5.311$ MHz/$\mu$s, yielding an approximate timing tolerance of only $1.88 \mu$s. 
This narrow window highlights the need for accurate timing estimation and \red{continuous} compensation.

\begin{figure}[t] 
    \centering
    \begin{subfigure}{\columnwidth}
        \centering
        \includegraphics[width=0.95\columnwidth]{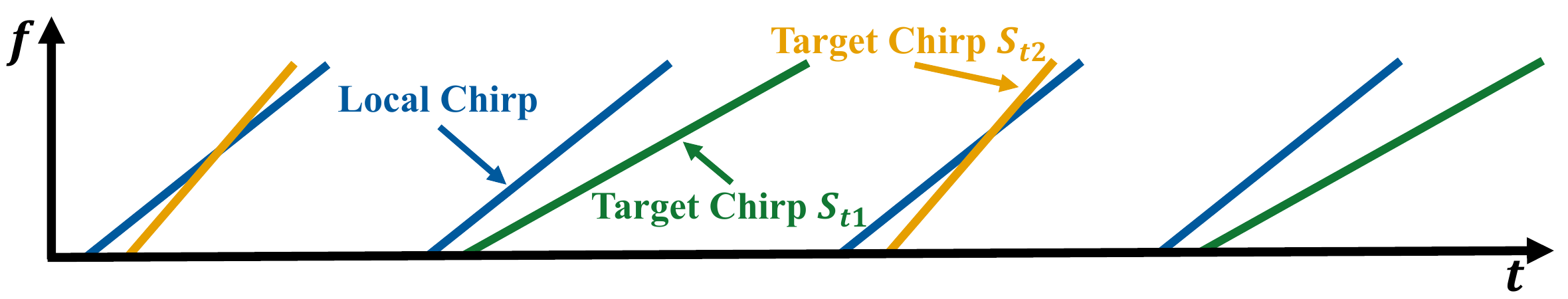}
        \vspace{-1mm}
        \caption{Dual-slope beacon design.}
        \label{fig:beacon_design_top}
    \end{subfigure}

    \begin{subfigure}{\columnwidth}
        \centering
        \includegraphics[width=0.95\columnwidth]{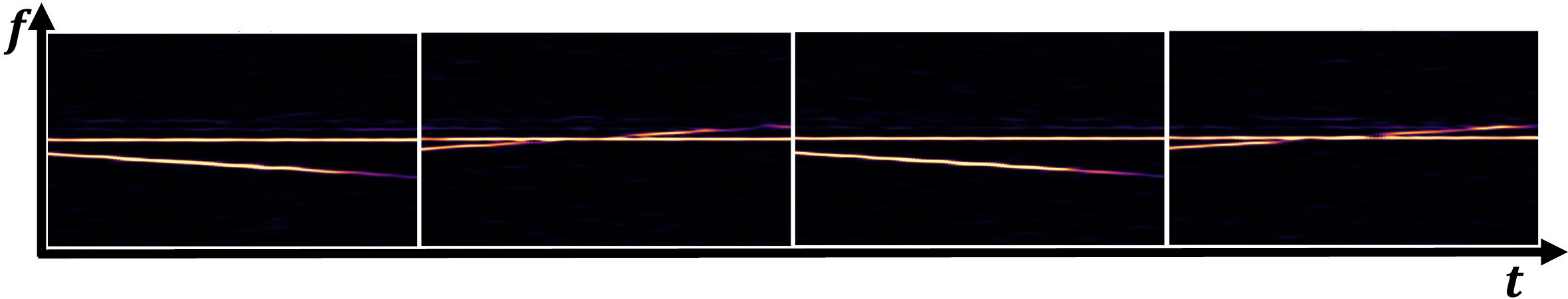}
        \vspace{-1mm}
        \caption{STFT of the local radar's IF signal.}
        \label{fig:beacon_stft}
    \end{subfigure}

    \begin{subfigure}{\columnwidth}
        \centering
        \includegraphics[width=0.95\columnwidth]{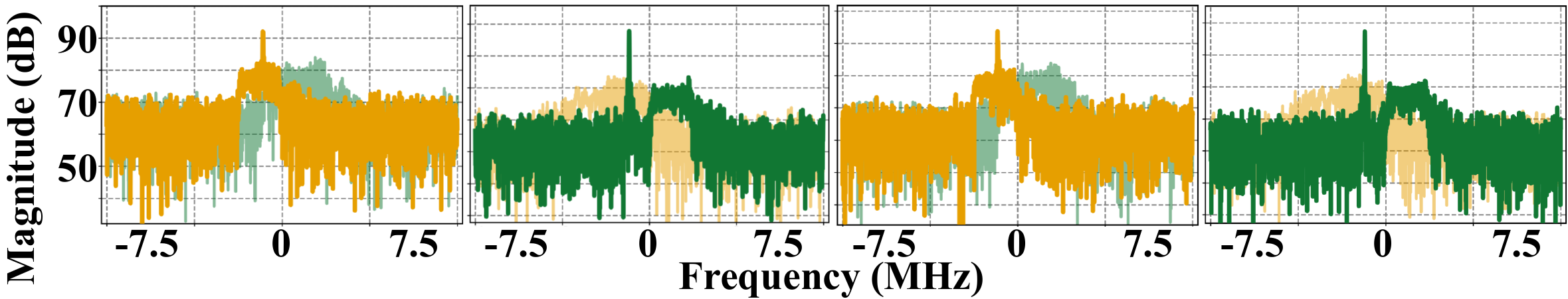}
        \caption{Alternating Peaks after dechirp-FFT.}
        \label{fig:alternative_stft}
    \end{subfigure}

    \begin{subfigure}{\columnwidth}
        \centering
        \includegraphics[width=0.95\columnwidth]{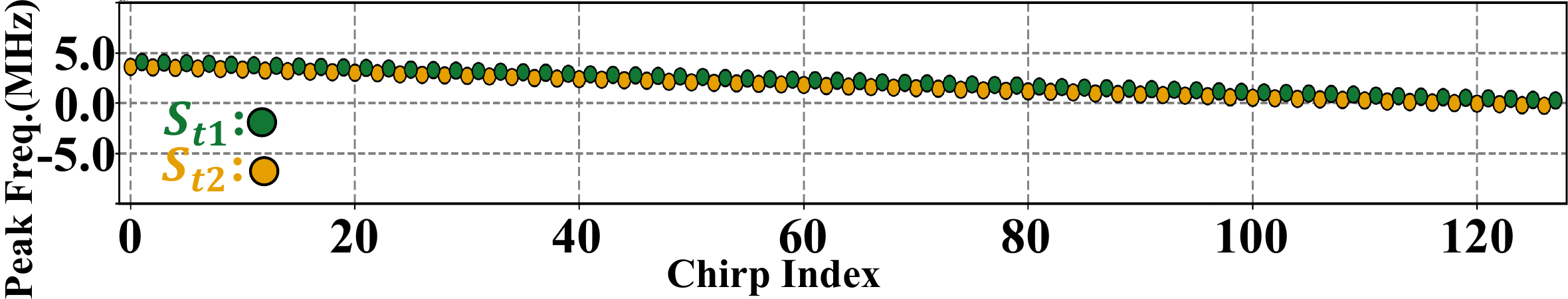}
        \caption{Extracted peak-frequency trajectories.}
        \label{fig:peak_sequence}
    \end{subfigure}
    \vspace{-5mm}
    \caption{Beacon chirp design and local radar observations.}
    \label{fig:beacon_design}
\end{figure}

$\bullet$ \textbf{\red{Dual-Slope Synchronization} Beacon Chirp Design}:
Reliable observation of the target radar requires microsecond-level synchronization accuracy, \red{depending on accurate estimation of } 
three parameters: the initial time offset $t_o$, the per-chirp time drift $t_d$ caused by SFO, and the CFO $\Delta f_c$. However, as shown in Eq.~\eqref{eq:drift}, a single-slope beacon sequence cannot uniquely determine $t_o$ from $\Delta f_c$, since both terms contribute to the peak-frequency intercept. To resolve this ambiguity, RadarCom introduces a dual-slope beacon sequence, where the target radar alternately transmits beacon chirps with slopes $S_{t1}$ and $S_{t2}$, chosen slightly below and above the local radar's slope $S_l$, respectively, as illustrated in Fig.~\ref{fig:beacon_design}(a).  According to Eq.~\eqref{if_chirp}, when each received beacon chirp is mixed with the local chirp, the slope mismatch produces a residual IF chirp with slope $S_l-S_{t1}$ or $S_l-S_{t2}$. Since $S_{t1}$ and $S_{t2}$ are on opposite sides of $S_l$, the resulting residual IF chirps appear as alternating upward and downward frequency sweeps in the received IF signal, as shown in Fig.~\ref{fig:beacon_design}(b). To extract them, the local radar applies dechirp-FFT using two dechirping references corresponding to $S_l-S_{t1}$ and $S_l-S_{t2}$. As illustrated in Fig.~\ref{fig:beacon_design}(c), the two dechirp branches produce alternating spectral peaks across chirps 
\red{according to the transmitted beacon slope}. 
\red{By jointly observing these alternating peaks, RadarCom obtains sufficient information to calculate  $t_o$, $t_d$, and $\Delta f_c$, as discussed next, thereby enabling accurate synchronization between independent radars.}

$\bullet$ \textbf{Synchronization Parameter Estimation}: 
\red{Once alternating peaks from the two dechirp branches are observed across chirps, the local radar confirms successful reception of the dual-slope synchronization beacon.}
Fig.~\ref{fig:beacon_design}(d) shows the extracted peak frequencies across chirps within one frame. The alternating beacon slopes produce two interleaved peak sequences that drift linearly over time due to SFO. Each peak observation satisfies $-S_{t}(t_o + n\cdot t_d) - \Delta f_c = f_{peak}$,
where $S_t \in \{S_{t1}, S_{t2}\}$. Collecting observations across chirps yields an overdetermined linear system. 
\red{Stacking the observations from all \green{$n =0,\ldots,N-1$}  beacon chirps within $i$th frame produces the following overdetermined linear system $\mathbf{A}\mathbf{x}=\mathbf{b}$, where:}

\begin{equation}
\vspace{-1mm}
\small
\mathbf{A} =
\begin{bmatrix}
-S_{t1} & -S_{t1} \cdot 0 & -1 \\
-S_{t2} & -S_{t2} \cdot 1 & -1 \\
-S_{t1} & -S_{t1} \cdot 2 & -1 \\
\vdots & \vdots & \vdots \\
-S_{t2} & -S_{t2} \cdot (N-1) & -1
\end{bmatrix},
\mathbf{x} =
\begin{bmatrix}
t_o \\
t_d \\
\Delta f_c
\end{bmatrix},
\mathbf{b} =
\begin{bmatrix}
f_{peak,0} \\
f_{peak,1} \\
f_{peak,2} \\
\vdots \\
f_{peak,N-1}
\end{bmatrix}.
\vspace{-1mm}
\end{equation}
\red{RadarCom then calculates the synchronization parameters using least squares:} $\hat{\mathbf{x}} = (\mathbf{A}^T\mathbf{A})^{-1}\mathbf{A}^T\mathbf{b}$.

It then predicts the timing offset of upcoming chirps relative to the current chirp as:
\begin{equation}
{\Delta \hat t}(n) = \hat{t}_o + n \cdot \hat{t}_d,
\end{equation}
\green{and compensates the predicted timing offset by adjusting the chirp idle time.} It also compensates the estimated CFO by applying an opposite shift to the chirp start frequency. Because the relative timing drift and CFO may vary over time due to oscillator instability and hardware condition changes~\cite{TI_SWRA794B_2025}, RadarCom periodically \red{re-estimates the synchronization parameters } 
to maintain stable synchronization. 
\red{To enable bidirectional observability, the target radar performs the same dechirp-based peak detection on its received IF samples. Synchronization is confirmed only when both radars observe stable beacon peaks over multiple consecutive frames. Once synchronization is established, the two radars proceed to the communication phase.}

\subsection{Differential Coding on Range-FFT Outputs}
\label{sec:differential_coding}

\noindent Once synchronization is established (Sec.~\ref{sec:synchronization}), the two radars enter the communication phase, where the target radar acts as the transmitter and the local radar acts as the receiver. At this stage, \red{the key challenge is how to encode/decode information using only the native sensing waveform and processing pipeline, without requiring dedicated communication waveforms or hardware modifications.}
\red{RadarCom addresses this challenge by leveraging}
the \emph{parallel interference} phenomenon in FMCW radar systems~\cite{yang2022interference}. Prior studies show that when two radars operate with identical chirp slopes, the interfering chirp can mix down to a constant-frequency tone in the receiver's IF band, \red{producing} 
a ghost peak in the range-FFT output at a range bin unrelated to any physical target. 
Unlike prior work that treats this effect as harmful interference, RadarCom intentionally creates 
this parallel-interference condition after synchronization by configuring the two radars with matched chirp slopes. Once synchronization keeps the transmitter's signal within the receiver's IF band, the resulting ghost peak can serve as a \red{controllable} communication carrier \red{that can be directly observed} in the receiver's range-FFT output.

\begin{figure}[t]
    \centering
    \includegraphics[width=\columnwidth]{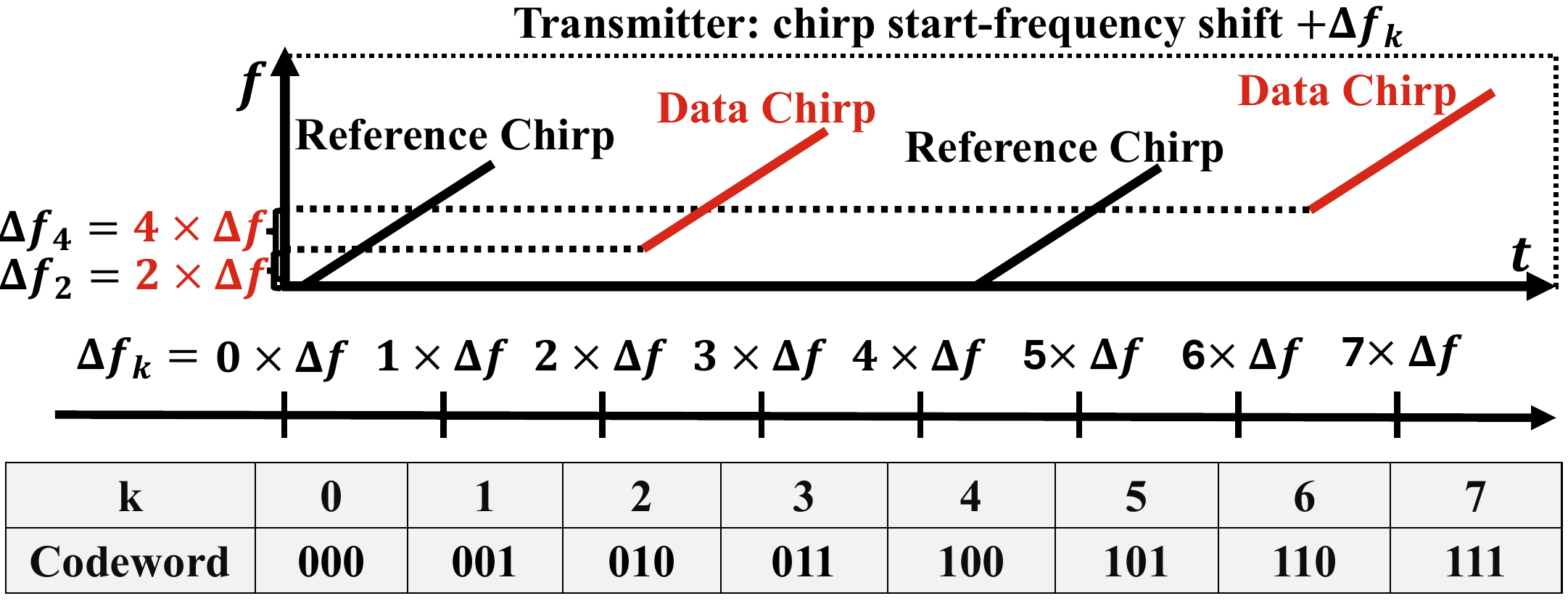}
    \caption{Differential Start-Frequency Encoding.}
    \vspace{-2mm}
    \label{fig:rangefft_differential_coding}
\end{figure}
\begin{figure}[t]
    \centering
    \includegraphics[width=\columnwidth]{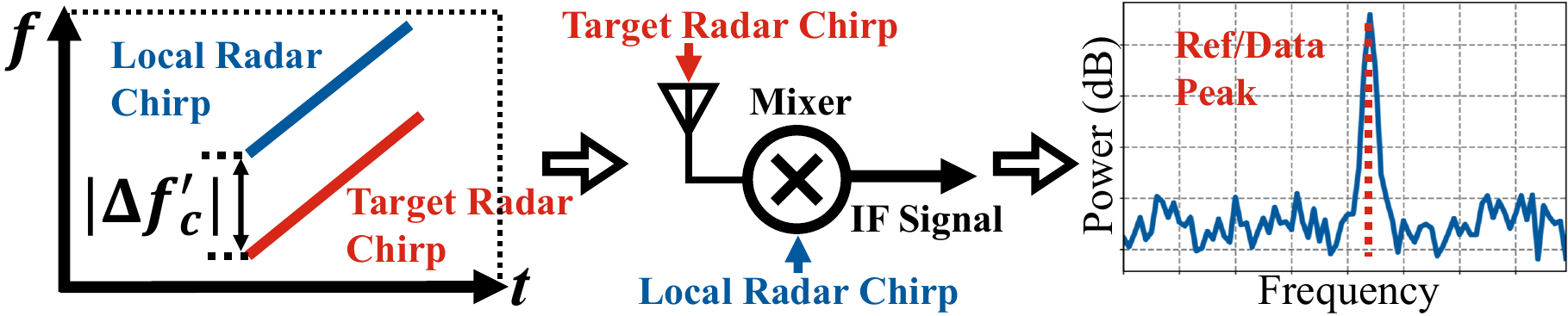}
    \caption{Matched-Slope Ghost Peak Generation.}
    \vspace{-2mm}
    \label{fig:receiver}
\end{figure}

$\bullet$ \textbf{Signal Model:}
After the beacon-based synchronization stage, the target radar reconfigures its chirp slope to match the local radar's chirp slope, yielding $S_l = S_t = S$. Before transmitting data, RadarCom performs a brief post-synchronization confirmation step.\footnote{After switching to the matched-slope configuration, the local radar checks whether the expected slope-matched pattern appears in subsequent frames. If not, the local radar returns to beacon discovery.} 
Once the matched-slope configuration is confirmed, the quadratic term in Eq.~\eqref{if_chirp} cancels. Let $\Delta t^{\prime}$ and $\Delta f_c^{\prime}$ denote the post-synchronization residual time offset and carrier-frequency offset, respectively. The received IF signal then reduces to a pure tone:
\begin{equation}
s_{IF}(t) = \exp\left(j\left(-2\pi(S\Delta t^{\prime} + \Delta f_c^{\prime})t - 2\pi(f_c + \Delta f_c^{\prime})\Delta t^{\prime}\right)\right).
\end{equation}
After range-FFT, this tone produces a single spectral peak at:
\begin{equation}
f_{ghost} = -(S\Delta t^{\prime} + \Delta f_c^{\prime}),
\label{eq:ghost_freq}
\end{equation}
which we refer to as the \emph{ghost peak}. The ghost peak position is therefore determined jointly by the residual timing term $\Delta t^{\prime}$ and the residual frequency-offset term $\Delta f_c^{\prime}$.

$\bullet$ \textbf{Differential Start-Frequency Encoding:}
\red{Although synchronization compensates for SFO-induced timing drift, the residual timing offset $\Delta t^{\prime}$ in Eq.~\eqref{eq:ghost_freq} still contains a propagation-delay component that depends on the inter-vehicle distance and may vary over time in dynamic driving scenarios. As a result, the absolute ghost-peak position is not a reliable carrier for communication. To address this challenge, we design a differential coding scheme that encodes information into the \emph{relative} spectral displacement between consecutive chirps, thereby canceling distance-dependent effects during decoding.}

\red{Specifically,}
after matched-slope reconfiguration, the target radar encodes information by alternately transmitting \emph{reference chirps} and \emph{data chirps}. The reference chirp uses the baseline start frequency, while the data chirp applies a controlled start-frequency shift relative to the reference chirp. Each reference-data chirp pair carries one symbol. Specifically, to transmit symbol $k$, the target radar shifts the start frequency of the data chirp by:
\begin{equation}
\Delta f_k = k \cdot \Delta f, \quad k = 0, 1, \ldots, M-1,
\label{eq:freq_shift}
\end{equation}
where $\Delta f$ denotes the start-frequency step between adjacent symbols, and $M$ is the modulation order, \red{corresponding to $M$ distinct frequency-shift levels.} Each symbol $k$ maps to a $\log_2 M$-bit codeword. For example, for $M=8$ case shown in Fig.~\ref{fig:rangefft_differential_coding}, the frequency shifts $\Delta f_k = k\Delta f$ for $k = 0, 1, \ldots, 7$ correspond to the 3-bit codewords $000, 001, \ldots, 111$. The upper part of Fig.~\ref{fig:rangefft_differential_coding} illustrates an example transmission 
over four consecutive chirps, where the two data chirps use frequency shifts $\Delta f_2 = 2\Delta f$ and $\Delta f_4 = 4\Delta f$ \red{to carry $010$ and $100$}, respectively.

\begin{figure}[t]
    \centering
    \includegraphics[width=\columnwidth]{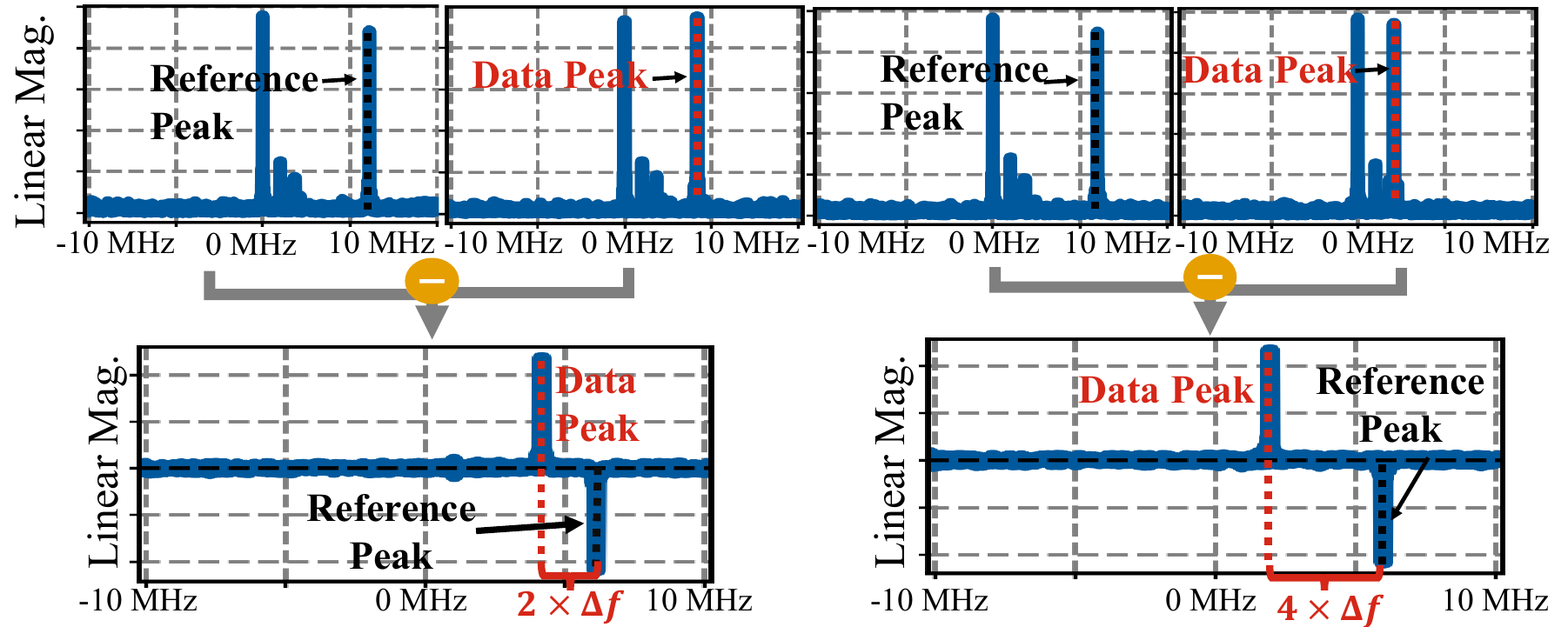}
    \caption{Adjacent-Chirp Spectrum Subtraction.}
    \vspace{-2mm}
    \label{fig:decoder}
\end{figure}

\begin{figure}[t]
    \centering
    \includegraphics[width=\columnwidth]{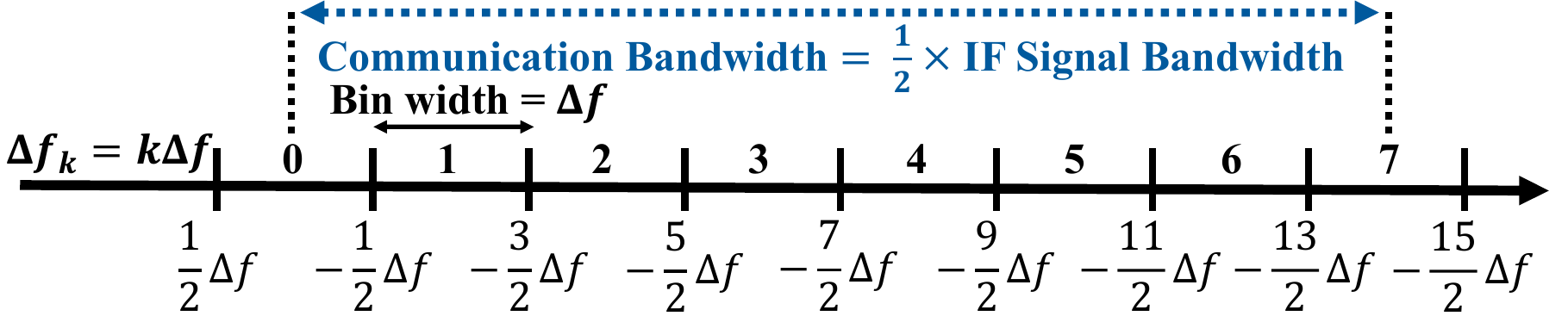}
    \caption{Demodulation Bins and Boundaries.}
    \vspace{-1mm}
    \label{fig:decoding_map}
\end{figure}

$\bullet$ \textbf{Differential Decoding on Range-FFT:}
At the local radar, each received chirp is mixed with the local chirp and processed through standard range-FFT processing, yielding a ghost peak in the spectrum. As illustrated in Fig.~\ref{fig:receiver}, the chirp from the target radar combines with the local chirp at the mixer and produces an IF signal whose range-FFT exhibits a ghost peak. Both reference and data chirps produce such peaks. 
The upper part of Fig.~\ref{fig:decoder} shows the resulting range-FFT spectra for four consecutive chirps (two reference-data pairs), where the data peak appears at different positions according to the encoded symbol.

Following Eq.~\eqref{eq:ghost_freq}, the ghost peak positions of the reference chirp at index $n$ and the data chirp at index $n+1$ are:
\begin{align}
f_{ref}(n) &= -(S \Delta t^{\prime}(n) + \Delta f_c^{\prime}), \\
\label{coding}
f_{data}(n+1) &= -(S \Delta t^{\prime}(n+1) + \Delta f_c^{\prime} + \Delta f_k),
\end{align}
where $\Delta f_k$ is the start-frequency shift applied to the data chirp. Since the two chirps are transmitted adjacently, the inter-vehicle distance and residual timing offset change negligibly within one chirp period, so $\Delta t^{\prime}(n+1) \approx \Delta t^{\prime}(n)$. Computing the difference between the two range-FFT spectra gives a differential spectrum:
\begin{equation}
\Delta \mathbf{F}(n) = \mathbf{F}_{{data}}(n+1) - \mathbf{F}_{{ref}}(n),
\label{eq:diff_spectrum}
\end{equation}
in which the common timing-dependent term $S\Delta t^{\prime}$ and the residual frequency-offset term $\Delta f_c^{\prime}$ are canceled. The reference and data peak frequencies therefore satisfy:
\begin{equation}
f_{data}(n+1) - f_{ref}(n) \approx -\Delta f_k.
\label{eq:peak_separation}
\end{equation}
The lower part of Fig.~\ref{fig:decoder} shows two example differential spectra. Since environmental reflections are nearly unchanged across adjacent reference and data chirps, subtracting the two spectra largely cancels these common reflections and leaves the modulated communication peaks clearly observable. 
The decoder then calculates 
the frequency gap between the reference and data peaks and maps it to the nearest candidate shift $k\Delta f$. As shown in Fig.~\ref{fig:decoding_map}, the candidate shifts are
\red{uniformly spaced along the frequency axis}
with interval $\Delta f$, \red{enabling RadarCom to decode the transmitted symbol index $k$ and its corresponding codeword from the measured frequency gap. }
\red{The differential start-frequency coding scheme enables reliable communication using the native radar sensing waveform and range-FFT pipeline. However, its performance is bounded by three practical considerations. First, the encoded frequency shifts must remain within the receiver's observable IF band to ensure reliable decoding. Second, the achievable data rate is constrained by the radar frame structure and the reference-data chirp pairs used for differential decoding. Third, start-frequency coding design slightly perturbs the sensing waveform and may affect sensing accuracy. We next analyze these three aspects.}

\begin{figure}[t]
    \centering
    \includegraphics[width=0.9\columnwidth]{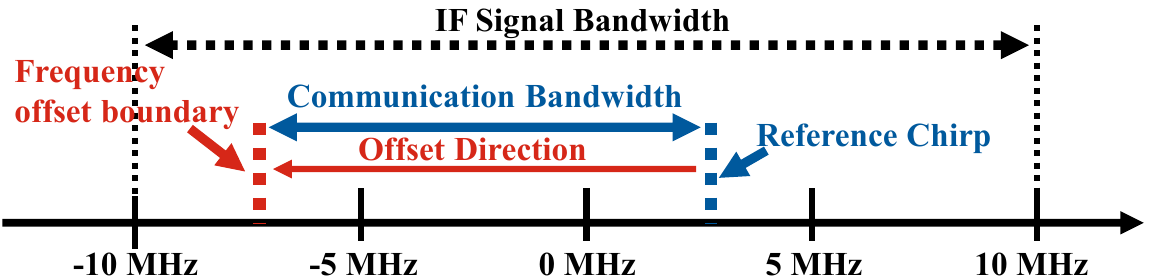}
    \caption{Communication Bandwidth Constraints.}
    \label{fig:com_bandwidth}
    \vspace{-2mm}
\end{figure}

$\bullet$ \textbf{Communication Bandwidth Constraint:}
To keep the transmitter's signal observable, \red{the transmitter's ghost peak must remain within the receiver's IF band.}
As shown in Fig.~\ref{fig:com_bandwidth}, \red{to achieve this,} before receiving data, the RadarCom receiver keeps all its chirps at the reference configuration and slightly increases its chirp start frequency relative to the received reference chirp from the transmitter. After I/Q mixing and low-pass filtering, this makes the transmitter's reference ghost peak appear on the positive-frequency side of the receiver's IF band. The transmitter then encodes data by increasing each data chirp's start frequency relative to its reference chirp. Thus, the frequency shift between the reference and data peaks observed at the receiver is:
\begin{equation}
f_{data} - f_{ref} = -\Delta f_k .
\end{equation}
\red{To prevent the data peak from moving outside the observable IF band, RadarCom limits the effective modulation bandwidth to half of the IF bandwidth. Let \(B_{\mathrm{IF}}\) denote the full two-sided IF bandwidth of the receiver:}
\begin{equation}
0 \leq \Delta f_k \leq \frac{1}{2}F_s.
\end{equation}
\red{This constraint helps maintain the modulated peaks within the observable IF band and reduces decoding uncertainty near the IF-band boundaries.}

$\bullet$ \textbf{Data Rate Analysis:}
\red{The achievable data rate is constrained by both the radar frame characteristics and the reference-data chirp pair based differential coding design.}
Let $T_f$ be the radar frame duration, $F_r=1/T_f$ the frame rate, and $N$ the number of chirps per frame. 
\red{Since RadarCom uses one reference-data chirp pair to convey a symbol, only half of the chirps in a frame carry information-related symbols.} 
In addition, RadarCom performs beacon-based re-synchronization once every $I$ frames, reserving one frame for synchronization and leaving $I-1$ frames for data transmission. The frame-limited data rate is therefore
\begin{equation}
\label{data_rate}
R_{\mathrm{frame}} =
\frac{I-1}{I}\cdot
\frac{N F_r}{2}\cdot
\log_2(M),
\end{equation}
where $M \leq N_{\mathrm{adc}}$ 
with \green{$N_{\mathrm{adc}}$ denoting the number of ADC samples per chirp.}  
\red{This expression gives the maximum achievable throughput }
limited by the radar frame structure, reference-data pairing coding design, and periodic re-synchronization requirement.

$\bullet$ \textbf{Sensing \red{Compatibility} under Start-Frequency Coding:}
RadarCom transmitter encodes data by shifting the start frequency of data chirps. For symbol $k$, the chirp start frequency becomes $f_c+\Delta f_k$, where $\Delta f_k=0$ for reference chirps and $\Delta f_k=k\Delta f$ for data chirps. Since the transmitter knows $\Delta f_k$ for every chirp, the induced sensing perturbation is deterministic.

Specifically, the range estimate can be calculated as $R = c f_b T_c / (2(f_{\max}-(f_c+\Delta f_k)))$, where $f_b$ is the beat frequency, $T_c$ is the chirp duration, and $f_{\max}$ is the chirp end frequency. Similarly, the velocity estimate becomes $v = c f_d / (2(f_c+\Delta f_k))$, where $f_d$ is the Doppler frequency. In practice, $\Delta f_k$ is constrained by the receiver's IF bandwidth, which is typically only several to tens of MHz, and RadarCom further keeps $\Delta f_k$ within half of the observable IF band. This shift is 
much smaller than both 
the GHz-level FMCW sweep bandwidth and the 77 GHz carrier frequency. Therefore, the resulting range and velocity distortion is negligible in practical deployments, \red{demonstrating that RadarCom remains compatible with native sensing operation. If higher sensing accuracy is desired, these effects can be further compensated on a per-chirp basis using the known $\Delta f_k$.}

\begin{table}[t]
  \centering
  \small
  \setlength{\tabcolsep}{2pt}
  \renewcommand{\arraystretch}{1.05}
  \resizebox{\columnwidth}{!}{%
  \begin{tabular}{lcccc}
    \toprule
    Method & FP/Frame & Precision & Recall & F1 \\
    \midrule
    Baseline           & 11.43 & 20.1\% & 81.0\% & 32.2\% \\
    Reconstruction Only &  1.8  & 58.6\% & 69.7\% & 63.7\% \\
    SubFrame Only      &  3.0  & 50.0\% & 83.8\% & 62.6\% \\
    Reconstruction + SubFrame &  0.90 & 76.9\% & 84.5\% & 80.5\% \\
    \bottomrule
  \end{tabular}}
  \caption{Ablation study of interference suppression.}
  \label{tab:detection_performance}
  \vspace{-2mm}
\end{table}

\vspace{-2mm}
\subsection{Sensing-Aware Interference Suppression}
\label{sec:interference_suppression}

\noindent \red{The compatibility analysis above shows that the deterministic sensing bias introduced by the proposed start-frequency based coding scheme is negligible and can be simply compensated. However, a more fundamental challenge arises from the communication signal transmitted by the other radar.}
As discussed in Sec.~\ref{sec:preliminary}, such inter-radar signals can generate ghost targets, elevate the noise floor, and degrade detection accuracy, as shown in Fig.~\ref{fig:combined_analysis}. Therefore, the native sensing process must suppress communication-induced interference while preserving real target detections.

To quantify the impact of interference, we conduct experiments in a static environment, as shown in Fig.~\ref{fig:experimental_setup}. We first operate the local radar alone to obtain ground-truth detections using range-Doppler FFT and OS-CFAR, and then repeat the experiment with the communicating target radar active. As shown in Table~\ref{tab:detection_performance}, 
\red{communication-induced interference results in}
an average of 11.43 false detections per frame. Although the recall remains 81.0\%, the precision drops to 20.1\%, yielding an F1 score of only 32.2\%. 
\red{These results demonstrate that the communication-induced ghost detections make the native sensing output unreliable, highlighting the need for interference suppression.}

\begin{figure*}[t]
    \centering
    \begin{minipage}[t]{\dimexpr3\figw+2\figsep\relax}
        \centering
        \vspace{0pt} %
        \begin{subfigure}[t]{\figw}
            \centering
            \includegraphics[width=\linewidth]{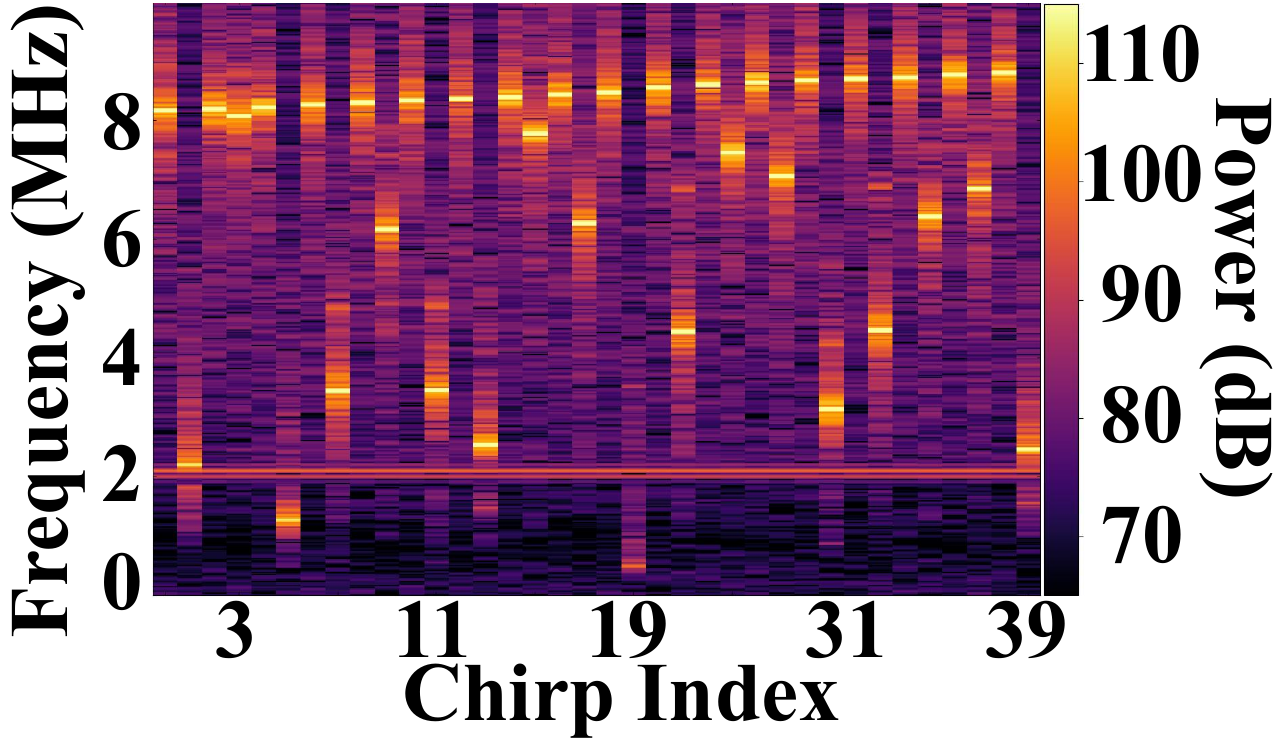}
            \caption{no \green{suppression}}
            \label{fig:stft_wo_mitigation}
        \end{subfigure}%
        \hspace{\figsep}%
        \begin{subfigure}[t]{\figw}
            \centering
            \includegraphics[width=\linewidth]{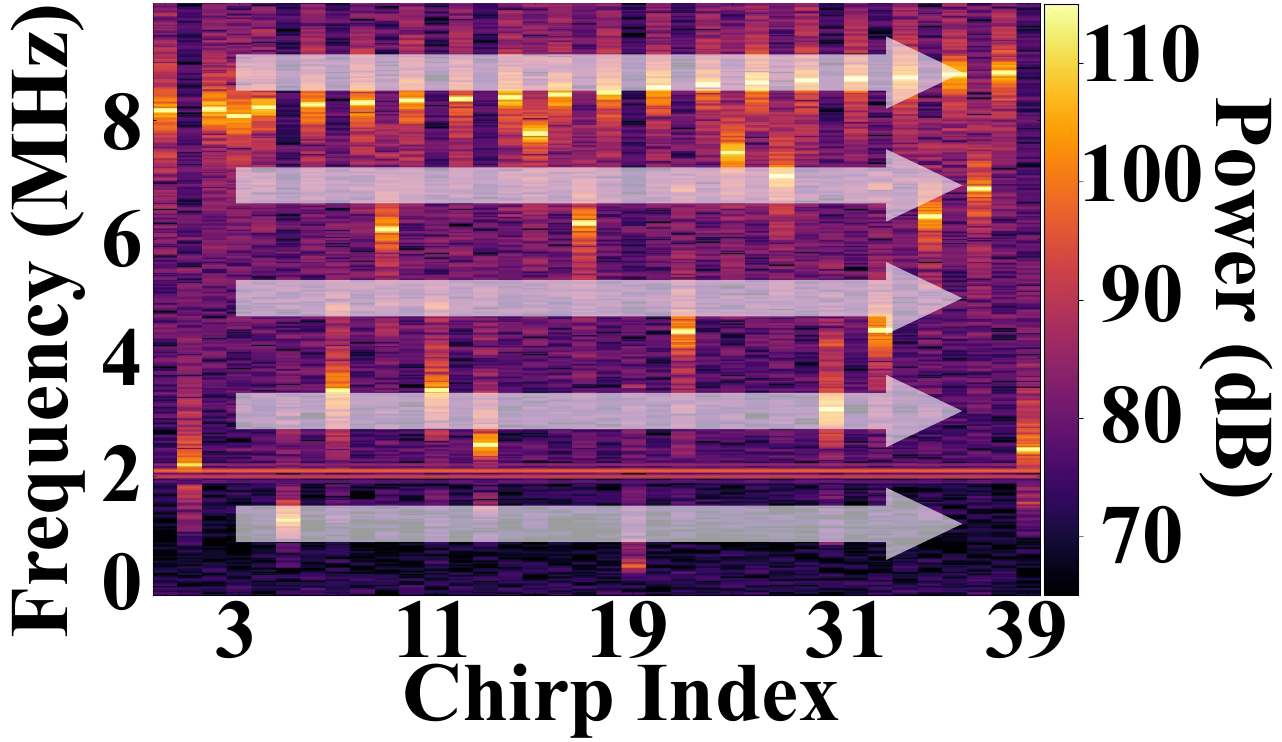}
            \caption{\green{suppression}}
            \label{fig:mitigation}
        \end{subfigure}%
        \hspace{\figsep}%
        \begin{subfigure}[t]{\figw}
            \centering
            \includegraphics[width=\linewidth]{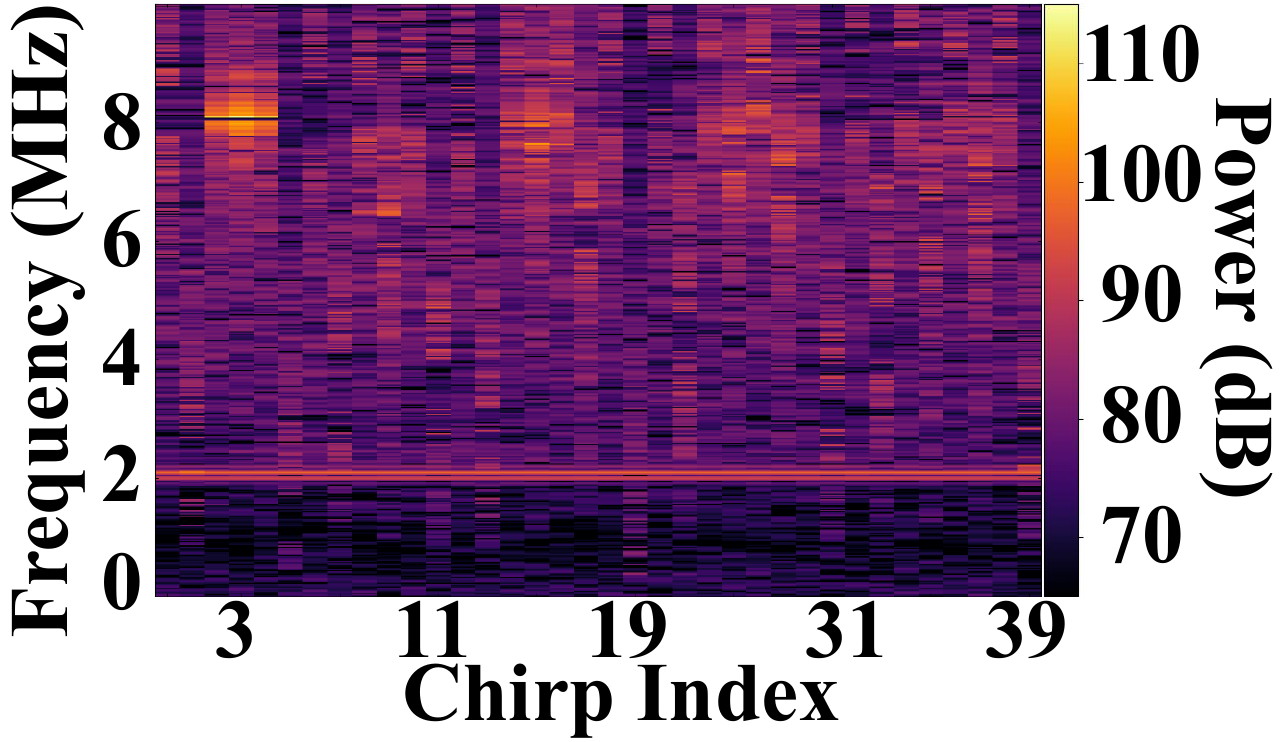}
            \caption{with suppression}
            \label{fig:stft_w_mitigation}
        \end{subfigure}

        \vspace{0mm}
        \caption{Frame STFT examples before/after interference suppression.}
        \label{fig:stft_comparison}
    \end{minipage}%
    \hspace{\figsep}%
    \begin{minipage}[t]{\figw}
        \centering
        \vspace{0mm} %
        \includegraphics[width=\linewidth]{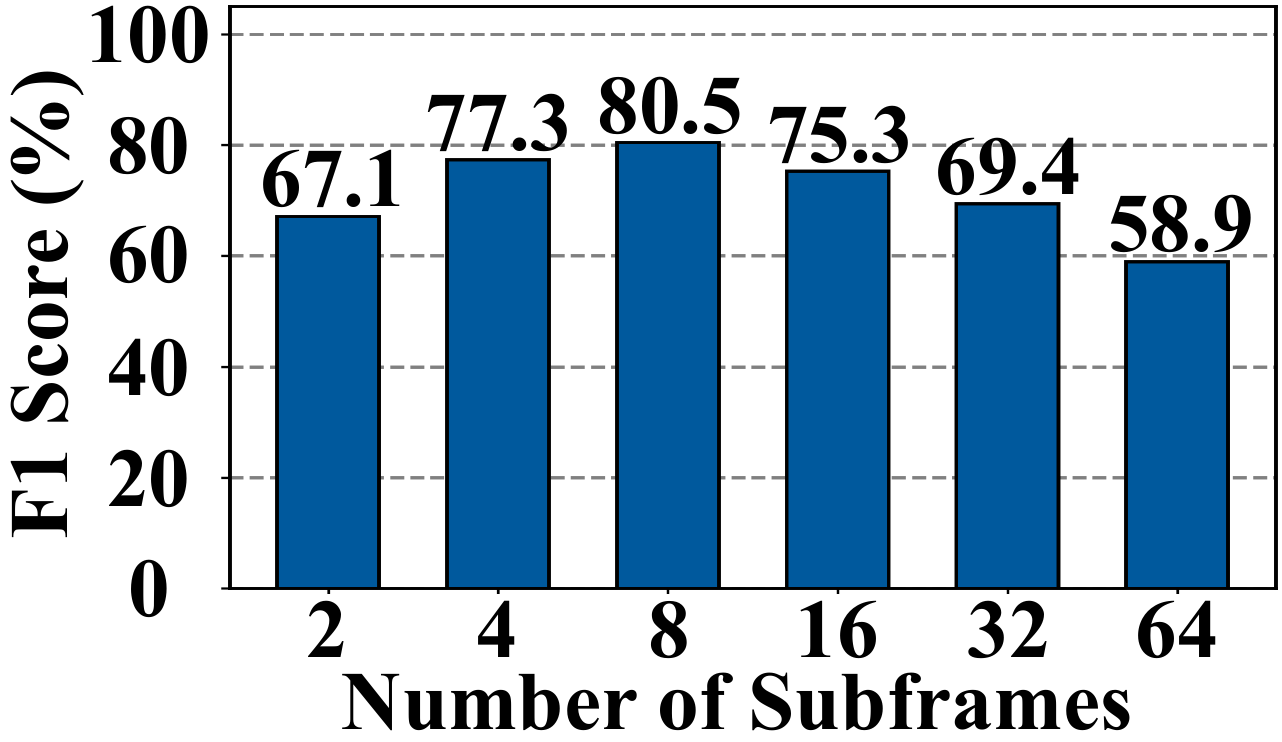}
        \vspace{-4mm} %
        \caption{Sensing F1 score vs. subframe count.}
        \label{fig:subframe_sweep}
    \end{minipage}
    \vspace{2mm}
\end{figure*}

\red{To address this challenge, RadarCom introduces a two-stage sensing-aware interference suppression framework. The design is motivated by a key observation: the ghost peaks generated from the differential coding scheme, where reference and data chirps use different start frequencies, shift abruptly between adjacent chirps, whereas real target reflections vary much more smoothly across chirps within a frame, even in the presence of moving targets, \blue{as shown in Fig.~\ref{fig:stft_comparison}(a)}. Building on this observation, RadarCom suppresses communication-induced interference through two stages: \textbf{(i)} \green{Power Jump Detection and Sample Reconstruction}, which removes dominant interference components in the range-FFT domain, and \textbf{(ii)} \green{Residual False Detection Removal}, which further eliminates residual interference in the range-Doppler domain, as detailed next.}

$\bullet$ \textbf{\green{Power Jump Detection and Sample Reconstruction.}}
The first stage suppresses the dominant communication-induced interference through cross-chirp reconstruction in the range-FFT domain. After computing the range-FFT for each chirp, RadarCom obtains a sequence of per-chirp range profiles. 
\red{It then examines each range/frequency bin along the chirp dimension to identify interference by detecting sudden upward power jumps, \blue{as illustrated in Fig.~\ref{fig:stft_comparison}(b)}.}

\begin{figure}[t]
    \centering
    \begin{subfigure}[t]{0.32\columnwidth}
        \centering
        \includegraphics[width=\linewidth,height=0.9\linewidth,keepaspectratio]{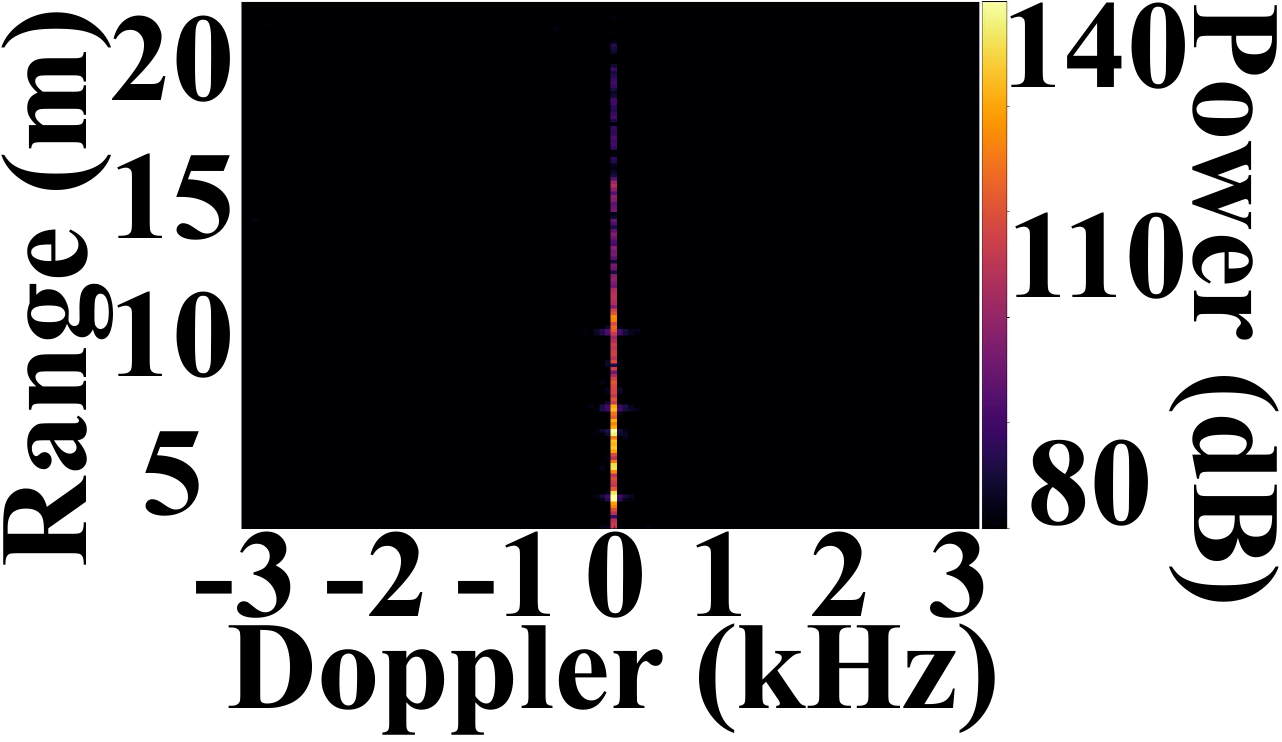}
        \caption{RD Ground Truth.}
        \label{fig:rd_gt}
    \end{subfigure}
    \hspace{-0.6mm}
    \begin{subfigure}[t]{0.32\columnwidth}
        \centering
        \includegraphics[width=\linewidth,height=0.9\linewidth,keepaspectratio]{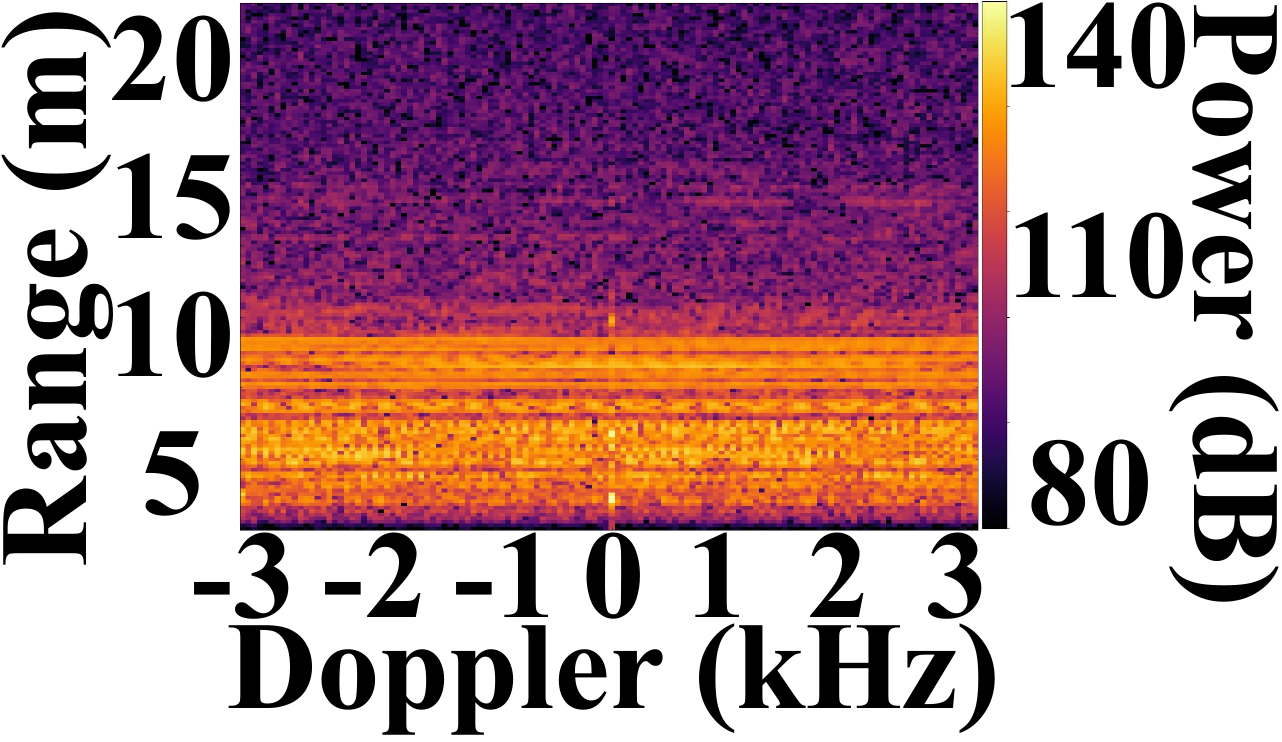}
        \caption{no suppression.}
        \label{fig:rd_wo_mitigation}
    \end{subfigure}
    \hspace{-0.6mm}
    \begin{subfigure}[t]{0.32\columnwidth}
        \centering    \includegraphics[width=\linewidth,height=0.9\linewidth,keepaspectratio]{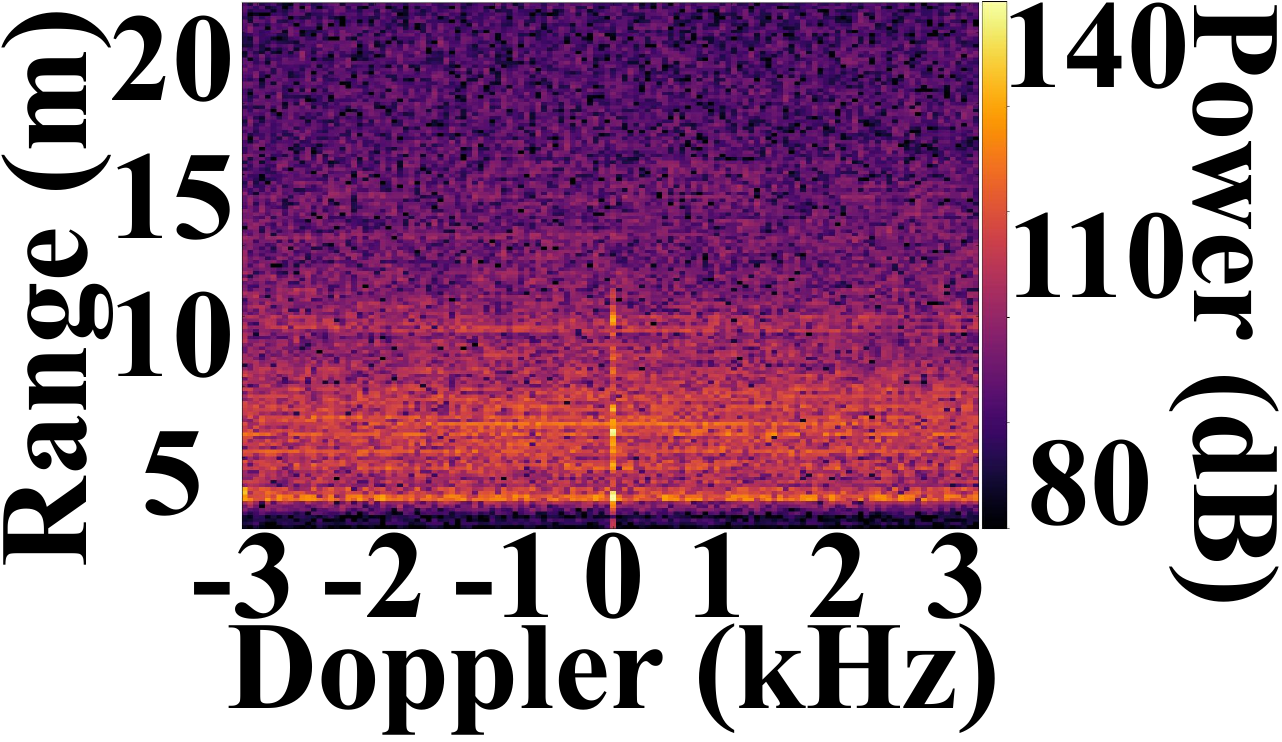}
        \caption{with suppression.}
        \label{fig:rd_w_mitigation}
    \end{subfigure}
    \vspace{-1mm}
    \caption{Range-Doppler comparison.}
    \label{fig:after_mitigation}
    \vspace{-2mm}
\end{figure}

Let $P(n,r)$ denote the power at range bin $r$ for chirp $n$. \green{RadarCom computes the power change between adjacent chirps as:}
\begin{equation}
\Delta P(n,r) = P(n,r) - P(n-1,r), \quad n=1,\ldots,N-1.
\end{equation}
A sample is first marked as a candidate interference point if
\begin{equation}
\Delta P(n,r) > \mu_{\Delta} + \lambda\sigma_{\Delta},
\label{eq:interp_threshold}
\end{equation}
\green{where $\mu_{\Delta}$ and $\sigma_{\Delta}$ are the mean and standard deviation of $\Delta P(n,r)$ within the frame, and $\lambda$ is a thresholding coefficient that controls the sensitivity of interference detection.}
\red{When an upward jump is detected at $P(n,r)$, RadarCom marks the corresponding sample and adjacent samples at the same range bin that exhibit a similar elevated power level. The marked segment is then reconstructed using linear interpolation between the nearest preceding and following unmarked samples as $X'(n,r)$:}
\begin{equation}
X'(n,r) =
\begin{cases}
X(n^*,r), & \text{if marked},\\
X(n,r), & \text{otherwise},
\end{cases}
\end{equation}
where $X(n,r)$ is the complex range-FFT output, and $X(n^*,r)$ denotes the reconstructed value obtained through linear interpolation.  This reconstruction suppresses abrupt communication-induced interference while preserving the smooth slow-time evolution of real target reflections, as shown in Fig.~\ref{fig:stft_comparison}(c) and Fig.~\ref{fig:after_mitigation}.

\begin{figure}[t]
    \centering
    \includegraphics[width=0.95\columnwidth]{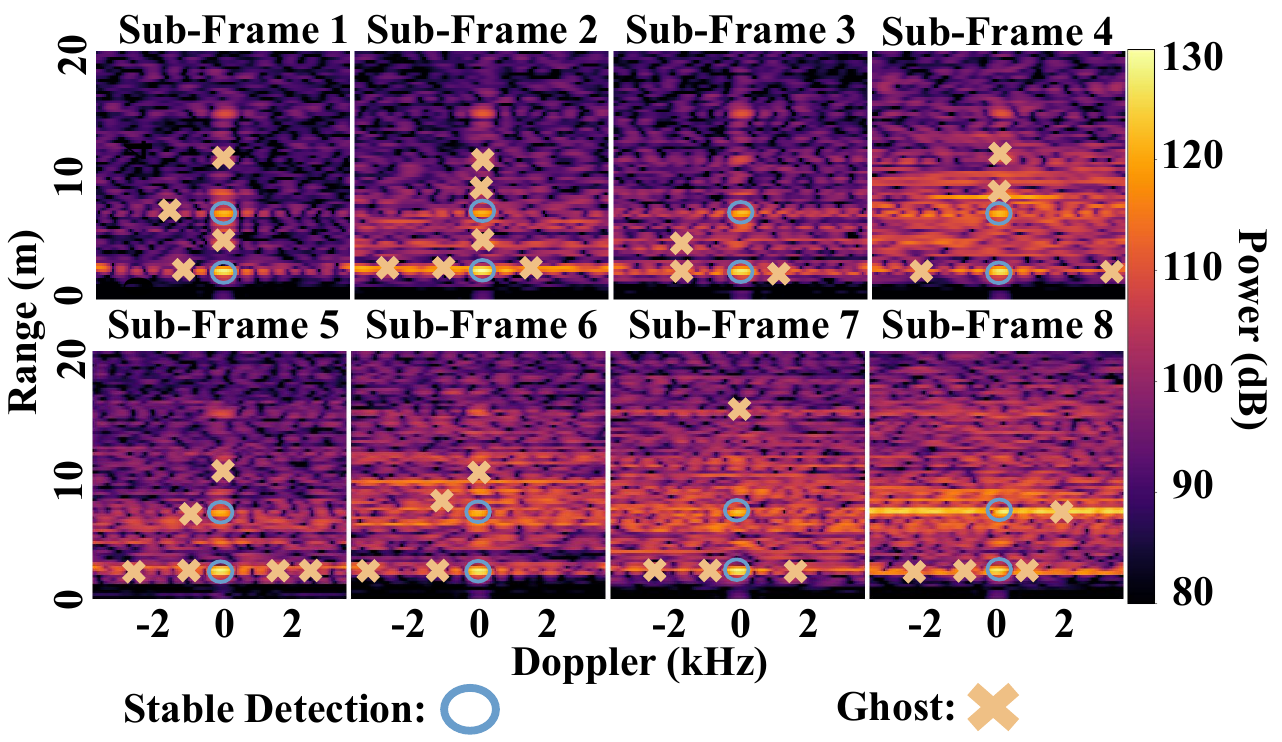}
    \vspace{-1mm}
    \caption{Sub-frame detection consistency.}
    \label{fig:sub_frame_comparison}
    \vspace{-3mm}
\end{figure}

$\bullet$ \textbf{\green{Residual False Detection Removal.}}
\green{After sample reconstruction, residual interference may still cause false detections in the RD map. To further remove them, RadarCom checks 
detection consistency within the frame. Specifically, RadarCom divides each 128-chirp frame into multiple equal-length sub-frames, where each sub-frame is a segment of consecutive chirps. RadarCom then independently applies RD-FFT and OS-CFAR to each sub-frame. A detection in the final CFAR result is kept only if a corresponding detection appears at a similar RD location in most sub-frames. This is based on the observation that real target reflections remain relatively stable within a frame, while residual communication-induced interference varies due to start-frequency coding and appears less consistently. Figure~\ref{fig:sub_frame_comparison} illustrates this difference.}

\green{Fig.~\ref{fig:subframe_sweep} shows that 8 sub-frames best balance residual false detection removal and target preservation, so we use this setting in our evaluation. As summarized in Table~\ref{tab:detection_performance}, sample reconstruction removes most high-energy communication interference, while residual false detection removal suppresses remaining false detections. Together, the two stages achieve the best sensing performance, with an F1 score of 80.5\%.}

\vspace{-1mm}
\section{Implementation and Evaluation}
\vspace{-1mm}
\subsection{Implementation Details}
\label{subsec:implementation_details}
\begin{figure}[t]
    \centering
    \begin{minipage}[t]{0.49\columnwidth}
        \centering
        \includegraphics[width=0.85\linewidth]{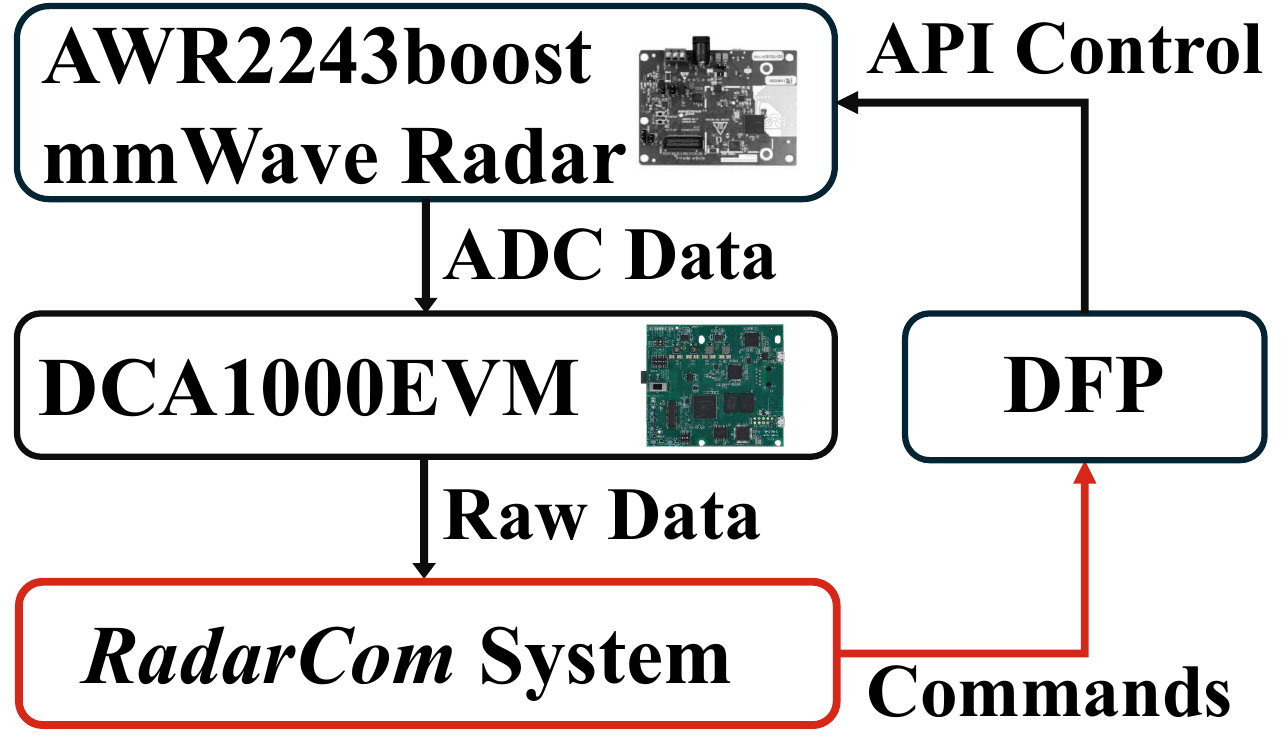}
        \caption{\small{System architecture.}}
        \label{fig:system_architecture}
    \end{minipage}\hfill
    \begin{minipage}[t]{0.49\columnwidth}
        \centering
        \includegraphics[width=0.85\linewidth]{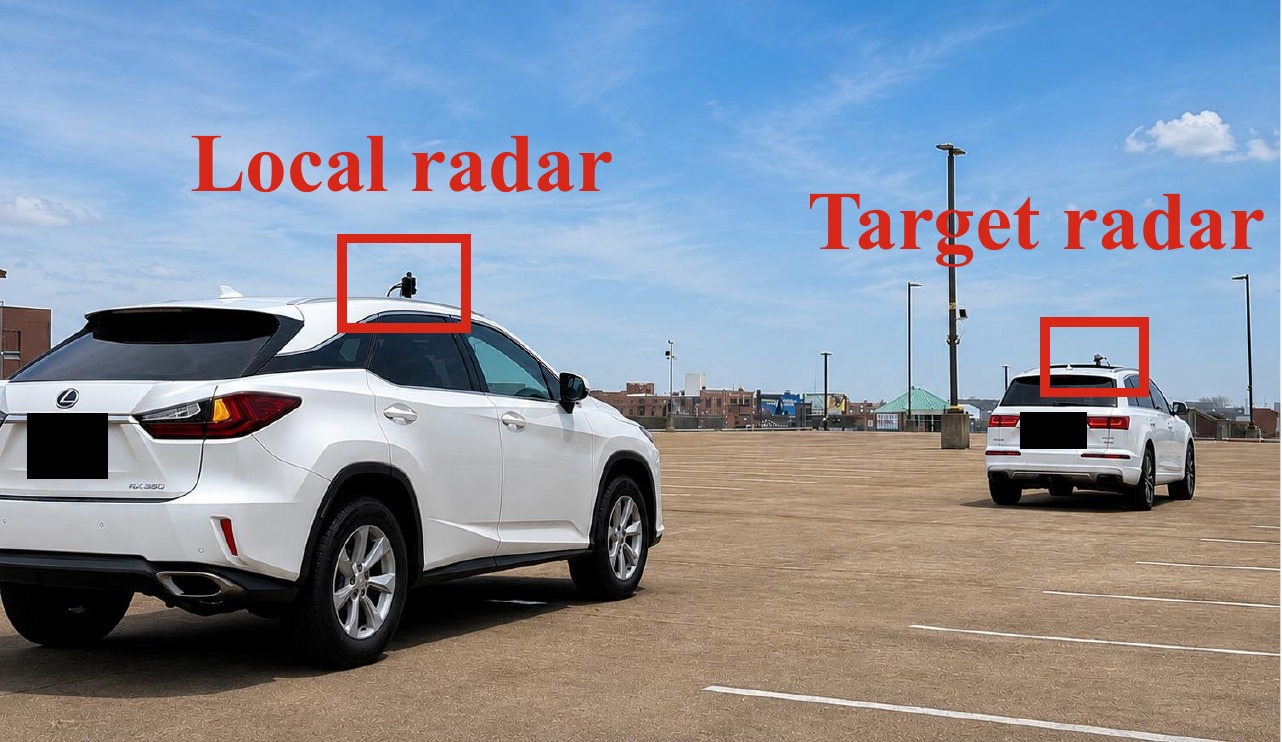}
        \caption{\small{Experimental setup.}}
        \label{fig:experimental_setup}
    \end{minipage}
    \vspace{-2mm}
\end{figure}

\noindent $\bullet$ \textbf{Hardware and Control Architecture.}
Figure~\ref{fig:system_architecture} illustrates the hardware and control architecture of the proposed RadarCom prototype. The system is built on a TI AWR2243BOOST mmWave radar connected to a DCA1000EVM capture card, which streams raw
ADC data to a host laptop through an Ethernet interface. 
\red{The host configures and controls the radar using TI's official mmWave Device Firmware Package (DFP v02\_02\_05\_00) and runs the RadarCom processing pipeline.}
The prototype logic is implemented in Python, including \red{target radar discovery, } synchronization, differential decoding, and sensing-aware interference suppression.

$\bullet$ \textbf{Experimental Setup.} 
Figure~\ref{fig:experimental_setup} shows the basic experimental setup. We deploy the RadarCom system on two vehicles equipped with mmWave radars, with the front vehicle serving as the target radar and the following vehicle as the local radar. The two vehicles communicate directly while performing sensing, without a shared timing reference or centralized synchronization.

In our prototype, each radar is controlled by an independent host using 
the vendor-provided dynamic chirp reconfiguration API~\cite{ti_mmwave_icd_2024} \red{to reconfigure} 
chirp parameters.
\red{We first measure the API reconfiguration latency and observe that each update takes an average delay of 94~ms.  This relatively high API latency forces (\textbf{i}) a long frame duration, where 
we configure a frame period of 101.592~ms with 128 chirps per frame, thus ensuring timely chirp reconfiguration for each frame; and (\textbf{ii}) a small chirp slope, where 
each chirp uses a 740~$\mu$s ramp time and a 7~$\mu$s idle time, resulting in a 747~$\mu$s chirp repetition interval. The corresponding chirp slopes are 5.311~MHz/$\mu$s for the local radar and 5.262/5.359~MHz/$\mu$s for the target radar's beacon chirps, thus providing the slope mismatch required for discovery and synchronization.}

\red{Under this configuration, RadarCom uses differential coding on range-FFT with 64 data chirps per communication frame and performs beacon-based re-synchronization once every 50 frames. The resulting data rate is $617.37\log_2(M)$~bps, reaching 5.56 kbps for $M=512$. Unless otherwise specified, all experiments use this API-constrained configuration. The long chirp duration also yields a native range resolution of approximately 0.55 m, while the 747~$\mu$s chirp repetition interval limits the maximum unambiguous Doppler velocity \SI{4.7}{\kilo\meter/\hour}. These limitations stem primarily from the high API reconfiguration latency rather than the RadarCom design itself. Integrating RadarCom directly into the radar firmware or vendor control stack could significantly reduce the reconfiguration latency, enabling shorter frame durations, higher communication throughput, finer range resolution, and larger unambiguous Doppler ranges.}

\vspace{-2mm}
\subsection{Basic Evaluation Results}
\label{sec:eval_basic}
\noindent \red{We evaluate RadarCom under controlled static and dynamic line-of-sight (LoS) conditions. In the static experiments, we vary the inter-radar distance to examine robustness to range, while in the dynamic experiments, we vary the relative velocity to examine robustness under mobility. Effective data rate, defined as the net application-layer throughput after accounting for periodic re-synchronization overhead and symbol errors, is used as the communication performance metric. For sensing, we use F1 score of CFAR outputs, range RMSE, \bluee{angle MAE}, and velocity MAE to quantify detection, ranging, \bluee{angle-estimation,} and velocity-estimation accuracy, respectively.} \bluee{\textcolor{red}{To further evaluate target detection performance,}
we \textcolor{red}{study} 
the target miss rate and the average number of residual ghost targets per frame.}

\blue{$\bullet$ \textbf{Static Scenario.}
We first evaluate RadarCom under LoS conditions, where both radars remain stationary. The inter-radar distance is varied from \SIrange{10}{400}{\meter} in \SI{10}{\meter} increments.} 

\begin{figure*}[t]
    \centering

    \begin{minipage}[t]{0.747\linewidth}
        \centering

        \begin{subfigure}[t]{0.328\linewidth}
            \centering
            \includegraphics[width=\linewidth]{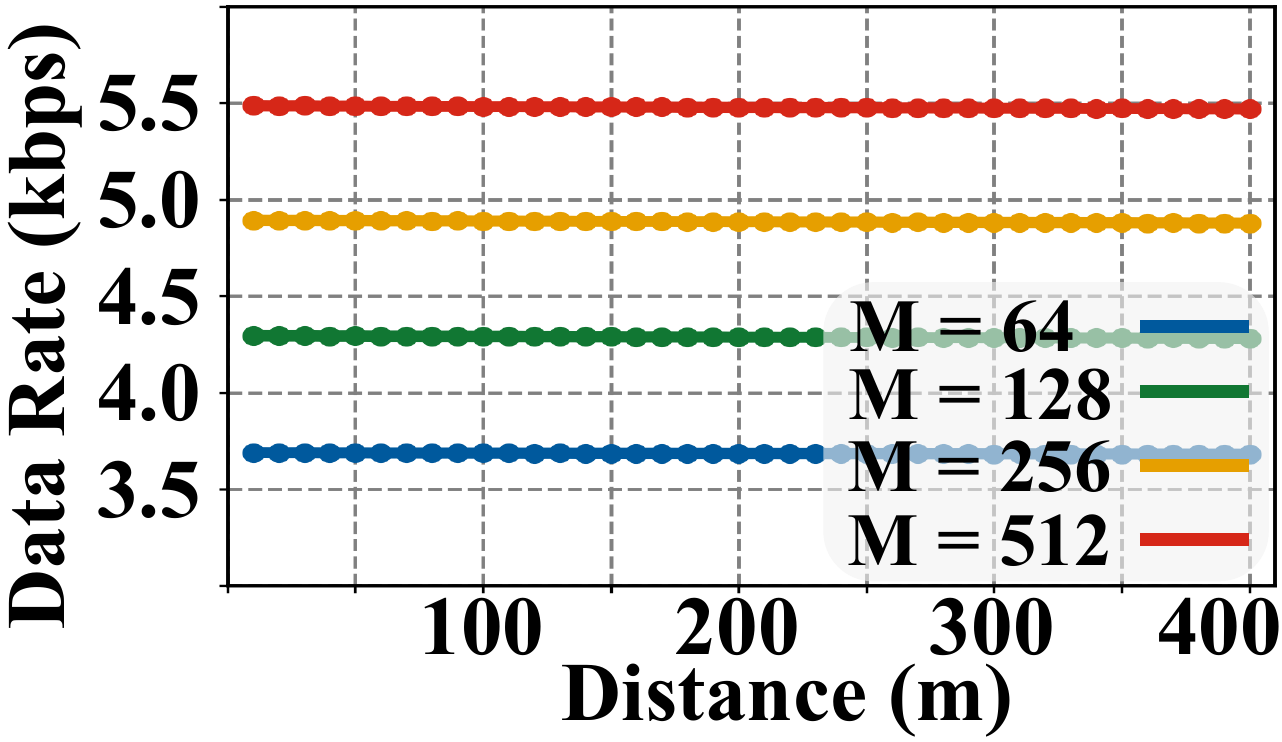}
            \subcaption{Distance vs.~rate}
            \label{fig:comm_static_rate}
        \end{subfigure}%
        \hfill
        \begin{subfigure}[t]{0.328\linewidth}
            \centering
            \includegraphics[width=\linewidth]{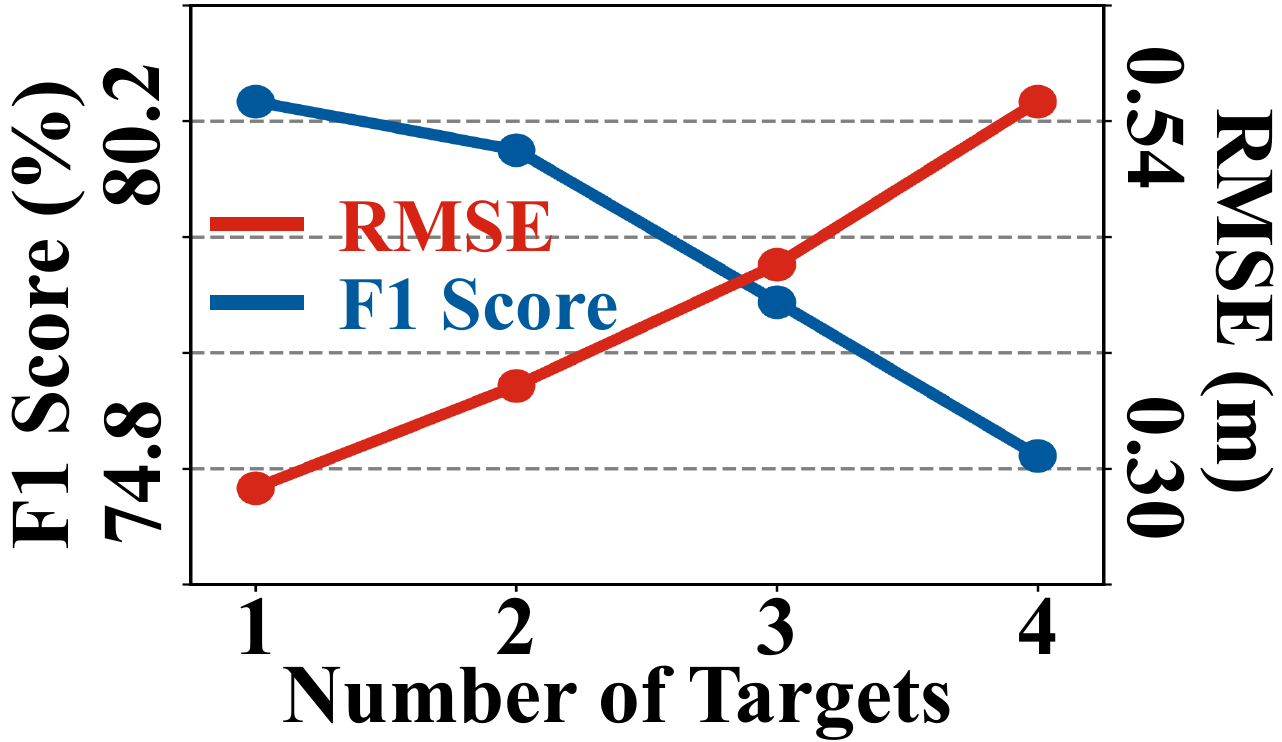}
            \subcaption{Detection and ranging}
            \label{fig:sensing_static}
        \end{subfigure}%
        \hfill
        \begin{subfigure}[t]{0.328\linewidth}
            \centering
            \includegraphics[width=\linewidth]{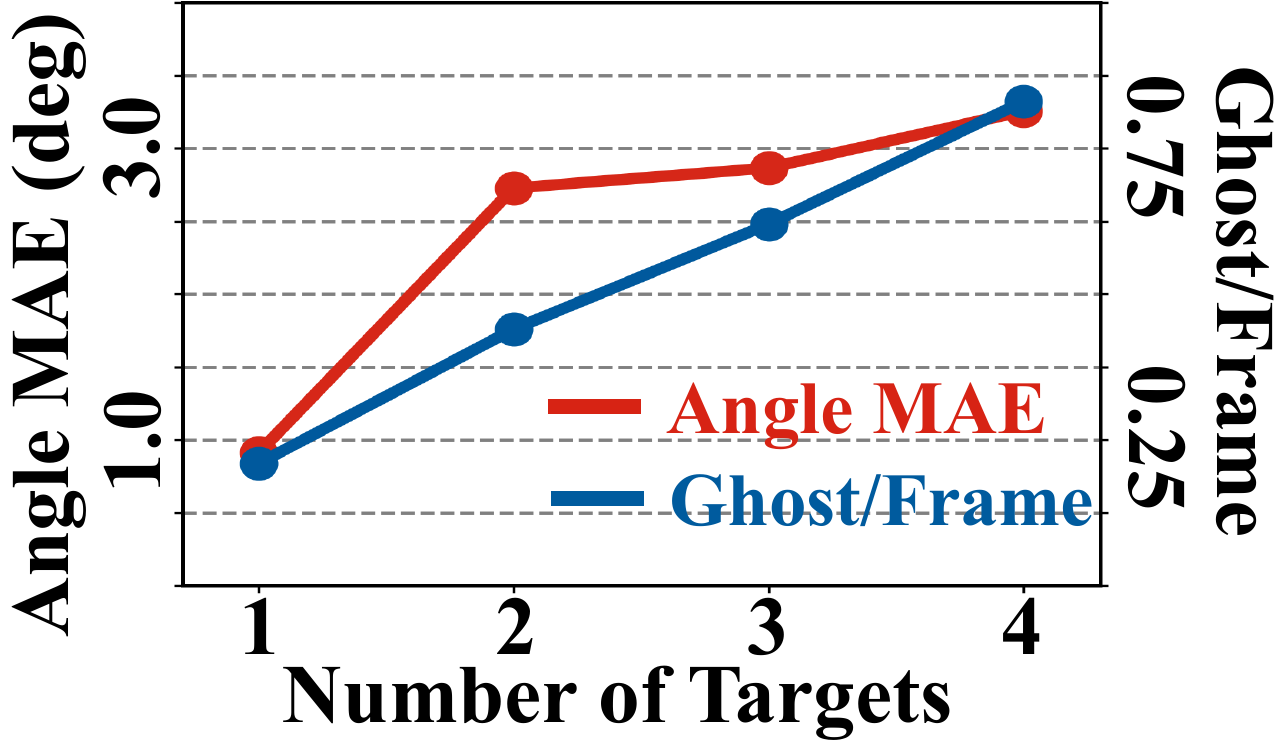}
            \subcaption{\bluee{Angle error and residual ghosts}}
            \label{fig:sensing_angle_ghosts}
        \end{subfigure}

        \vspace{-1mm}
        \caption{\bluee{Performance under static conditions.}}
        \label{basic_dynamic}
    \end{minipage}%
    \hfill
    \begin{minipage}[t]{0.245\linewidth}
        \centering
        \includegraphics[width=\linewidth]{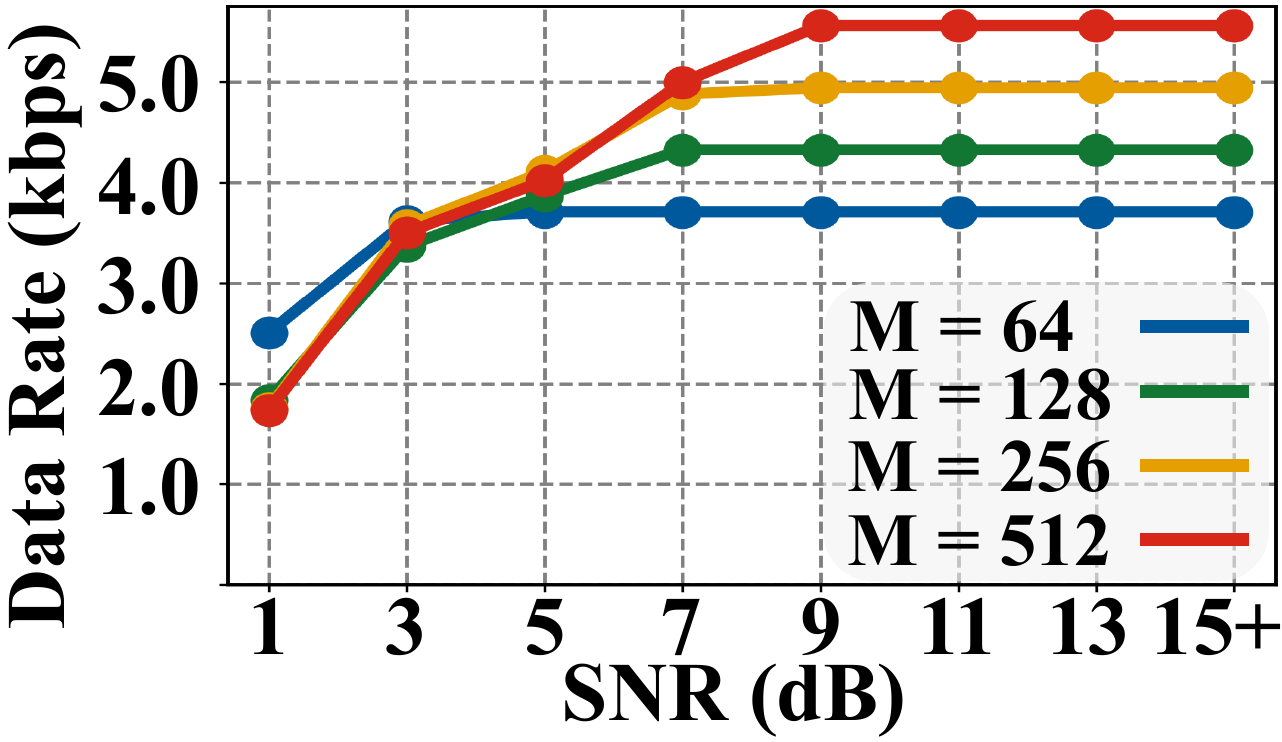}

        \vspace{-1mm}
        \caption{SNR vs.~data rate.}
        \label{fig:snr_ser}
    \end{minipage}

    \vspace{-1mm}
\end{figure*}

\blue{\textit{(i) Communication Performance:}
\red{We first study the effective data rate. 
As shown in Fig.~\ref{basic_dynamic}(a), the effective data rate remains stable across distances from \SI{10}{\meter} to \SI{400}{\meter}, demonstrating robust LoS communication. As expected, larger modulation orders provide higher throughput, increasing the modulation order from $M=64$ to $M=512$ raises the effective data rate from approximately 3.7~kbps to 5.5~kbps. The $400$~m distance is limited by our test site.}
}

\begin{figure}[t]
    \centering
    \subfloat[1 target.]{\includegraphics[width=0.24\columnwidth]{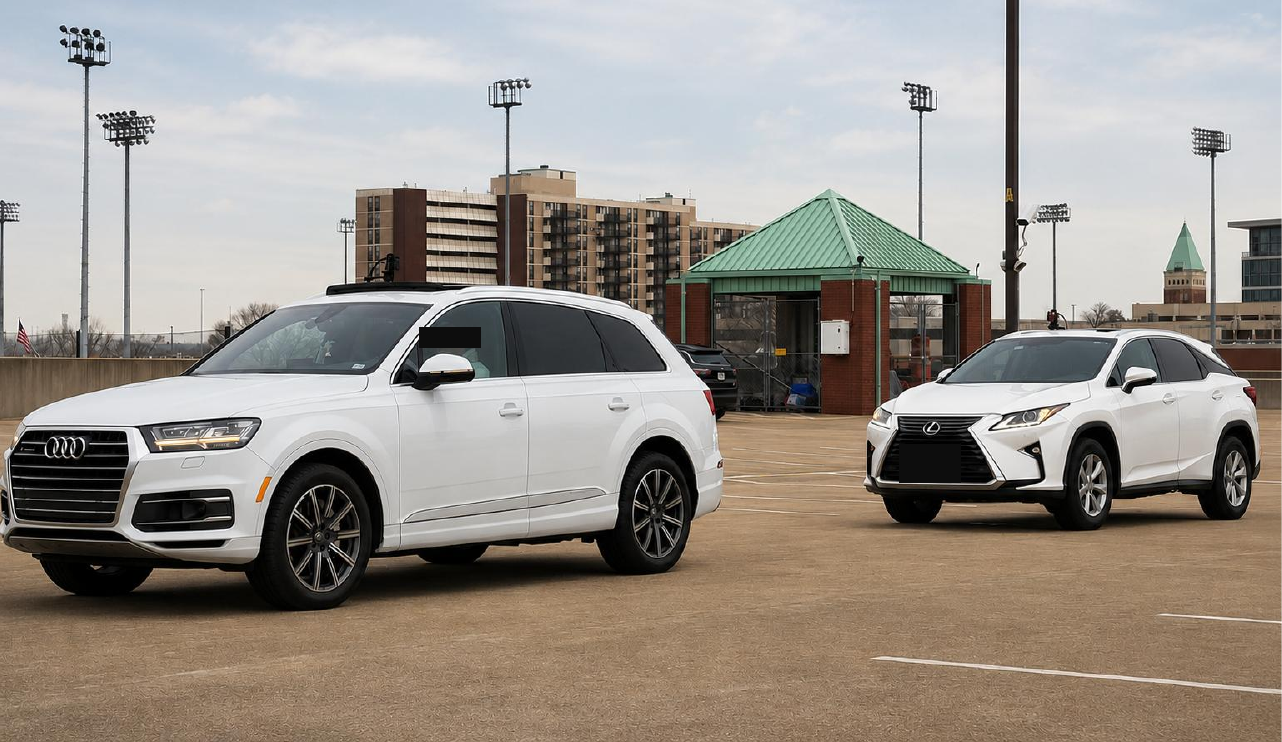}\label{fig:scene_a}}\hfill
    \subfloat[2 targets.]{\includegraphics[width=0.24\columnwidth]{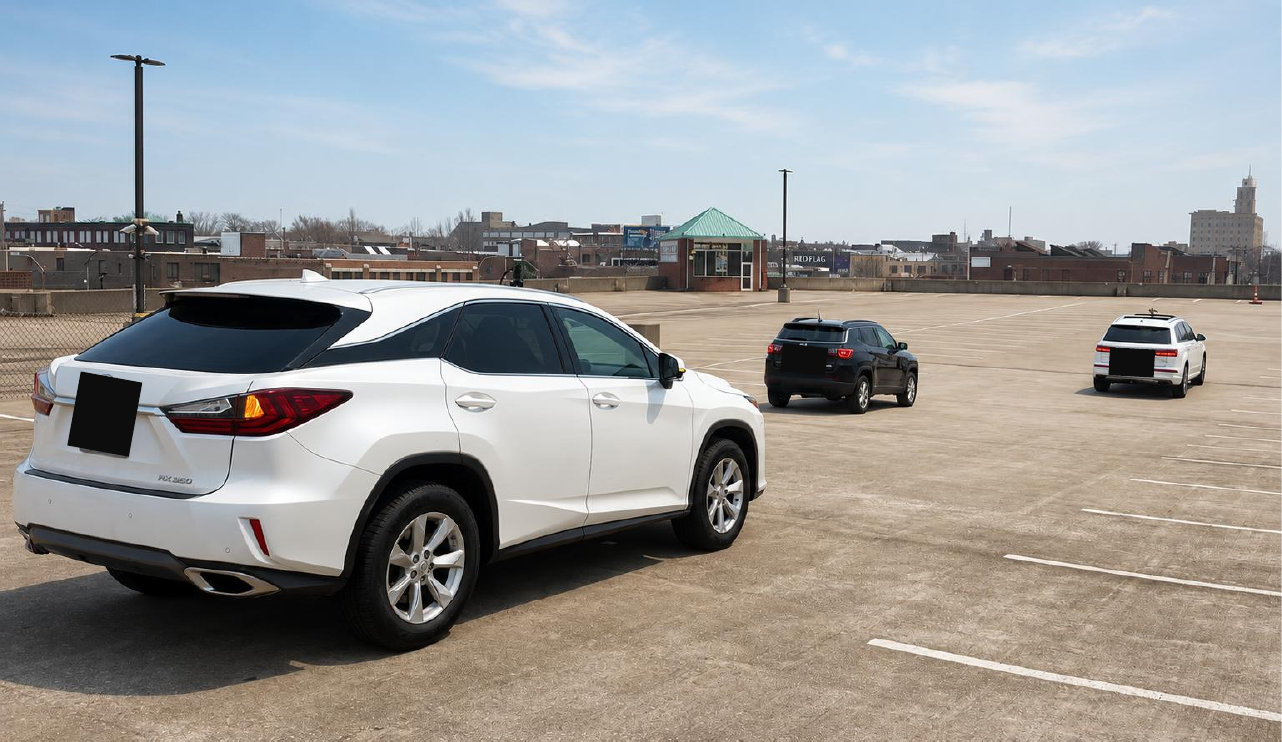}\label{fig:scene_b}}\hfill
    \subfloat[3 targets.]{\includegraphics[width=0.24\columnwidth]{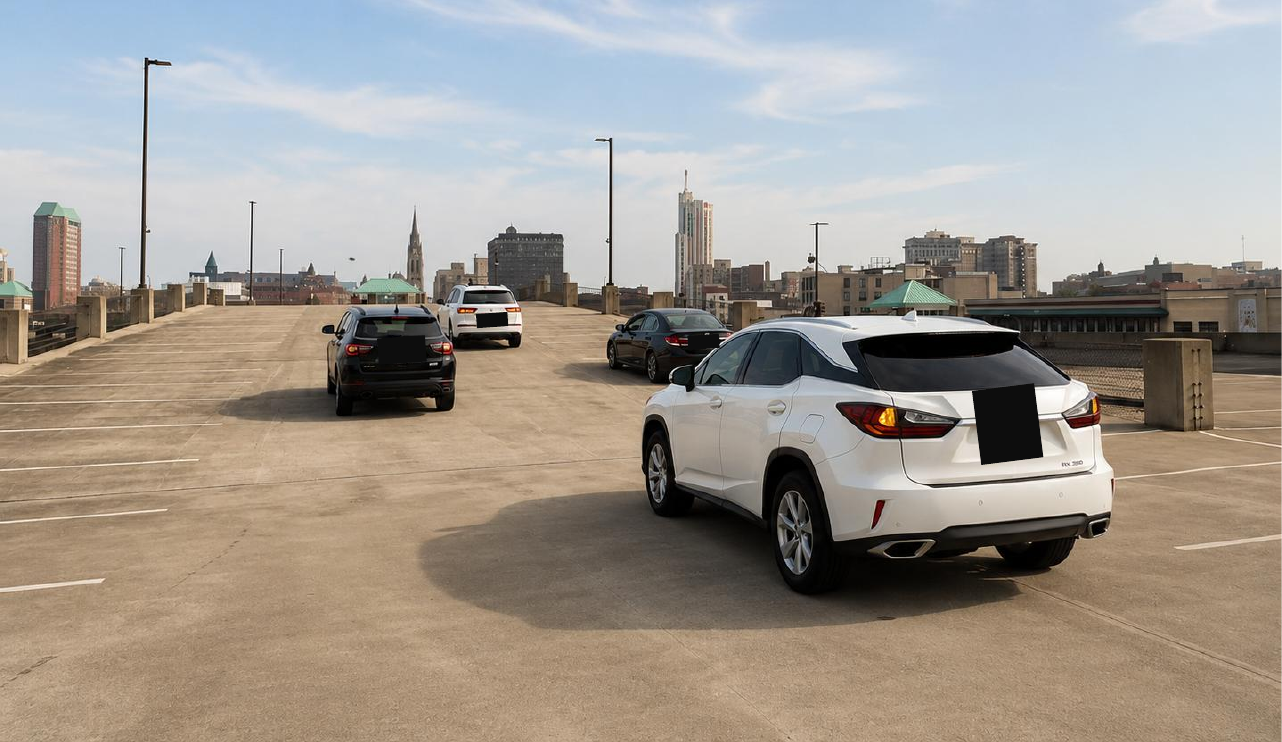}\label{fig:scene_c}}\hfill
    \subfloat[4 targets.]{\includegraphics[width=0.24\columnwidth]{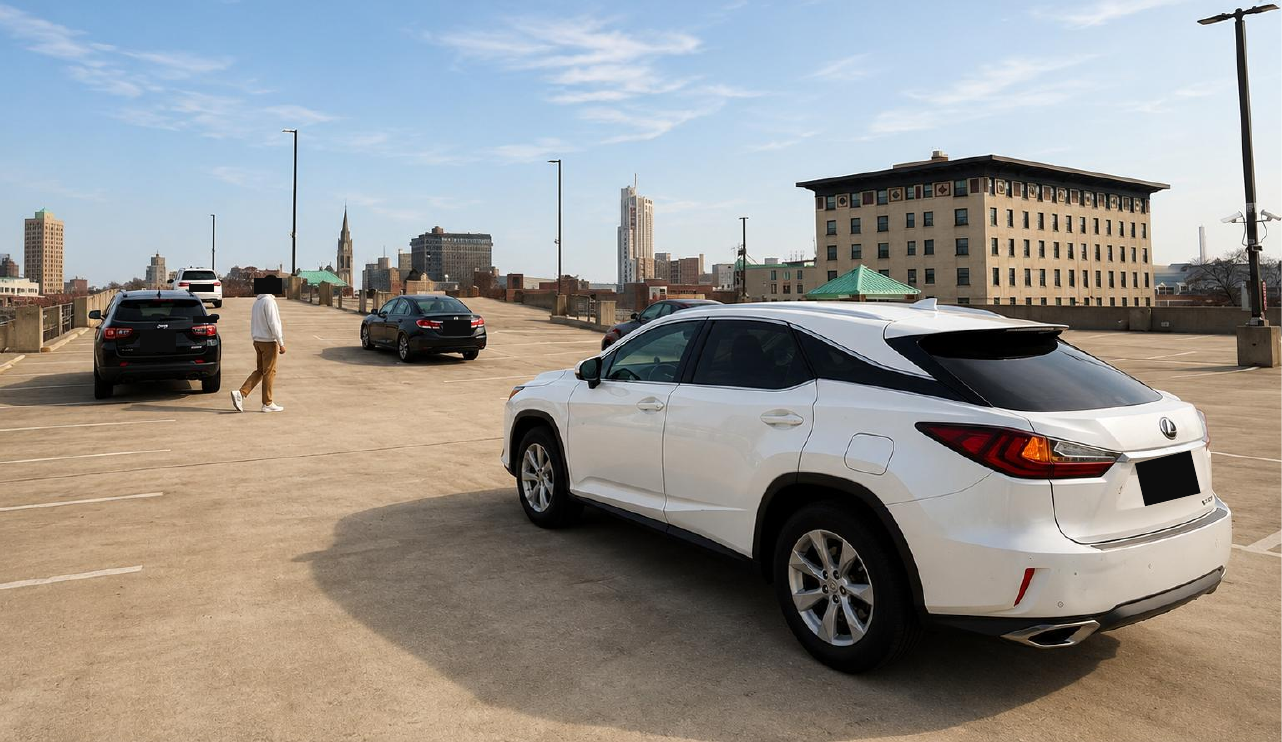}\label{fig:scene_d}}
    \vspace{-1mm}
    \caption{Sensing scenes with increasing complexity.}
    \label{fig:sensing_static_scenes}
    \vspace{-2mm}
\end{figure}

\textit{(ii) Sensing Performance:}
We next investigate sensing performance in four scenes with increasing complexity,
as shown in Fig.~\ref{fig:sensing_static_scenes}: (1) the target \red{vehicle} 
only, (2) the target vehicle with one interfering vehicle, (3) the target vehicle with two interfering vehicles, and (4) the target vehicle with two interfering vehicles and one pedestrian. 
We first measure a reference sensing result with the target radar's communication disabled. We then enable RadarCom communication while sensing.
Fig.~\ref{basic_dynamic}(b) shows the F1 score and range RMSE. As scene complexity increases, the F1 score remains above 74\% across all scenes with a moderate decrease due to stronger multi-path effects. Meanwhile, the range RMSE remains below 0.55~m, around one range-resolution bin in our configuration. \bluee{Fig.~\ref{basic_dynamic}(c) further shows the angle MAE of matched targets and the residual ghost targets per frame. The angle MAE increases from $0.92^\circ$ to $3.26^\circ$, well below the $\approx 29^\circ$ angular resolution of our 4-element RX array. Meanwhile,
\textcolor{red}{the average number of residual ghost targets remains low, ranging from $0.21$ to $0.83$ per frame, while the target miss rate stays below $0.5\%$ across all scenes.}} These results demonstrate that RadarCom achieves reliable \bluee{detection, ranging, and angle estimation} during simultaneous communication, even in increasingly complex environments.

\blue{$\bullet$ \textbf{Dynamic Scenario.}
We further evaluate RadarCom under relative motion between the two radars. In each experiment, one radar remains stationary while the other moves along the LoS at a speed of 5 to 25 km/h. During the motion, the inter-radar distance varies from approximately \SI{5}{\meter} to \SI{30}{\meter}.}

\begin{figure}[t]
    \centering

    \begin{subfigure}[t]{0.49\linewidth}
        \centering
        \includegraphics[width=\linewidth]{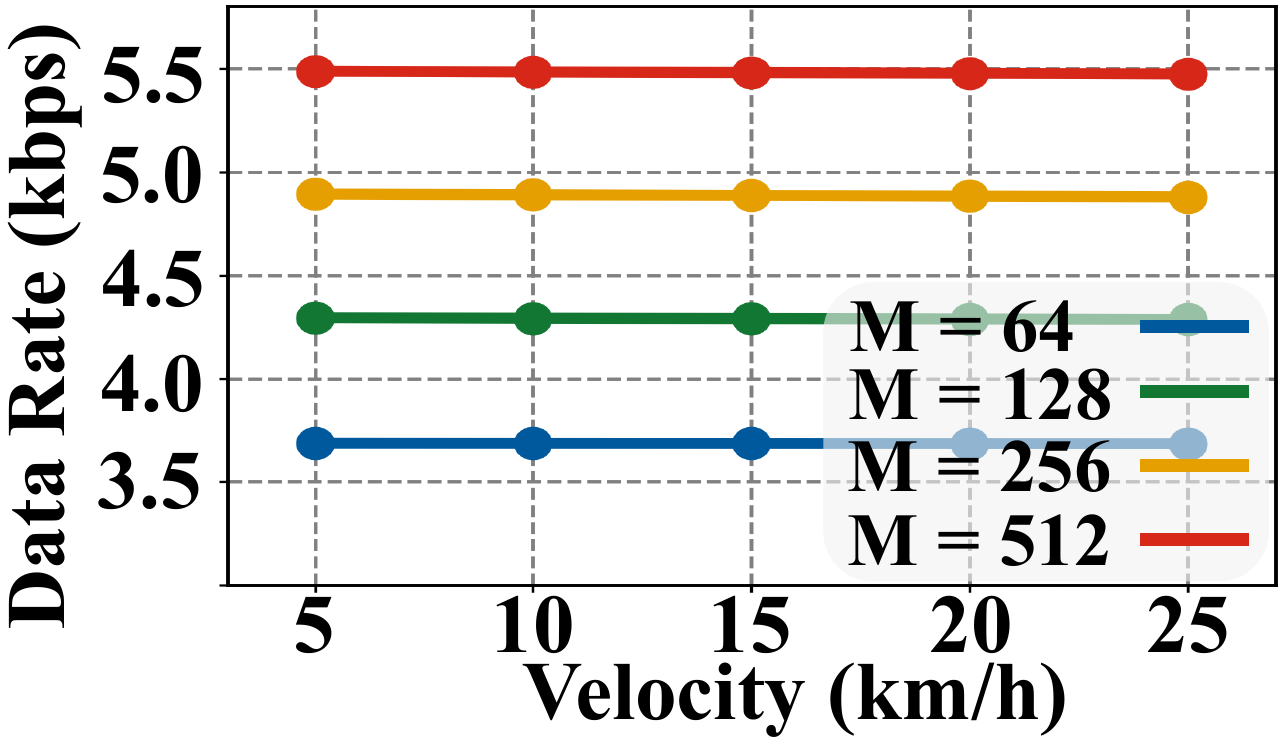}
        \subcaption{Speed vs.~rate}
        \label{fig:dynamic_communication}
    \end{subfigure}%
    \hfill
    \begin{subfigure}[t]{0.49\linewidth}
        \centering
        \includegraphics[width=\linewidth]{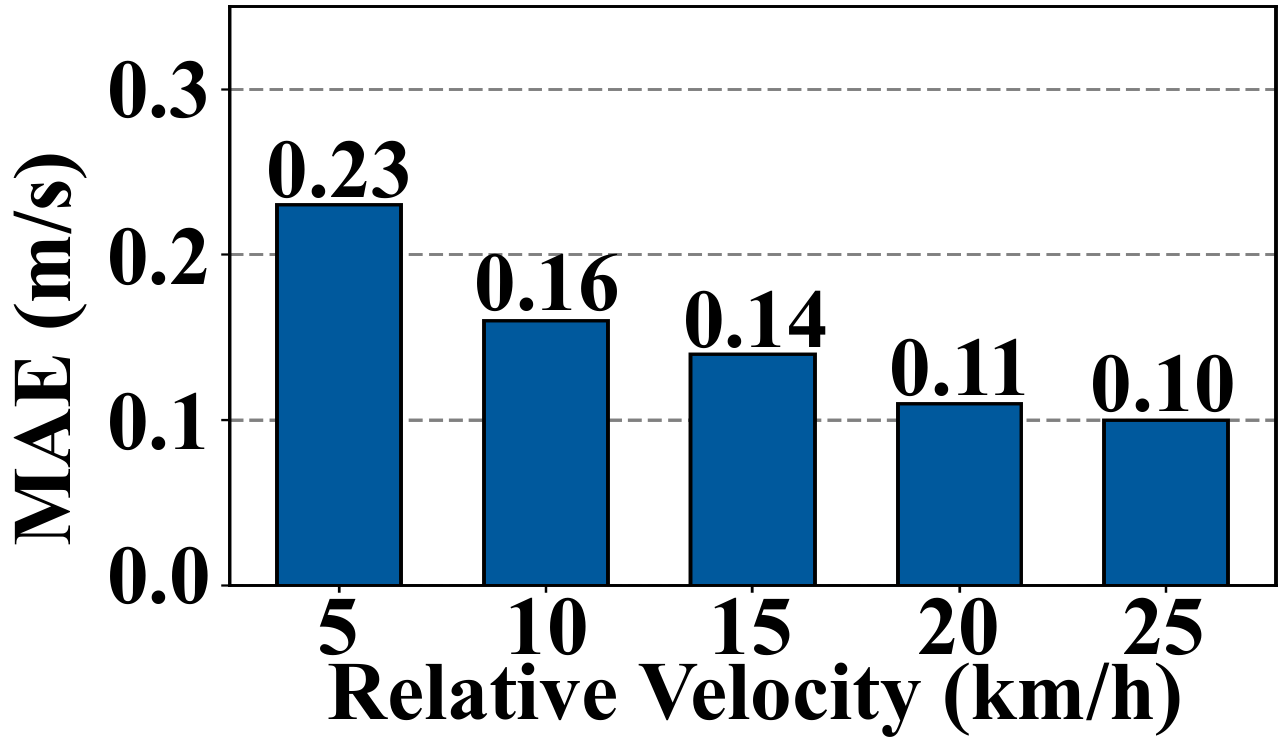}
        \subcaption{Velocity MAE}
        \label{fig:dynamic_sensing}
    \end{subfigure}

    \vspace{0mm}
    \caption{Performance under relative motion.}
    \label{fig:dynamic_evaluation}
    \vspace{-2mm}
\end{figure}

\blue{\textit{(i) Communication Performance:}
\red{We next evaluate the effective data rate under relative mobility. As shown in Fig.~\ref{fig:dynamic_evaluation}(a), results are consistent with the static scenario: the effective data rate remains nearly constant across the tested speed range, while increasing the modulation order from $M=64$ to $M=512$ raises the throughput from approximately 3.7~kbps to 5.5~kbps. Even for the highest modulation order ($M=512$), the SER increases only slightly from 1.27\% to 1.51\% as the relative speed rises from 5 to 25~km/h, resulting in negligible throughput degradation. This robustness stems from the differential coding scheme, which cancels the distance-dependent shift between adjacent reference and data chirps.}
}


\blue{\textit{(ii) Sensing Performance:}}
\red{We next evaluate velocity estimation under relative motion. We use an LRF30M90HS laser distance sensor to obtain the ground-truth velocity \SIrange{5}{30}{\meter} and
derive the reference velocity from its temporal change.}
\red{Since the tested speeds (5 - 25~km/h) exceed the Doppler-unambiguous velocity supported by our radar configuration (\SI{4.7}{\kilo\meter/\hour}) in Sec.~\ref{subsec:implementation_details}, which is limited by the API reconfiguration latency, we do not rely on Doppler FFT. Instead, after interference suppression (Sec.~\ref{sec:interference_suppression}), we estimate radial velocity from the target peak's range displacement across consecutive frames.}
\blue{We compare the radar-estimated velocity with the laser-captured ground
truth \red{using} 
velocity MAE. As shown in Fig.~\ref{fig:dynamic_evaluation}(b),
the MAE remains below \SI{0.25}{\meter/\second} across all tested speeds. The MAE is slightly higher at lower speeds because the frame-to-frame range change is small and is therefore more sensitive to peak-localization errors and residual clutter. These
results show that RadarCom can suppress communication-induced sensing interference
and maintain reliable vehicle-level relative-motion tracking during
simultaneous communication and sensing.}

\blue{$\bullet$ \textbf{Impact of SNR.}
\red{We next evaluate the impact of received SNR on communication.
Using the TI radar's built-in transmit-power attenuation, we create different received-SNR conditions and measure the resulting effective data rate. Since symbol decoding requires both the reference and data chirps, we define the symbol SNR as $\mathrm{SNR}=\min(\mathrm{SNR}_{\text{ref}},\mathrm{SNR}_{\text{data}})$.}
}\blue{Figure~\ref{fig:snr_ser} shows the effective data rate versus SNR (measured at the receiving end). \red{At low SNR (e.g., below 2~dB), symbol decoding errors increases thus reducing the effective throughput, particularly for higher modulation orders. For example, the effective data rate of $M=512$ drops to 1.74~kbps at $\mathrm{SNR}\approx1$~dB, below its maximum achievable rate of 5.56~kbps under our implementation constraints, i.e., the high reconfiguration latency of the TI radar API. As SNR increases, the symbol error rate decreases, enabling the effective data rate to rapidly approach the maximum achievable rate. Above 8~dB, all tested modulation orders nearly achieve their maximum rates, ranging from 3.70~kbps ($M=64$) to 5.56~kbps ($M=512$). These results highlight RadarCom's throughput-robustness tradeoff: higher modulation orders provide higher throughput, while lower modulation orders offer better robustness under low-SNR conditions.}
}
\vspace{-2mm}

\begin{figure*}[t]
    \centering
    \begin{minipage}[t]{0.235\textwidth}
        \centering
        \vspace{0pt}
        \includegraphics[width=0.9\linewidth]{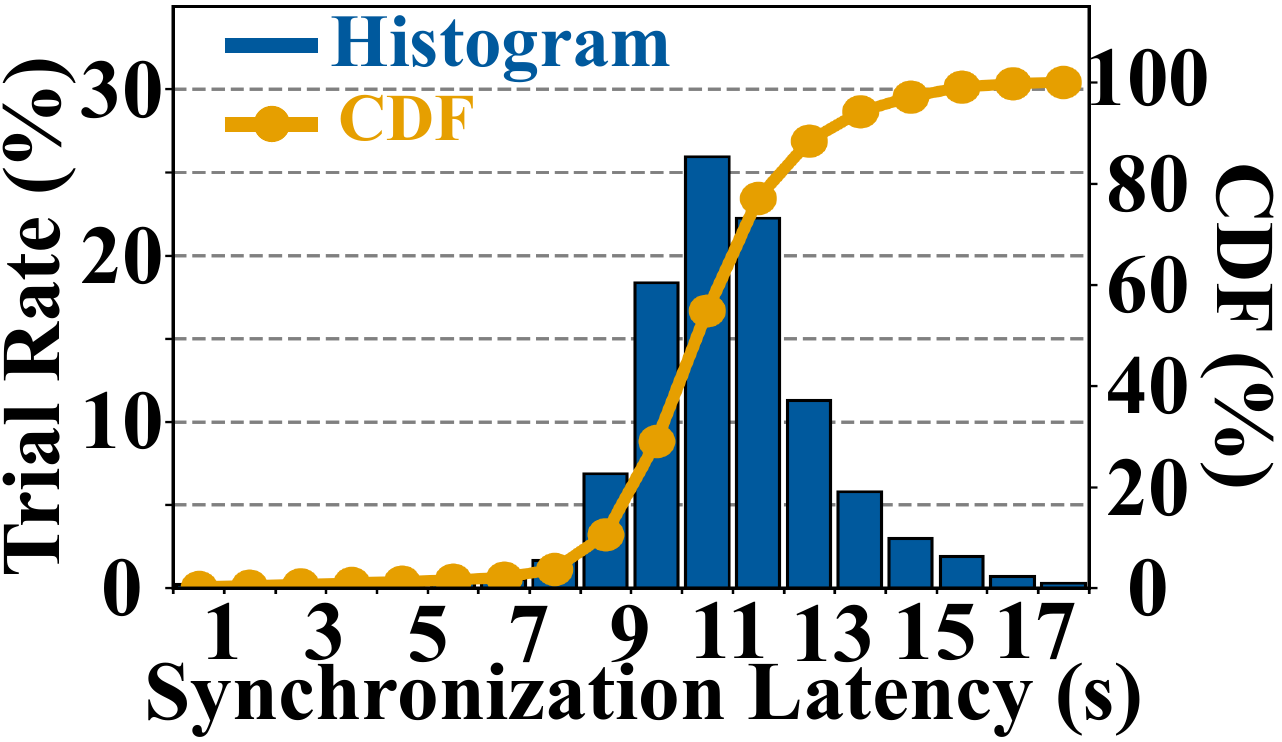}
        \caption{Synchronization latency under cold start.}
        \label{fig:cold_start_distribution}
    \end{minipage}%
    \hspace{0.02\textwidth}%
    \begin{minipage}[t]{0.745\textwidth}
        \centering
        \vspace{0pt}
        \begin{subfigure}[t]{0.315\linewidth}
            \centering
            \includegraphics[width=0.9\linewidth]{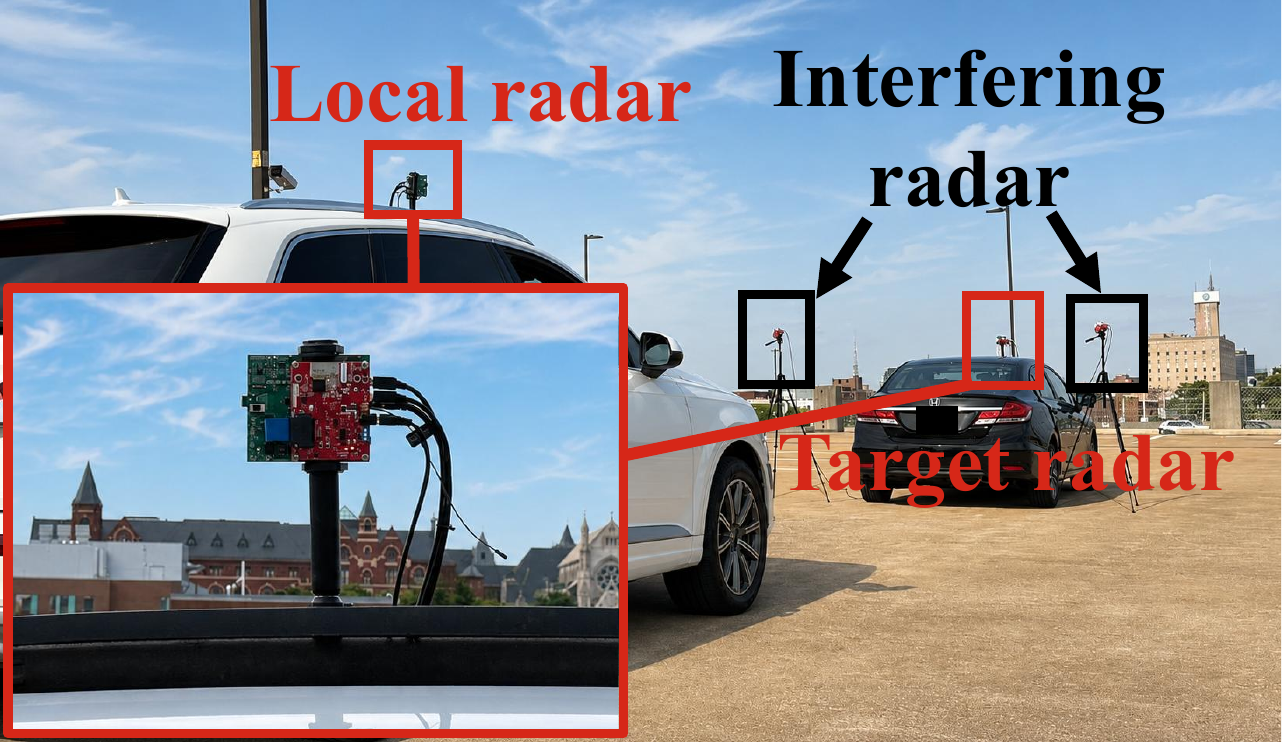}
            \caption{Interference layout.}
            \label{fig:interfering_layout}
        \end{subfigure}\hfill
        \begin{subfigure}[t]{0.315\linewidth}
            \centering
            \includegraphics[width=0.9\linewidth]{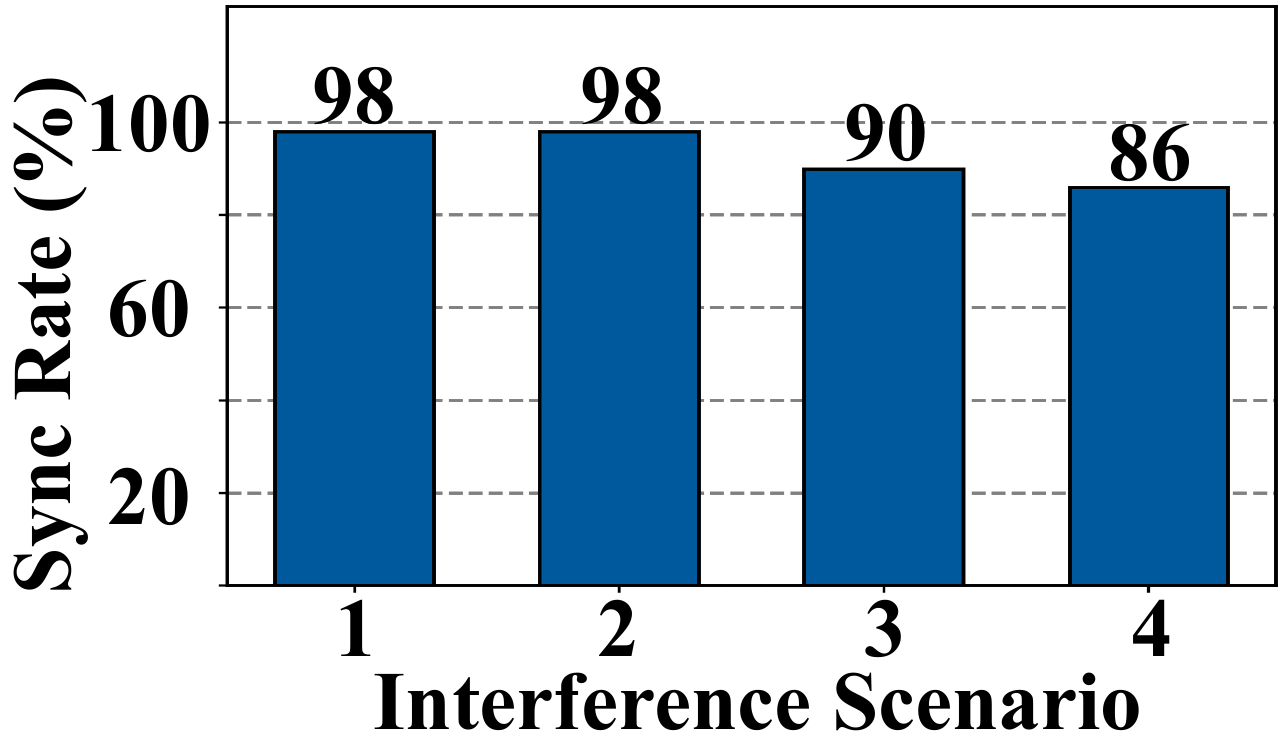}
            \caption{Synchronization rate.}
            \label{fig:int_success}
        \end{subfigure}\hfill
        \begin{subfigure}[t]{0.315\linewidth}
            \centering
            \includegraphics[width=0.9\linewidth]{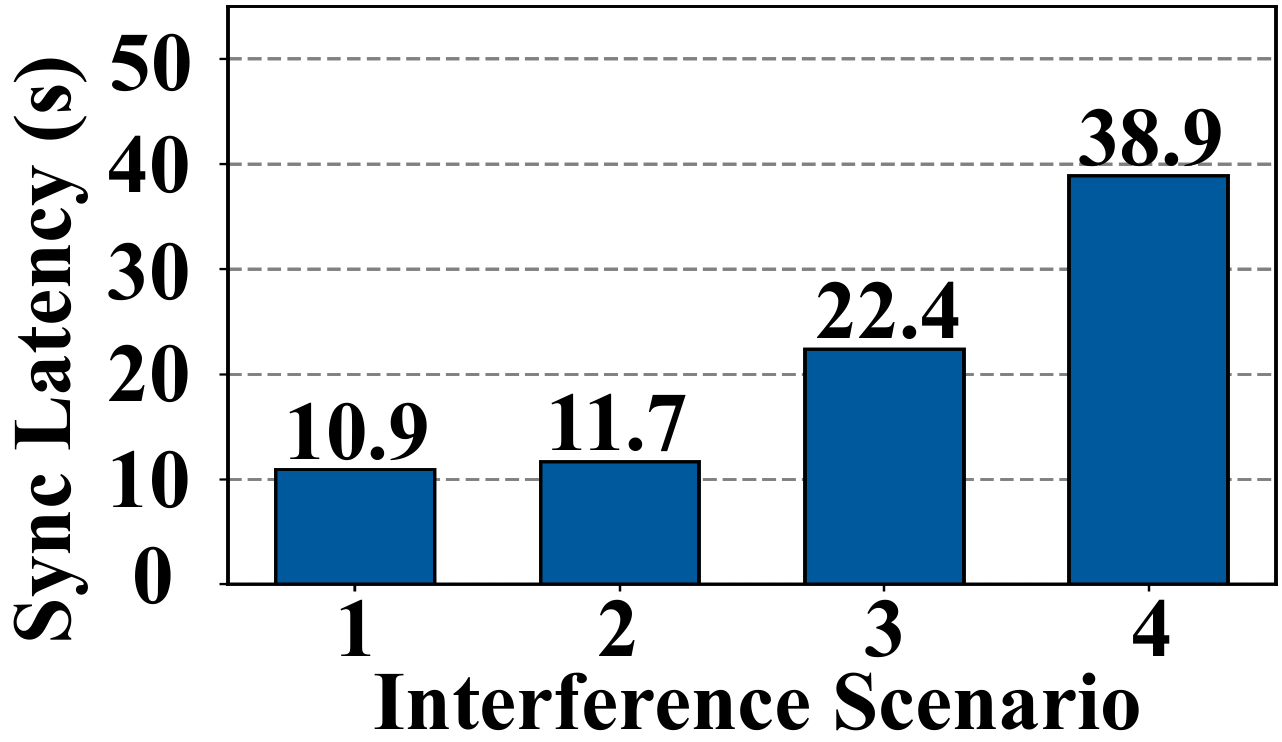}
            \caption{Synchronization latency.}
            \label{fig:int_latency}
        \end{subfigure}
        \vspace{-1mm}
        \caption{Synchronization robustness under environmental radar interference.}
        \vspace{0mm}
        \label{fig:int_results}
    \end{minipage}
            \vspace{-5mm}
\end{figure*}

\vspace{0mm}
\subsection{Robustness Evaluation Results}
\label{subsec:robustness}
\noindent 
\red{While the previous experiments evaluate RadarCom under nominal operating conditions, we next investigate its robustness under practical deployment challenges, including 
independent cold starts, external radar interference, and diverse deployment environments\bluee{, and further examine the impact of API latency}. Besides the communication and sensing performance metrics evaluated above,  we also study synchronization success rate, defined as the percentage of trials that successfully complete discovery, chirp synchronization, and data communication. \blue{For brevity, we denote this metric as \emph{Sync Rate} in the result figures.}}

\vspace{+0.2mm}
$\bullet$ \textbf{Impact of Random Cold Start.}
\red{A practical decentralized deployment requires two radars to autonomously discover and synchronize after independent cold starts, without shared triggers, clocks, or centralized coordination. To evaluate this capability, we conduct approximately 300 outdoor experiments under clear LoS conditions. In each trial, the two radars
are powered independently, and we measure the latency from cold start to
successful synchronization.}
Figure~\ref{fig:cold_start_distribution} shows the histogram and cumulative
distribution function of the synchronization latency. 
\red{Most trials complete within 9 - 12 seconds, with 25\% falling between 10 and 11 seconds. More than 54\% of trials synchronize within 11 seconds, and over 94\% within 14 seconds. The overall synchronization success rate reaches 99\%. These results demonstrate that RadarCom can reliably establish radar-to-radar links after independent cold starts with both high reliability and accurate synchronization latency prediction.}

\vspace{+0.2mm}
$\bullet$ \textbf{Impact of Multi-Radar Interference.}
\label{sec:eval_interference}
\red{We next evaluate RadarCom under multi-radar interference scenario. In practical deployments, multiple nearby radars may operate simultaneously in the same spectral band, potentially interfering with beacon discovery and synchronization. As shown in Fig.~\ref{fig:int_results}(a), we consider four increasingly challenging interference scenarios. The interfering radars either transmit \emph{local-like chirps}, i.e., conventional FMCW chirps without RadarCom beacons, or \emph{target-like beacons}, i.e., RadarCom beacons that follow the same synchronization design as the target radar and can therefore create competing synchronization candidates. Specifically, we evaluate: (1) a single interfering radar transmitting local-like chirps; (2) two interfering radars transmitting local-like chirps; (3) a single interfering radar transmitting target-like beacons; and (4) two interfering radars transmitting target-like beacons. For each scenario, we perform 50 trials to obtain the synchronization success rate and the synchronization latency, where synchronization attempts taking longer than 1 minute are considered failures.
}

\vspace{+0.2mm}
\red{Figure~\ref{fig:int_results}(b) shows that the synchronization success rate remains high across all scenarios, achieving 98\% in Scenarios 1 and 2, 90\% in Scenario 3, and 86\% in Scenario 4. Figure~\ref{fig:int_results}(c) illustrates the synchronization latency, where we observe that local-like chirp interference has little impact, yielding average latencies of 10.9~s and 11.7~s, because it does not generate valid synchronization candidates. In contrast, target-like beacon interference increases the latency to 22.4~s and 38.9~s in Scenarios 3 and 4, respectively. Such an increase is expected because the local radar must distinguish the intended target from competing beacon sources during the discovery phase when multiple RadarCom pairs operate simultaneously. Nevertheless, RadarCom still achieves synchronization success rate of 90\% and 86\%, enabled by the \blue{post-synchronization verification mechanism} (Sec.~\ref{sec:differential_coding}), which reliably identifies the intended target.
Overall, multi-radar interference primarily introduces a modest increase in discovery latency rather than causing synchronization failures.}

\begin{figure*}[t]
    \centering

    \begin{minipage}[t]{0.392\linewidth}
        \centering
        \vspace{0pt}

        \begin{subfigure}[t]{0.49\linewidth}
            \centering
            \vspace{0pt}
            \includegraphics[width=\linewidth]{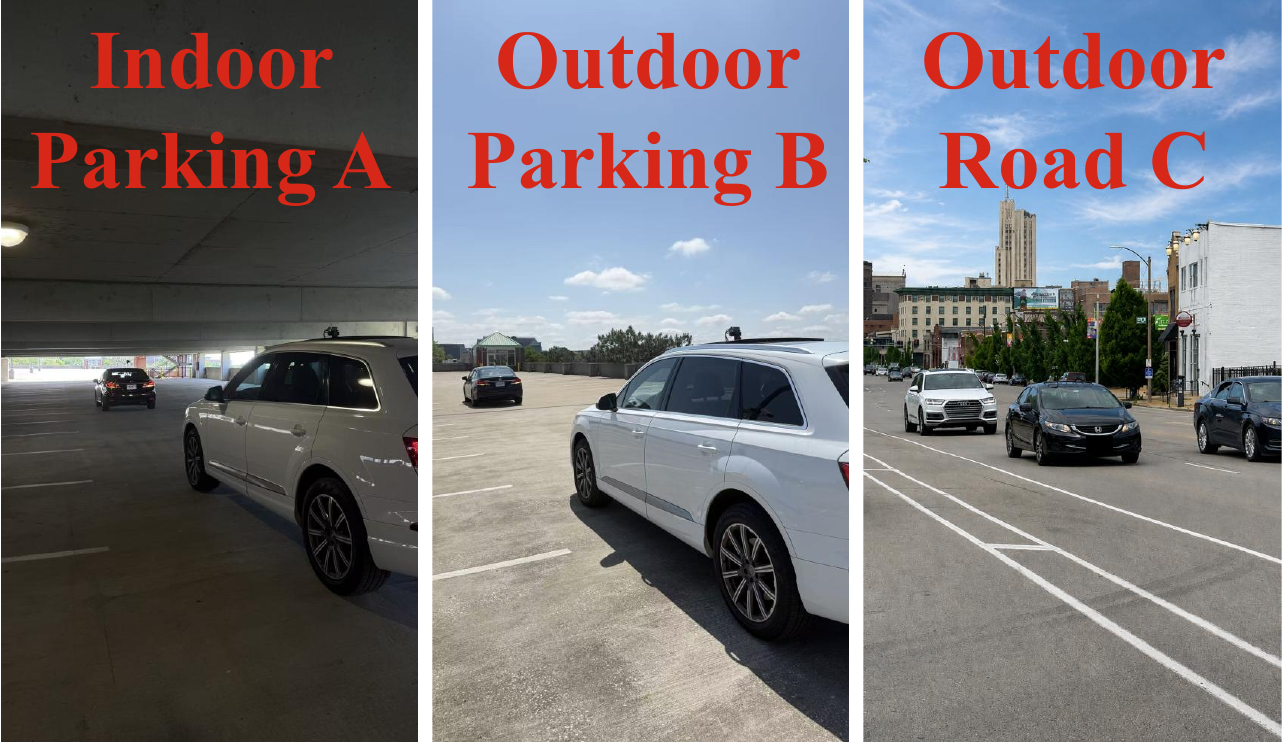}
            \subcaption{Different scenarios.}
            \label{fig:different_env_a}
        \end{subfigure}%
        \hfill
        \begin{subfigure}[t]{0.495\linewidth}
            \centering
            \vspace{0pt}
            \includegraphics[width=\linewidth]{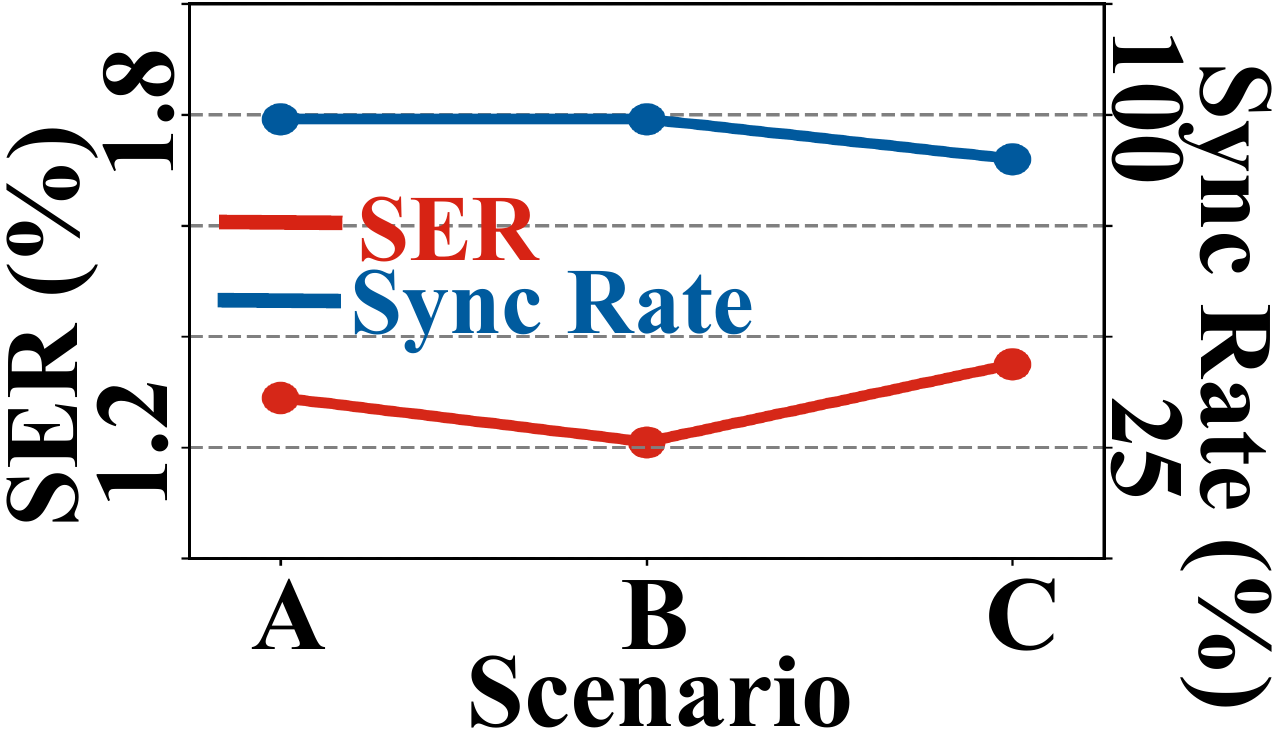}
            \subcaption{Impact of scenarios.}
            \label{fig:different_env_b}
        \end{subfigure}

        \vspace{-1mm}
        \caption{Cross-environment evaluation.}
        \label{fig:different_env}
    \end{minipage}%
    \hspace{0.01\linewidth}%
    \begin{minipage}[t]{0.196\linewidth}
        \centering
        \vspace{0pt}

        \includegraphics[width=\linewidth]{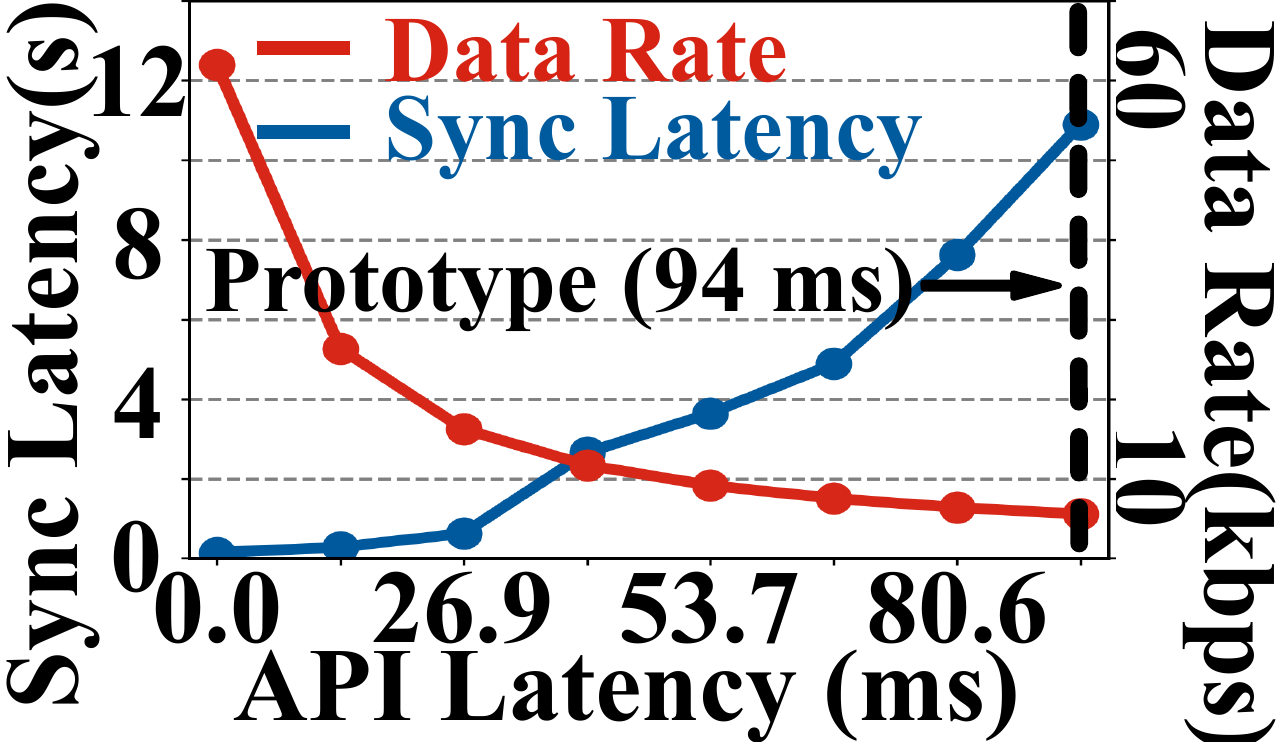}

        \vspace{-1mm}
        \caption{\bluee{Impact of API latency.}}
        \label{fig:api_latency}
    \end{minipage}%
    \hspace{0.01\linewidth}%
    \begin{minipage}[t]{0.392\linewidth}
        \centering
        \vspace{0pt}

        \begin{subfigure}[t]{0.49\linewidth}
            \centering
            \vspace{0pt}
            \includegraphics[width=\linewidth]{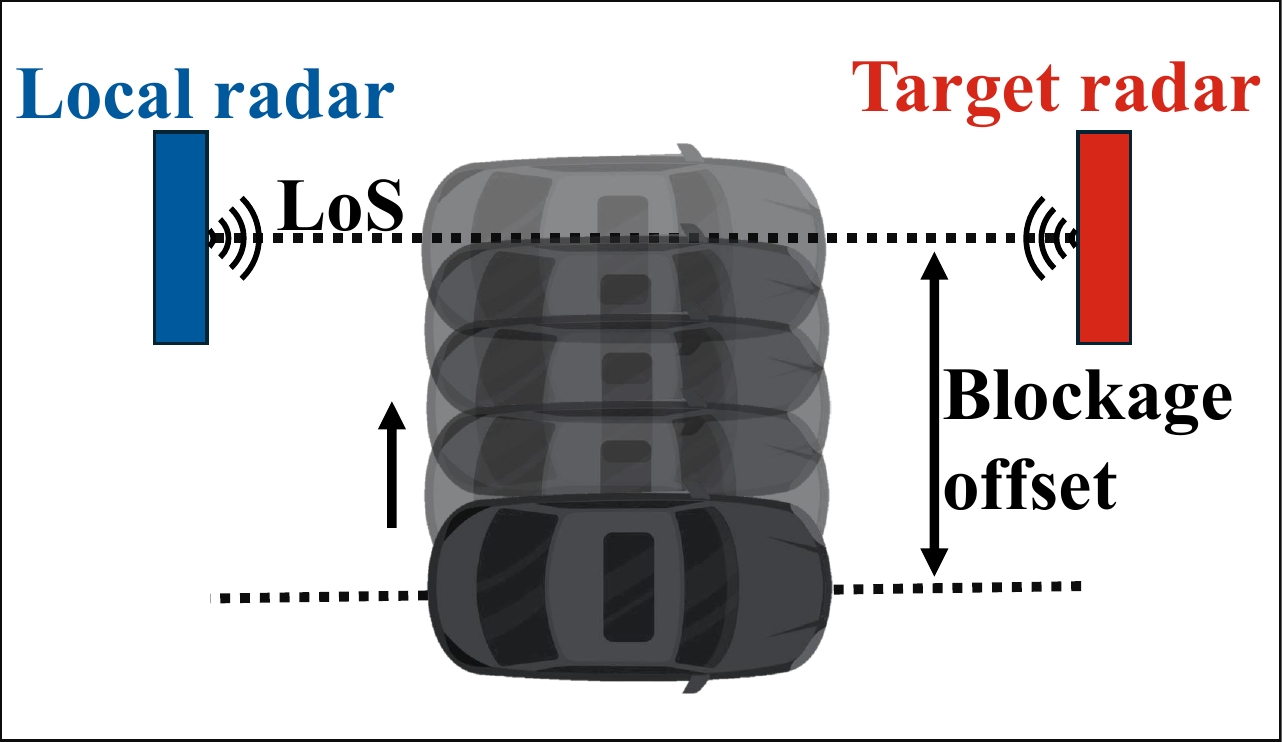}
            \subcaption{Blockage layout.}
            \label{fig:blockage_layout}
        \end{subfigure}%
        \hfill
        \begin{subfigure}[t]{0.49\linewidth}
            \centering
            \vspace{0pt}
            \includegraphics[width=\linewidth]{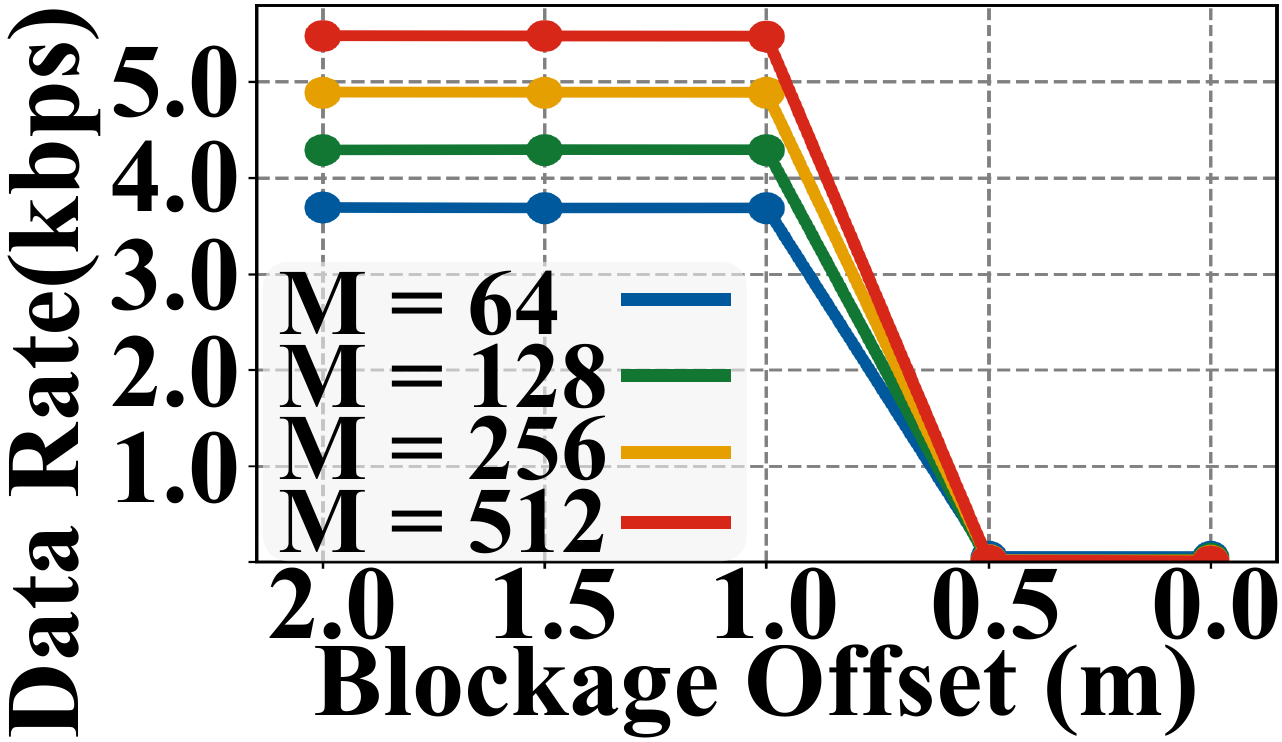}
            \subcaption{Data Rate.}
            \label{fig:block_comm}
        \end{subfigure}

        \vspace{-1mm}
        \caption{Impact of Non-LoS.}
        \label{fig:block_results}
    \end{minipage}

    \vspace{-1mm}
\end{figure*}

$\bullet$ \textbf{Impact of Different Deployment Scenarios.}
\label{sec:eval_different_env}
We further evaluate RadarCom in different deployment environments to study its end-to-end performance under more realistic driving conditions. Unlike controlled static experiments, both the local and target vehicles move during the test, introducing relative motion and time-varying propagation conditions. As shown in Fig.~\ref{fig:different_env}(a), we consider three representative scenarios: Indoor Parking A, Outdoor Parking B, and Outdoor Road C. Indoor Parking A emulates enclosed environments such as underground parking lots or tunnels, where conventional infrastructure-assisted communication may be unavailable. Outdoor Parking B represents an open-area LoS environment with relatively limited surrounding traffic. Outdoor Road C evaluates RadarCom on a real road with more realistic vehicle motion and environmental dynamics.
Figure~\ref{fig:different_env}(b) shows the SER and synchronization success rate in the three scenarios. RadarCom maintains low SER across all scenarios, achieving 1.29\%, 1.21\%, and 1.35\% in Indoor Parking A, Outdoor Parking B, and Outdoor Road C, respectively. 
The synchronization success rate remains 99\% in both parking scenarios and decreases to 90\% on the outdoor road. 
\red{This reduction is expected due to stronger environmental dynamics and less predictable vehicle motion during the discovery and synchronization phases.}
Overall, RadarCom maintains low SER across all scenarios, indicating that real-road operation primarily affects initial synchronization rather than communication reliability after synchronization.

$\bullet$ \textbf{\bluee{Impact of} API Latency.}
\bluee{It is worthy to note that, the performance of RadarCom across all the above experiments is fundamentally bounded by the prototype platform. \red{As discussed in Sec.~\ref{subsec:implementation_details}, each frame requires approximately 94~ms for API updates and 7~ms for host computation, resulting in a frame period of 101.592~ms. With a fixed number of chirps per frame, this extended frame period also results in a longer chirp period.} Meanwhile, the ADC sampling window is only 51.2~$\mu$s  per chirp, covering less than 7\% of the chirp period. This sparse observation reduces opportunities to capture an asynchronous beacon, leading to long cold-start latency, while the extended frame period limits the data rate.}

\bluee{Since the API latency cannot be directly varied, \textcolor{red}{we emulate different API latencies by preloading the chirp configurations into radar memory, allowing the radar to \greenn{perform beacon detection} without real-time API calls.} 
\greenn{We then adjust the frame periods to vary the emulated API latency from 0 to 94 ms. In each trial, both radars are cold-started, and we measure the synchronization latency as the time from cold start}
\textcolor{red}{until the beacon is successfully detected for the first time.}}
\bluee{For communication, 
\greenn{we evaluate the effective data rate under the same emulated API latencies by comparing the received symbols with the known transmitted sequence.}
As shown in Fig.~\ref{fig:api_latency}, reducing the emulated API latency from 94 ms to 0 ms decreases cold-start synchronization latency from 10.9 s to 0.15 s, while increasing the effective data rate from 5.55 kbps to 61.92 kbps.}

\vspace{-2mm}
\section{Discussion}
\noindent $\bullet$ \textbf{\bluee{Potential Use Cases.}} 
\greenn{The above evaluation demonstrates that RadarCom can reliably exchange data between vehicles while preserving native radar sensing. This capability enables RadarCom to support a variety of practical V2V applications, \textcolor{red}{such as exchanging} 
vehicle speed, safety warnings, and road condition information using the same COTS radar hardware. One representative example is the Basic Safety Message (BSM), which vehicles periodically broadcast to share their position, speed, heading, and brake status~\cite{automotive2021study}. Each BSM is approximately 300 bytes, \textcolor{red}{and its broadcast frequency can be reduced to 2 Hz in dense freeway traffic to avoid channel congestion, requiring a data rate of $4.8$ kbps. The current RadarCom prototype achieves 5.55 kbps, sufficient to support this BSM rate. As shown in Sec.~\ref{subsec:robustness}, eliminating the emulated API latency increases the effective data rate to 61.92 kbps, exceeding the 24 kbps required for BSM transmission at the full 10 Hz broadcast rate ~\cite{automotive2021study}.}}

$\bullet$ \textbf{\bluee{Operation under Non-LoS.}}
RadarCom currently relies on direct LoS propagation and therefore faces a common challenge in mmWave radar systems~\cite{he2024see,mehrotra2024hydra}: performance can degrade significantly under severe blockage because mmWave signals are strongly attenuated by surrounding vehicles and metallic obstacles. \greenn{As shown in Fig.~\ref{fig:block_results}, RadarCom maintains near-LoS performance under partial blockage, but the effective data rate drops significantly when the blocking vehicle obstructs the direct path.} \bluee{One potential solution is to establish alternative communication links through environmental reflections, inspired by recent mmWave sensing systems that exploit ground-reflected paths~\cite{he2024see} and multi-bounce scattering for beyond-FoV sensing
\cite{mehrotra2024hydra}. Unlike these sensing systems, which rely on round-trip reflections, 
\textcolor{red}{RadarCom requires only one-way signal propagation between the two radars, potentially reducing propagation loss over reflected paths.}
\textcolor{red}{Moreover, RadarCom's differential coding scheme can mitigate path-dependent \greenn{delay} variations, as the reference and data chirps propagate through the same path, allowing their common \greenn{frequency} shifts to be canceled during decoding.} }

\begin{figure*}[t]
    \centering

    \begin{subfigure}[t]{0.192\linewidth}
        \centering
        \vspace{0pt}
        \includegraphics[width=\linewidth]{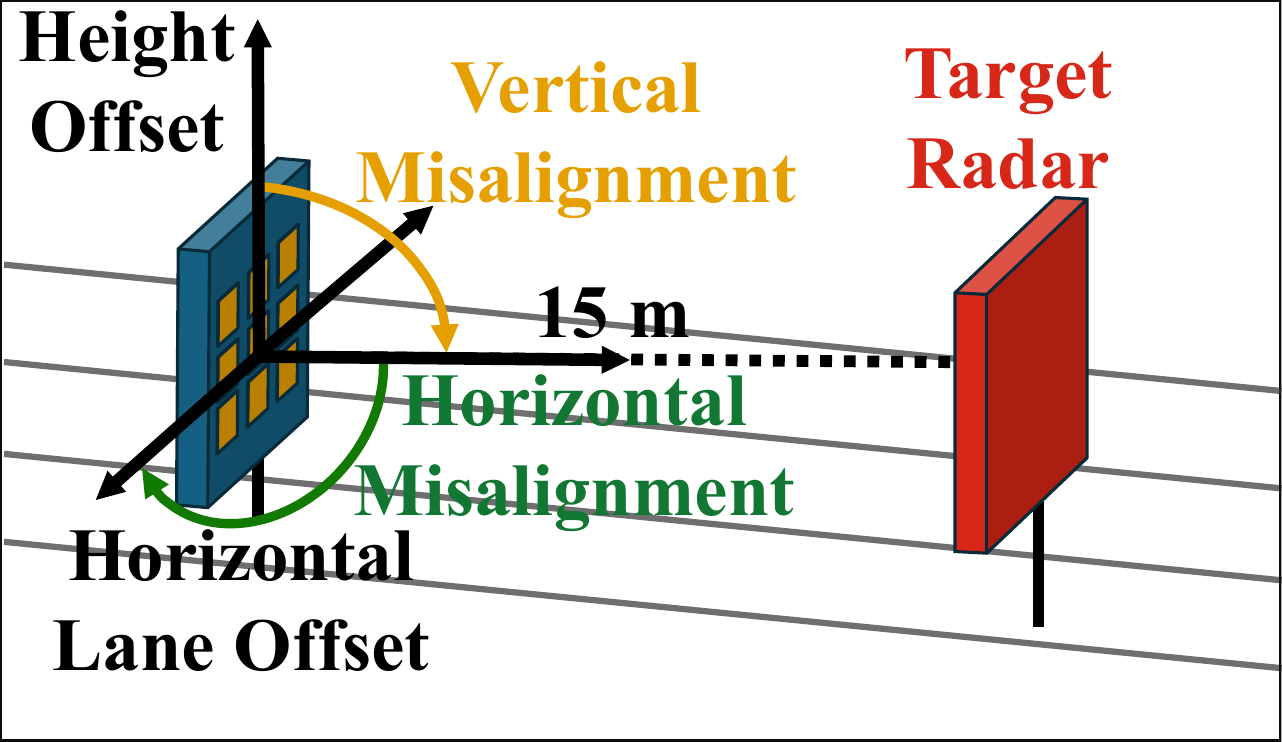}
        \subcaption{\bluee{Experiment layout.}}
        \label{fig:angle_layout}
    \end{subfigure}%
    \hfill
    \begin{subfigure}[t]{0.192\linewidth}
        \centering
        \vspace{0pt}
        \includegraphics[width=\linewidth]{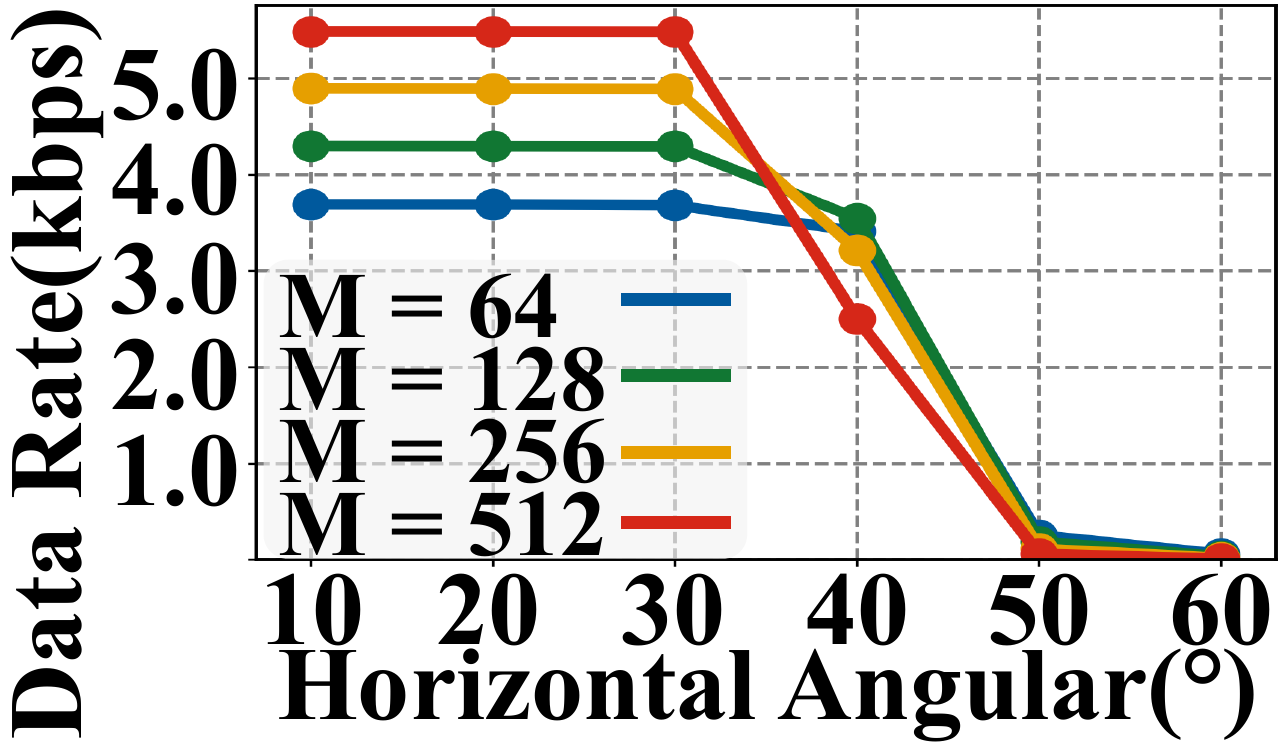}
        \subcaption{Horizontal angle.}
        \label{fig:angle_comm_rotation}
    \end{subfigure}%
    \hfill
    \begin{subfigure}[t]{0.192\linewidth}
        \centering
        \vspace{0pt}
        \includegraphics[width=\linewidth]{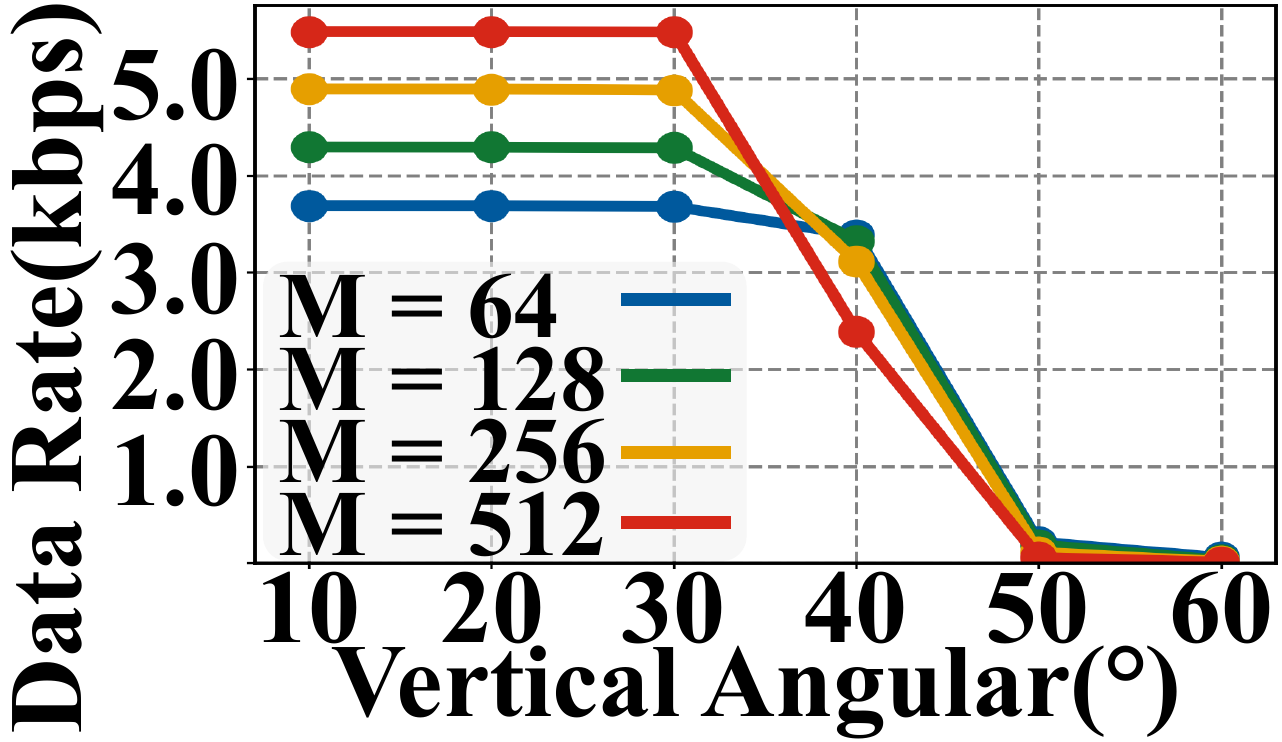}
        \subcaption{\bluee{Vertical angle.}}
        \label{fig:vertical_angular}
    \end{subfigure}%
    \hfill
    \begin{subfigure}[t]{0.192\linewidth}
        \centering
        \vspace{0pt}
        \includegraphics[width=\linewidth]{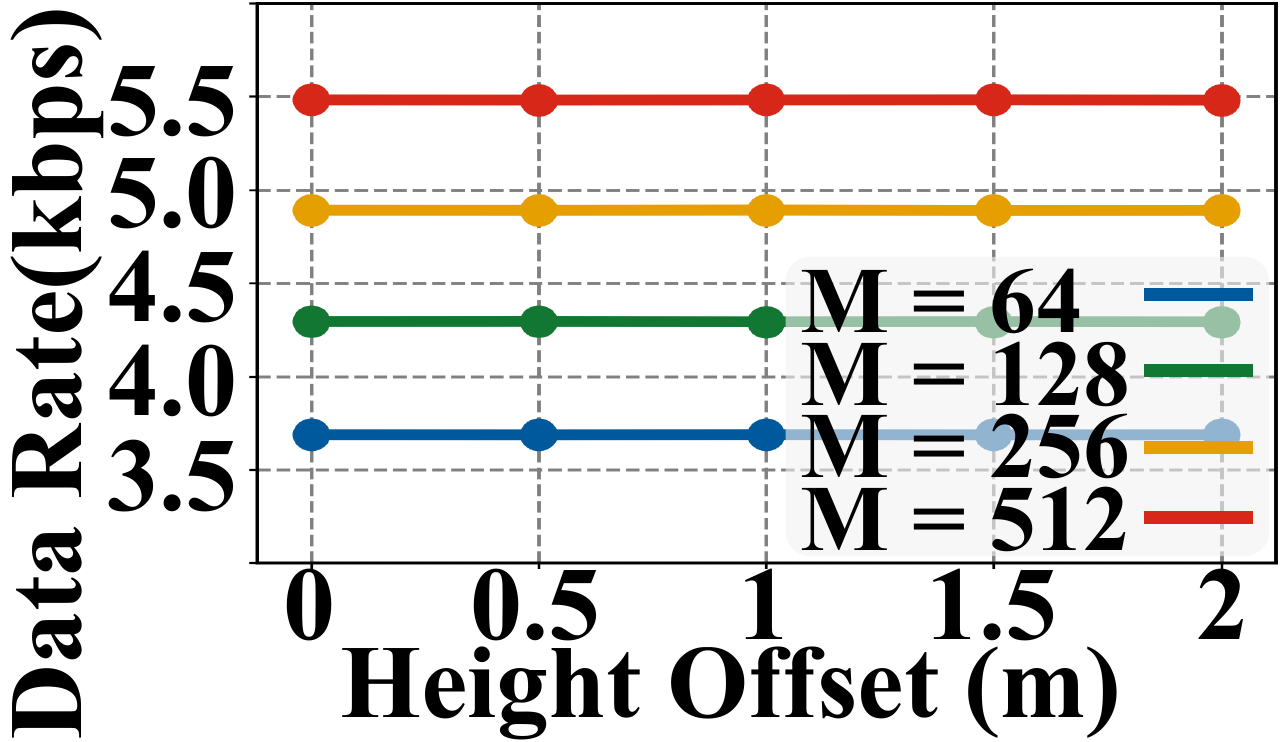}
        \subcaption{\bluee{Height offset.}}
        \label{fig:vertical_offset}
    \end{subfigure}
    \begin{subfigure}[t]{0.192\linewidth}
        \centering
        \vspace{0pt}
        \includegraphics[width=\linewidth]{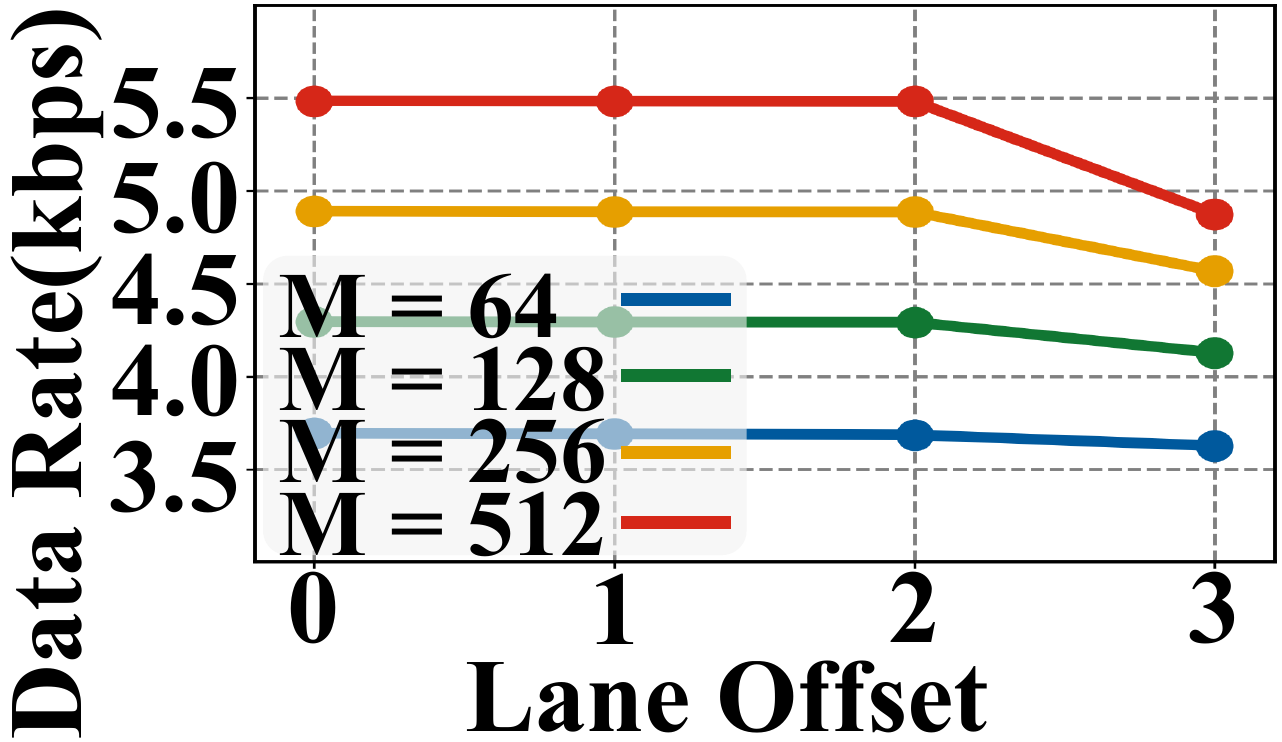}
        \subcaption{\bluee{Lane offset.}}
        \label{fig:horizontal_offset}
    \end{subfigure}%
    \hfill

    \vspace{-1mm}
    \caption{\bluee{Performance under different radar placements.}}
    \label{fig:angle_results}
    \vspace{0mm}
\end{figure*}

$\bullet$ \textbf{\bluee{Operation under Diverse Placements.}} \bluee{As shown in Fig.~\ref{fig:angle_results}(a), we further evaluate RadarCom under diverse radar placements, keeping the two radars 15~m apart along the LoS while varying the horizontal and vertical angular misalignment, the mounting height offset, and the horizontal lane offset. Fig.~\ref{fig:angle_results}(b) and~\ref{fig:angle_results}(c) show similar trends for horizontal and vertical angular misalignment: 
the effective data rate remains close to the LoS case within $30^\circ$, decreases significantly at $40^\circ$, and drops to nearly zero at $50^\circ$ as the local radar moves outside the FoV. 
\textcolor{red}{Higher-order modulation schemes exhibit greater sensitivity to angular misalignment. In contrast, Fig.~\ref{fig:angle_results}(d) shows that mounting height offsets of up to $2$~m have a negligible impact on the effective data rate.}
Finally, Fig.~\ref{fig:angle_results}(e) evaluates a practical deployment where the two vehicles drive in different lanes. \textcolor{red}{The effective data rate remains stable for horizontal lane offsets of up to two lanes, with only a slight reduction at a three-lane offset of approximately $11$~m.} 
}

$\bullet$ \textbf{Duplex and Multi-Radar Communication.}
The current prototype focuses on validating the physical-layer feasibility of RadarCom. Bidirectional communication is supported by alternating the transmitting role between the local and target radars, resulting in a half-duplex, time-division design. \bluee{Moreover, the existing mechanisms of RadarCom already lay the foundation for multi-radar communication. During discovery, the SFO-based target identification (Sec.~\ref{discovery}) distinguishes multiple radars even when they use the same beacon pattern. As shown in Sec.~\ref{subsec:robustness}, RadarCom reliably identifies and synchronizes with the target radar even in the presence of two additional interfering radars. After synchronization, signals from different radars can be separated by angle using the native receiver antenna array~\cite{ti_awr2243_cascade}. When angular separation is insufficient, 
e.g., for \greenn{signals} from different radars arriving from similar directions, RadarCom can adopt time-division multiple access (TDMA) \greenn{, i.e., allocate different time slots to frames from different radars, naturally} 
extending current time-division design. \greenn{Alternatively, \textcolor{red}{RadarCom can employ frequency-division multiple access (FDMA) by assigning different frequency sub-bands within the receiver's intermediate-frequency band to different radar communication links.}} Similar time-frequency coordination methods among automotive radars have been explored for interference mitigation~\cite{alland2019interference, 9337437}, \textcolor{red}{suggesting a promising direction for extending RadarCom to support multi-radar communication.} }

\vspace{-2mm}
\section{Related Work}
\noindent Existing ISAC and radar communication systems face several deployment barriers on COTS automotive radars. Hardware- or waveform-centric designs \cite{okubo2024integrated,bae2022omniscatter,gaofeng2024co,faghih2025combining,xu2024dual} require customized architectures incompatible with mass-produced platforms. Cooperative radar systems \cite{9209195,meng2024cooperative,alemdar2021rfclock,zhang2022integration} assume centralized time-frequency synchronization (e.g., GPS), which is impractical across independent vehicular radars. Hybrid approaches that pair radar sensing with Wi-Fi or mmWave networking \cite{aydogdu2020distributed,choi2016millimeter,hossain2019hybrid,kim2019experimental,coll2018evaluation} introduce additional radios or infrastructure, limiting decentralized scalability. A recent prototype \cite{sanchez2025radar} demonstrates radar-to-radar communication on commercial devices but relies on dedicated modulation and separates sensing from communication. In contrast, our system enables fully integrated sensing and communication directly on unmodified COTS automotive radars, without hardware modification, external synchronization, or additional radios.

\vspace{-2mm}
\section{Conclusion}
\vspace{0pt}
\noindent \red{In this paper, we present RadarCom, a software-defined system that enables direct radar-to-radar communication on unmodified COTS automotive mmWave radars while preserving native sensing functionality. By repurposing mutual inter-radar interference as a communication carrier and introducing mechanisms for discovery, synchronization, communication, and sensing-aware interference suppression, RadarCom establishes reliable communication between independent radars without hardware modification, external synchronization support, or additional communication radios. We implement RadarCom on TI AWR2243 automotive radars and evaluate it under static, dynamic, and real-world conditions. Experimental results show that RadarCom prototype achieves an average effective data rate of 5.55~kbps while maintaining high sensing accuracy, demonstrating the feasibility of transforming widely deployed automotive radars into dual-purpose communication and sensing devices.}

\FloatBarrier 

\bibliographystyle{ACM-Reference-Format}

\bibliography{reference}

\end{document}